\documentclass[epj]{svjour}

\usepackage{amsmath}
\usepackage{amssymb}
\usepackage{color}
\usepackage{graphicx}
\usepackage{hyperref}
\usepackage{tablefootnote}
\usepackage{booktabs}

\begin{document}
\title{Linear polarization--direction correlations in $\gamma$-ray scattering experiments}
\author{Christian Iliadis\inst{1,2} \and Udo Friman-Gayer\inst{1,2}
}

\institute{
  Department of Physics \& Astronomy, University of North Carolina at Chapel Hill, NC 27599-3255, USA
  \and Triangle Universities Nuclear Laboratory (TUNL), Duke University, Durham, North Carolina 27708, USA
  }

\date{Received: date / Revised version: date}

\abstract{
Scattering measurements with incident linearly polarized $\gamma$ rays provide information on spins, parities, and $\gamma$-ray multipolarity mixing coefficients, and, therefore, on the nuclear matrix elements involved in the transitions. We present the general formalism for analyzing the observed angular correlations. The expressions are used to compute three-dimensional radiation patterns, which are important tools for optimizing experimental setups. Frequently, $\gamma$-ray transitions can proceed via two multipolarities that mix coherently. In such cases, the relative phases of the nuclear matrix elements are important when comparing results from different measurements. We discuss different phase conventions that have been used in the literature and present their relationships. Finally, we propose a basic experimental geometry consisting of detectors located at four different spatial locations. For this geometry, we present the measured anisotropies of the emitted $\gamma$ rays in graphical format as an aid in the data analysis.
}

\maketitle

\tableofcontents

\section{Introduction} \label{sec:intro}
Scattering measurements with incident linearly polarized $\gamma$ rays provide valuable information on spins, parities, branching ratios, and $\gamma$-ray multipolarity mixing coefficients, and, thus, on the matrix elements involved in the transition between two levels. Such resonant $\gamma$-ray scattering (or Nuclear Resonance Fluorescence \cite{Metzger:2015uu,Alder:1975ug,kneissl96}) measurements have surged over the past decade because of the availability of linearly polarized quasi-monoenergetic $\gamma$-ray beams provided by laser Compton $\gamma$-ray facilities (see Ref.~\cite{ying20} for a review). 

Detector locations are crucial for measuring the radiation anisotropies containing the essential information about quantum numbers and mixing ratios mentioned above. While the formalism for $\gamma\gamma$ angular correlations has been presented before \cite{devonsgoldfarb57,Fagg:1959wo,biedenharn60,Rose:1967vq}, three-dimensional visualizations of the radiation patterns have only been published for the simplest cases when $\gamma$-ray transitions proceed via a unique multipolarity (see, {\it e.g.}, Refs.~\cite{Tonchev2005,Ur2016}). Such visualizations are important tools for the optimization of experimental setups.  
  
When more than one multipolarity is involved in a given transition, the multipolarities mix coherently and the total radiation pattern is sensitive to the mixing ratio. However, different phase conventions for the reduced matrix elements connecting the initial and final nuclear states have been adopted in different formalisms. This problem has been pointed out since the 1950's \cite{huby54}, but is unfortunately still causing confusion in the published literature.

In this review, we are pursuing several goals. First, the formalism of $\gamma\gamma$ angular correlations with incident linearly polarized $\gamma$ rays is reviewed and the issue of different phase conventions is discussed. Second, three-dimensional representations of the radiation patterns in resonant $\gamma$-ray scattering experiments are presented for a range of spins, parities, and $\gamma$-ray multipolarity mixing ratios. Third, a detection geometry consisting of a minimum of four detectors is proposed and graphs are presented for analyzing the measured anisotropies.

We will summarize the formalism in Sect. \ref{sec:formalism}, comment on the confusing problem of the phase convention in Sect.~\ref{sec:phases}, discuss an illustrative example in Sect.~\ref{sec:example}, and present three-dimensional representations of radiation patterns in Sect.~\ref{sec:results}. The angular correlation formalism for the case of an unobserved intermediate $\gamma$ ray is given in Sect.~\ref{sec:unobs}. Detector configurations and data analysis are discussed in Sect.~\ref{sec:results2} and Sect.~\ref{sec:analyzing_powers_use}, respectively. A concluding summary is provided in Sect.~\ref{sec:summary}. Radiation patterns for the most common spin sequences are shown in App.~\ref{app:pattern}. Graphs displaying analyzing powers are presented in App.~\ref{app:powers} as an aid for the data analysis.

\section{Formalism for a two-step angular correlation} \label{sec:formalism}
\begin{figure}
  \includegraphics[width=1.0\columnwidth]{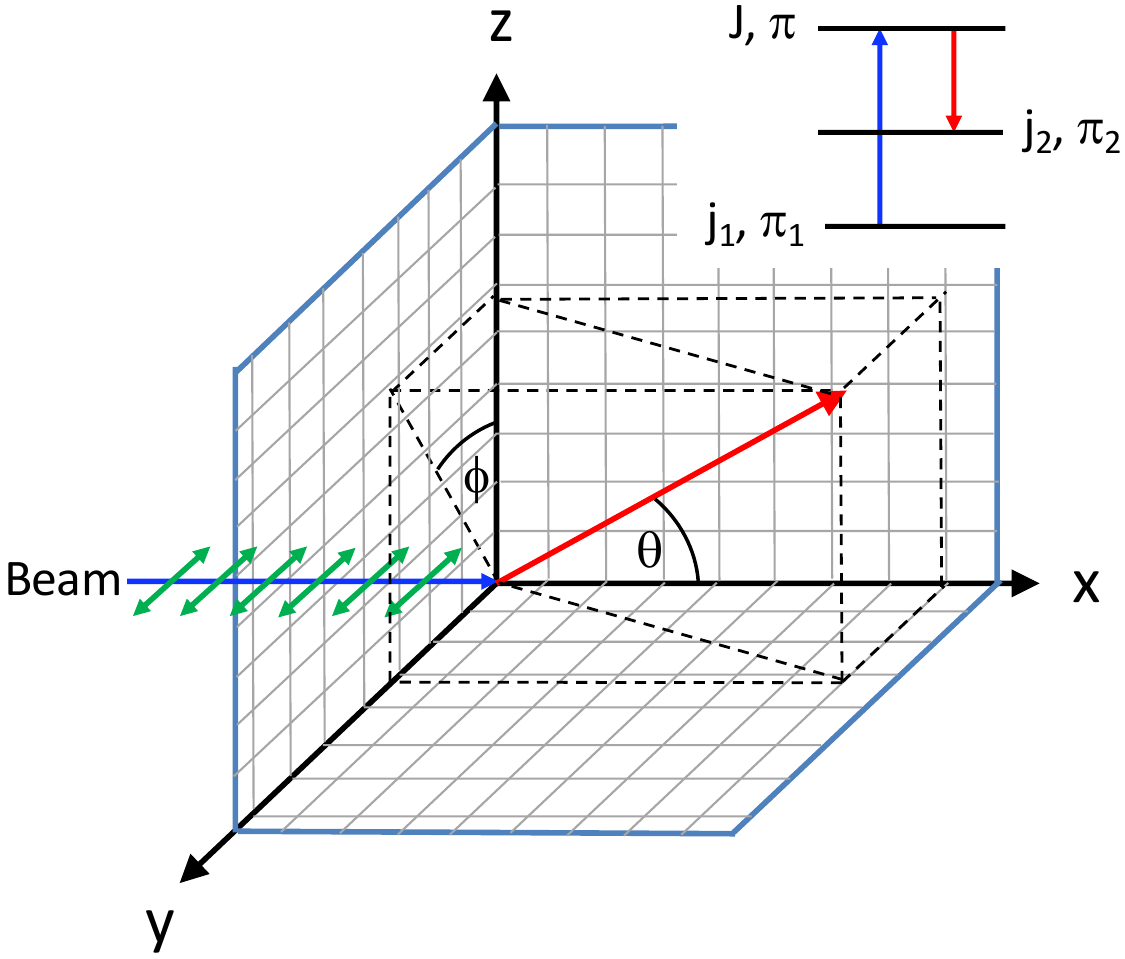}
  \caption{\label{fig:geo1} (Color online) Geometry of a $\gamma$-ray scattering experiment. A target nucleus with spin $j_1$ and parity $\pi_1$ is located at the origin. The incident linearly polarized $\gamma$-ray beam (shown in blue) moves along the positive $x$ direction with its polarization vector pointing parallel to the $y$-axis (green arrows). It is absorbed by the target nucleus, causing a transition to an excited level of spin $J$ and parity $\pi$. The decay to a lower-lying state with spin $j_2$ and parity $\pi_2$ proceeds via emission of another $\gamma$ ray (shown in red). The angle between the incident beam direction and that of the scattered $\gamma$ ray is $\theta$ (polar angle). The angle $\phi$ is defined between the $z$-axis and the projection of the scattered $\gamma$-ray direction onto the $y$--$z$ plane (azimuthal angle).
  }
\end{figure}

The spatial geometry is given in Fig.~\ref{fig:geo1}. A linearly polarized $\gamma$-ray beam (blue arrow) is incident along the positive $x$-axis on an assembly of target nuclei with spin $j_1$ and parity $\pi_1$ located at the origin of the coordinate system. The $\gamma$-ray polarization ({\it i.e.}, electric field) vector, $\vec{E}$, points parallel to the $y$-axis (green arrows). Absorption of an incident $\gamma$ ray will cause a transition from the ground state to an intermediate (excited) state with spin $J$ and parity $\pi$. This intermediate state subsequently decays by emission of a second $\gamma$ ray (shown in red) to a final state of spin $j_2$ and parity $\pi_2$ that is not necessarily the same as the initial state. We assume throughout this review that all nuclear states involved in this process are isolated, have unique spin values and well-defined parities, and that the target nuclei are randomly oriented. For cases where the intermediate state does not have a definite spin or parity, see Refs.~\cite{devonsgoldfarb57,biedenharn60,Rose:1953tw}.

The process just described results in general in an anisotropic emission of the second $\gamma$ ray. The reason can be understood conceptually by first considering an excited nuclear level that has all its magnetic substates populated equally by some process. In that case, each magnetic substate emits anisotropically, but on summing over these substates, with equal populations and random relative phases, the sum is incoherent and the total intensity becomes independent of emission angle. However, when the level is excited by an incident beam of precisely defined momentum and polarization, the process will single out directions in space, which gives different weights to the various substates of the excited nuclear level. Since the substates are now populated unequally (i.e., the excited nuclear level is ``oriented''), the $\gamma$ rays originating from the decaying state $J^\pi$ are emitted anisotropically, {\it i.e.}, their emission direction is correlated with the direction of the incident $\gamma$ rays. 

The observed angular correlation\footnote{In the literature, the terms ``angular correlation'' and ``angular distribution'' are sometimes used interchangeably, while at other times a distinction is being made. The former expression has been applied to the emission of successive radiations from excited states, while the latter is frequently used in connection with nuclear reactions \cite{devonsgoldfarb57}. As will be seen, the formalism presented here applies equally to the absorption and emission of radiation. Since ``angular correlation'' is the more general term, it will be used throughout this work.} is completely determined by the quantum mechanical rules for the coupling of angular momenta and parities. If the incident beam is unpolarized, the measured correlation is between the directions of the incident (absorbed) and emitted $\gamma$ rays. Therefore, this process is referred to as ``direction--direction correlation.'' On the other hand, if the incident beam is polarized, both its polarization as well as the direction of motion will impact the anisotropy of the emitted radiation. This process is called ``polarization--direction correlation.''

\begin{figure}
  \includegraphics[width=0.8\columnwidth]{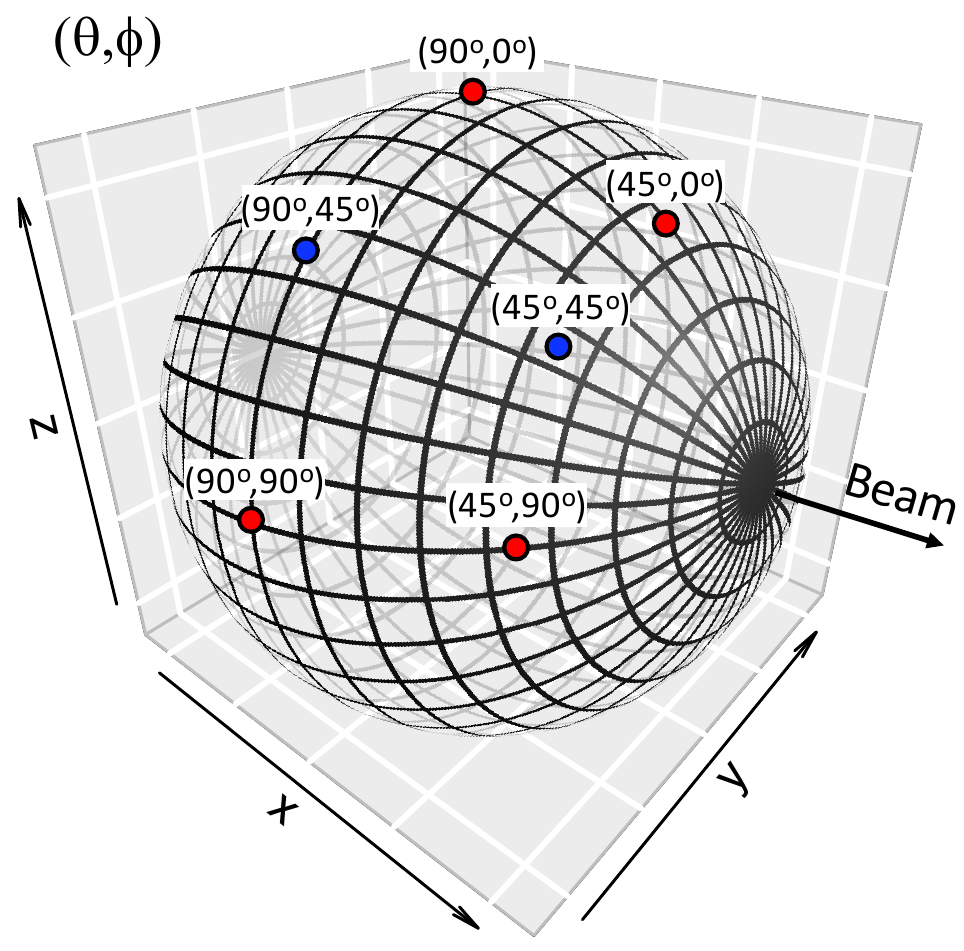}
  \caption{\label{fig:geo2} (Color online) Examples for angle combinations $(\theta,\phi)$ on the surface of a unit sphere. The incident linearly polarized $\gamma$-ray beam moves along the direction of the positive $x$-axis. Notice that $\phi$ is not the angle between the $z$-axis and the outgoing radiation direction, but between the $z$-axis and the projection of the outgoing radiation direction onto the $y$--$z$ plane (see Fig.~\ref{fig:geo1}). 
  }
  \end{figure}

The angle between the incident and emitted $\gamma$ rays is denoted by $\theta$ (polar angle). The angle $\phi$ is between the $z$-axis and the projection of the direction of the second $\gamma$ ray onto the $y$--$z$ plane (azimuthal angle). Or, stated differently, $\phi$ is the angle between the plane of the $\vec{E}$ vector of the incident $\gamma$ ray and the plane defined by the direction of this incident $\gamma$ ray and the normal to the scattering plane. The angular ranges are $\theta$ $=$ $0 \dots \pi$ and $\phi$ $=$ $0 \dots 2\pi$. Notice that our definition of the angle $\phi$ differs from that of, {\it e.g.}, Refs.~\cite{kneissl96,Fagg:1959wo}. Examples of $(\theta,\phi)$ combinations on the surface of a unit sphere are provided in Fig.~\ref{fig:geo2}. 

  Depending on the angular momenta and parities, each of the two transitions may proceed via different multipolarities, $L$ and $L^\prime$. Here, only the lowest two possible values for the multipolarity will be considered as these correspond to the cases of practical interest. We represent the two-step correlation symbolically by
  \begin{equation}
  j_1 \left(   \begin{array}{c}
         \overrightarrow{L_1} \\
         L_1^\prime  
        \end{array}  \right) J   \left(   \begin{array}{c}
         L_2 \\
         L_2^\prime  
        \end{array}  \right) j_2   \label{eq:steps}
  \end{equation}
  The arrow above the multipolarities $L_1$ and $L_1^\prime$ indicates that the first transition is initiated by a polarized $\gamma$-ray beam. The absence of an arrow above the second transition implies that the polarization of the emitted $\gamma$ rays is not measured.
  
  The radiation pattern for the polarization--direction correlation, $W(\theta\phi)$, is the same as the angular correlation for two sequentially emitted $\gamma$ rays when the polarization of the first $\gamma$ ray is measured. Following Biedenharn \cite{biedenharn60}, it is given by two terms 
  \begin{equation}
  W(\theta\phi) = W_{DD}(\theta) + W_{LP}(\theta\phi) \label{eq:wtotal}
  \end{equation}
  The first term, $W_{DD}(\theta)$, describes the direction--direction correlation obtained with an unpolarized incident $\gamma$-ray beam and depends only on the angle $\theta$. The second term, $W_{LP}(\theta\phi)$, describes the linear polarization correlation and depends on both $\theta$ and $\phi$.  These two contributions are given by
  \begin{equation}
  W_{DD}(\theta) = \sum_{n=0,2,...} A_n(1) A_n(2) P_n(\cos\theta) \label{eq:wdd}
  \end{equation}
  \begin{equation}
  W_{LP}(\theta\phi) = \sum_{n=2,4,...} E_n(1) A_n(2) P_n^{|2|}(\cos\theta) \cos(2\phi) \label{eq:wlp}
  \end{equation}
  where $P_n(\cos\theta)$ and $P_n^{|2|}(\cos\theta)$ are the unassociated and associated Legendre functions, respectively. The sums in the first and second expression start at $n$ $=$ $0$ and $n$ $=$ $2$, respectively. The summation is restricted to $n$ $=$ even since parity is a good quantum number and, in addition, we assume that the measurement does not distinguish between right- and left-handed coordinate systems. For example, odd values of $n$ would have entered the above sums in experiments with incident {\it circularly} polarized $\gamma$ rays.
  
  The coefficients $A_n(1)$, $A_n(2)$, and $E_n(1)$\footnote{An alternative formulation \cite{kneissl96,Fagg:1959wo} of $W_{LP}$ introduces ``polarization coefficients,'' $\kappa_n$, to avoid the more complicated definition of the coefficients $E_n$. Although not explicitly mentioned in these references,  $\kappa_n$ is restricted to $n>1$. We note that the definition of the $\kappa_n$ coefficient exhibits a division by a Clebsch-Gordan coefficient that is avoided in our formulation of the $E_n$ coefficients. For numerical computations, the use of $E_n$ coefficients is preferred over $\kappa_n$ coefficients because the latter are subject to a division by zero for vanishing terms.} for the first (1) or second (2) step in the sequence can be written as 
  \begin{eqnarray}
  \lefteqn{ A_n(1)  = } \nonumber \\
   & & F_n(L_1 L_1 j_1 J) + 2\delta_1 F_n(L_1 L_1^\prime j_1 J) + \delta^2_1 F_n(L_1^\prime L_1^\prime j_1 J) \label{eq:an1}
  \end{eqnarray}
  \begin{eqnarray}
  \lefteqn{ A_n(2)  = } \nonumber \\
  & &  F_n(L_2 L_2 j_2 J) + 2\delta_2 F_n(L_2 L_2^\prime j_2 J) + \delta^2_2 F_n(L_2^\prime L_2^\prime j_2 J) \label{eq:an2}
  \end{eqnarray}
  \begin{eqnarray}
  \lefteqn{E_n(1)  =  \frac{(n-2)!}{(n+2)!} ~\times} \nonumber \\ 
  & & \bigg[ (-1)^{\sigma_{L_1}} F_n(L_1 L_1 j_1 J) \frac{2n(n+1)L_1(L_1+1)}{n(n+1)-2L_1(L_1+1)} + \nonumber \\
  & &  2 \delta_1  (-1)^{\sigma_{L_1^\prime}} F_n(L_1 L_1^\prime j_1 J) (L_1^\prime - L_1) (L_1^\prime + L_1 + 1) + \nonumber \\
  & &  \delta_1^2  (-1)^{\sigma_{L_1^\prime}}  F_n(L_1^\prime L_1^\prime j_1 J) \frac{2n(n+1)L_1^\prime(L_1^\prime+1)}{n(n+1)-2L_1^\prime(L_1^\prime+1)} \bigg] \label{eq:en1}
  \end{eqnarray}
  where $\sigma_L$ $=$ $0$ for electric radiation and $\sigma_L$ $=$ $1$ for magnetic radiation. The quantities $\delta_1$ and $\delta_2$ are the $\gamma$-ray multipolarity mixing ratios for the first and second transition, respectively. They are defined, for each transition, as an intensity ratio by
  \begin{equation}
  \delta^2 \equiv \frac{\mathrm{Intensity~of~radiation~L^\prime}}{\mathrm{Intensity~of~radiation~L}} \label{eq:mix1}
  \end{equation}
  where $L^\prime$ denotes the higher multipolarity, {\it i.e.}, $L^\prime$ $=$ $L$ $+$ $1$. The definition of the mixing ratio in terms of reduced matrix elements is given in Sect.~\ref{sec:phases}. The coefficients $F_n$ are defined\footnote{The pure correlation coefficients, $F_n(LLjJ)$, are often written as $F_n(LjJ)$.} by \cite{biedenharn60}
  \begin{eqnarray}
  \lefteqn{F_n(L L^\prime j J) \equiv } \nonumber \\
  & &  (-1)^{j-J-1} \sqrt{(2L+1)(2L^\prime+1)(2J+1)}   \times  \nonumber \\
  & & (L 1 L^\prime -1 | n 0) W(JJLL^\prime;nj) 
  \label{eq:effcoeff}
  \end{eqnarray}
  where $(L 1 L^\prime -1 | n 0)$ and $W(JJLL^\prime;nj)$ denote Clebsch-Gordan and Racah coefficients, respectively. Numerical values for the $F$ coefficients can be found in Refs.~\cite{Rose:1953tw,appel68}. 
  
  To determine how many terms need to be taken into account in the sums of Eqs.~(\ref{eq:wdd}) and (\ref{eq:wlp}), it is useful to consider the symmetry properties of the $F$ coefficients, which follow directly from those of the Clebsch-Gordan and Racah coefficients. For given values of $L$, $L^\prime$, and $J$, the $F$ coefficient can be non-zero only for
  \begin{equation}
  | L - L^\prime | \leq n \leq \mathrm{min}(2J, L + L^\prime) \label{eq:selectionx}
  \end{equation}
  Therefore, the sums are restricted to\footnote{Because of the products $A_n(1)A_n(2)$ and $E_n(1)A_n(2)$ in Eqs.~(\ref{eq:wdd}) and (\ref{eq:wlp}), respectively, it is particularly important for mixed transitions to apply the correct conditions for the index $n$. The literature states sometimes confusing \cite{appel68} or wrong \cite{FerentzRosenzweig1955} conditions. In other work, the assumptions entering the conditions for $n$ are not explicitly stated \cite{Fagg:1959wo,Rose:1967vq,Krane:1973wr}. The general expression used in the present work is adopted from the review by Biedenharn and Rose \cite{Rose1953}. 
  }
  \begin{equation}
  0 \leq n \leq \mathrm{min}(2J, 2L_{1,max}, 2L_{2,max}) \label{eq:selection}
  \end{equation}
  where $L_{1,max}$ is the largest value assumed by $L_1$ or $L_1^\prime$, and similarly for $L_{2,max}$. For proper normalization, the total angular correlation, $W(\theta,\phi)$, in Eq.~(\ref{eq:wtotal}) must be divided by $(1+\delta_1^2)(1+\delta_2^2)$, {\it i.e.},
  \begin{equation}
    \int_0^{2 \pi} \mathrm{d} \phi \int_0^{\pi}  W(\theta \phi) \sin (\theta) \mathrm{d} \theta = (1+\delta_1^2)(1+\delta_2^2) \label{eq:normalization}
  \end{equation}
  The following comments result directly from the structure of the formalism presented above, as will be illustrated in Sect.~\ref{sec:results}: 
  
  (i) For an unpolarized incident $\gamma$-ray beam, the correlation function $W_{DD}(\theta)$ depends on the spins of the nuclear levels and the multipolarities of the transitions, but not on their parities; see Eqs.~(\ref{eq:wdd}), (\ref{eq:an1}), and (\ref{eq:an2}). 
  
  (ii) For the sequence $0$($L$)$J$($L$)$0$, the measurement of $W_{DD}(\theta)$ at only two angles ({\it e.g.}, $\theta$ $=$ $90^\circ$ and $127^\circ$) is sufficient to determine the multipolarity of the scattered radiation. 
  
  (iii) The Hermitian property of the matrix elements involved in the transitions requires that the correlation function, $W_{DD}(\theta)$, for $j_1$($L_1$)$J$($L_2$)$j_2$ be identical to that of $j_2$($L_2$)$J$($L_1$)$j_1$; {\it i.e.}, the initial and final states as well as the first and second radiations may be reversed without any change in the correlation. 
  
  (iv) The direction--direction correlation function, $W_{DD}(\theta)$, exhibits rotational symmetry about the incident beam direction. The polarization--direction correlation function, $W(\theta\phi)$, is reflection (mirror) symmetric about three orthogonal planes. 
  
  (v) The presence of terms with $(-1)^{\sigma_L}$ in Eqs.~(\ref{eq:en1}) and (\ref{eq:wlp}) implies that the linear polarization correlation, $ W_{LP}(\theta\phi)$, depends upon the parity of the radiation. Therefore, the total correlation function, $W(\theta\phi)$, is also sensitive to the parities of the levels involved in the scattering process. 
  
  (vi) For transitions proceeding via more than one multipolarity (``mixed transitions"), the total correlation function, $W(\theta\phi)$, will be subject to interference, as can be seen from the terms of $A_n(1)$, $A_n(2)$, and $E_n(1)$ that depend on the first power of the mixing ratio $\delta$; the presence of interference may change the correlation markedly and, thus, measuring the angular correlation is a sensitive method for determining the mixing ratios (Sect.~\ref{sec:phases}). 
  
  (vii) An isotropic angular correlation function results if all $F$ coefficients vanish for $n$ $\geq$ $2$, see Eq.~(\ref{eq:selection}). This is always the case when the spin of the intermediate level is $J$ $=$ $0$ or $1/2$, because then all magnetic substates are populated equally. However, for certain combinations of angular momenta the $F$ coefficients can also vanish when $J$ $>$ $1/2$ (Sect.~\ref{sec:visual3}). 
 
\section{Multipolarity mixing ratios and phase conventions} \label{sec:phases}
We already mentioned in Sect.~\ref{sec:formalism} that, if a transition can proceed via mixed multipolarities, $L$ and $L^\prime$, the polarization--direction angular correlation of Eq.~(\ref{eq:wtotal}) will depend on the relative intensity of these radiations. The $\gamma$-ray multipolarity mixing ratio, $\delta$, defined in Eq.~(\ref{eq:mix1}), quantifies this intensity ratio. Since the measured mixing ratios are eventually to be interpreted in terms of a nuclear model, it is important to state explicitly their definition in terms of the reduced nuclear matrix elements involved. 

Unfortunately, various authors have defined the mixing ratios differently, leading to confusing phase inconsistencies of the type $(-1)^{L-L^\prime}$. The reader may find the discussions at the end of Ref.~\cite{Rose:1967vq} and at the beginning of Ref.~\cite{ferguson65} illuminating. For a discussion of the advantages and disadvantages of the different conventions, see Refs.~\cite{Rose:1967vq,ferguson65}. Our goal in this section is to summarize the relationships between the different phase conventions so that numerical values encountered in the literature can be meaningfully compared.

Since we adopted in Sect.~\ref{sec:formalism} the formalism of Biedenharn \cite{biedenharn60}, we must also adopt his definition of the mixing ratio, which is given in terms of real reduced matrix elements by
\begin{equation}
\delta =  \frac{\left<j_x \left\| L^\prime \right\| J \right>}{\left< j_x \left\| L \right\| J \right>}   \label{eq:mix2}
\end{equation}
where $J$ labels the intermediate level and $j_x$ denotes either the initial or the final level, depending on the transition. The reduced matrix elements are defined by the Wigner-Eckart theorem in Ref.~\cite{biedenharn60} as
\begin{equation}
\left<j_x m_x | T(LM) | J m \right> \equiv   \left<j_x \left\| L \right\| J \right> (JLmM | j_x m_x) \label{eq:mix3}
\end{equation}
with $m_x$, $m$, and $M$ $=$ $\pm1$ denoting the magnetic substates corresponding to $j_x$, $J$, and $L$, respectively. Biedenharn \cite{biedenharn60} defines the multipole responsible for the transition by
\begin{equation}
T(LM) = \mathbf{j}_{op} \cdot \mathbf{A}^{e,m}(LM)^*        \label{eq:mix4}
\end{equation}
where $ \mathbf{j}_{op}$ is the nuclear current operator and the $\mathbf{A}^{e,m}(LM)$ symbols denote the electric or magnetic $(e,m)$ standing wave vector potentials. Without considering the structure of Eqs.~(\ref{eq:wtotal})--(\ref{eq:en1}) and the definitions of Eqs.~(\ref{eq:mix2})--(\ref{eq:mix4}) it is not possible to directly compare values of mixing ratios obtained by different authors using different formalisms.

Notice that, in Biedenharn's formalism \cite{biedenharn60,Rose:1953tw}, the intermediate state always appears on the right in the reduced matrix elements of Eq.~(\ref{eq:mix2}), regardless of whether it is the initial or final state involved in the transition. In other words, for the two-step correlation process of Eq.~(\ref{eq:steps}), the definition of Biedenharn yields
\begin{equation}
\delta_1 =  \frac{\left<j_1 \left\| L^\prime \right\| J \right>}{\left< j_1 \left\| L \right\| J \right>}  ~~ \mathrm{and} ~~ \delta_2 =  \frac{\left<j_2 \left\| L^\prime \right\| J \right>}{\left< j_2 \left\| L \right\| J \right>}  \label{eq:mix5}
\end{equation}
In the formalism of Rose and Brink \cite{Rose:1967vq}, the initial state for each transition stands always on the left, which yields the relations
\begin{equation}
\delta_1 = \delta_1(\mathrm{RB}) ~~ \mathrm{and} ~~  \delta_2 = - \delta_2(\mathrm{RB})\label{eq:mix6}
\end{equation}
On the other hand, Steffen and collaborators \cite{beckersteffen69,kranesteffen} use the convention that the initial state always appears on the right in the reduced matrix element, resulting in
\begin{equation}
\delta_1 = - \delta_1(\mathrm{S}) ~~ \mathrm{and} ~~  \delta_2 =  \delta_2(\mathrm{S})\label{eq:mix7}
\end{equation}
Needless to say that the different definitions have caused significant confusion when comparing numerical values of mixing ratios. For a list of additional phase conventions, see Table~I in Ref.~\cite{Martin1987}.

Consider as an example a ($\gamma$,$\gamma$) elastic scattering experiment on a $j_1$ $\neq$ $0$ target nucleus, with a spin sequence of $j_1$($L_1, L_1^\prime$)$J$($L_1, L_1^\prime$)$j_1$. Since this process starts and ends with the ground state, the first and second transitions are identical. In this case, Eq.~(\ref{eq:mix5}) yields immediately $\delta_1$ $=$ $\delta_2$. On the other hand, according to Eq.~(\ref{eq:mix7}), the phase convention of Steffen gives $\delta_1(S)$ $=$ $- \delta_2(S)$.

Multipolarity mixing ratios of mixed M1 and E2 transitions between low-lying levels in $^{11}$B were measured by Rusev et al. \cite{Rusev:2009eo} using resonant absorption of linearly polarized $\gamma$ rays. A similar measurement for low-lying states in $^{27}$Al has been reported by Shizuma et al. \cite{Shizuma:2019ie}.\footnote{Figure 4 in Ref.~\cite{Shizuma:2019ie} can only be reproduced assuming $\delta_1$(S) $=$ $-\delta_2$(S) when adopting the formalism of Steffen and collaborators. Therefore, their statement of $\delta_1$(S) $=$ $\delta_2$(S) is inconsistent with their reported results.} Both groups used the formalism of Steffen and collaborators, but quote $\delta_1(S)$ $=$ $\delta_2(S)$. This assumption, which is erroneous as explained above, does not only impact the sign of the mixing ratio extracted from the data, but also the magnitude of $W(\theta\phi)$ since the angular correlation function is sensitive to the phases. 

Clearly, a consistent phase convention must be employed in the data analysis and presentation of the results. The Evaluated Nuclear Structure Data File (ENSDF \cite{ensdf2021}) has adopted a policy of following the phase convention of Steffen and collaborators \cite{beckersteffen69,kranesteffen} (sometimes called ``Krane--Steffen  convention''), in which emission matrix elements are always used for the multipole operators. For a proper comparison of newly measured mixing ratios to other results, it is imperative to determine the sign convention used in the previous work. We also recommend that authors always state their adopted phase convention when reporting new results.

\section{Numerical example} \label{sec:example}
Let us next consider an illustrative example for the application of the formalism outlined in Sect.~\ref{sec:formalism}. A nucleus with spin-parity of $j_1^{\pi_1}$ $=$ $0^+$ absorbs a linearly polarized $\gamma$ ray of M1 multipolarity and character, populating an intermediate level of spin-parity $J^{\pi}$ $=$ $1^+$. The subsequent transition to a final state with $j_2^{\pi_2}$ $=$ $2^+$ proceeds via (unpolarized) mixed M1/E2 radiation. We write in symbolic notation
\begin{equation}
0 \left(  \overrightarrow{M1}  \right) 1   \left(   \begin{array}{c}
       M1 \\
       E2  
      \end{array}  \right) 2   
\end{equation}
The selection rule of Eq.~(\ref{eq:selection}) restricts the terms in the sums of Eqs.~(\ref{eq:wdd}) and (\ref{eq:wlp}) to $n$ $\leq$ $2$.

For the contribution of the direction--direction correlation, Eq.~(\ref{eq:wdd}) yields
\begin{eqnarray*}
\lefteqn{W_{DD}(\theta) =  } \\
& &  F_0(1101) [F_0(1121) + 2\delta_2 F_0(1221) + \delta_2^2 F_0(2221)] + \\
& & F_2(1101) [F_2(1121) + 2\delta_2 F_2(1221) + \delta_2^2 F_2(2221)] P_2(\cos\theta)  \\
& & = 1 [1 + 2\delta_2 0 + \delta_2^2 1]  P_0(\cos\theta)+ \\
& & 0.7071 [0.0707 + 2\delta_2 0.4743 + \delta_2^2 0.3535] P_2(\cos\theta)  \\
& & = (1 + \delta_2^2) + (0.0500 + 0.6708 \delta_2 + 0.2500 \delta_2^2) P_2(\cos\theta)
\end{eqnarray*}
The linear polarization--direction correlation, according to Eq.~(\ref{eq:wlp}), gives
\begin{eqnarray*}
\lefteqn{W_{LP}(\theta,\phi) =  (-1)^1 F_2(1101) \frac{2\cdot2\cdot3\cdot1\cdot2}{2\cdot3-2\cdot1\cdot2} \left(\frac{0!}{4!}\right) \times} \\
& &  [F_2(1121) + 2\delta_2F_2(1221) + \delta_2^2 F_2(2221) ] P^{|2|}_2(\cos\theta) \cos(2\phi) \\
& & = -0.7071 \cdot 12 \cdot \frac{1}{24} [0.0707 + 2\delta_2 0.4743 + \delta_2^2 0.3535] \times \\
& & P^{|2|}_2(\cos\theta) \cos(2\phi) \\
& & = - [0.0250 + 0.3354 \delta_2 + 0.1250 \delta_2^2] P^{|2|}_2(\cos\theta) \cos(2\phi) 
\end{eqnarray*}
The total polarization--direction correlation, given by Eq.~(\ref{eq:wtotal}), is then 
\begin{eqnarray*}
\lefteqn{W(\theta\phi) =  } \\
& & (1+\delta_2^2) + [0.0500 + 0.6708\delta_2 + 0.2500\delta_2^2 ] P_2(\cos\theta) - \\
& &  [0.0250 + 0.3354 \delta_2 + 0.1250 \delta_2^2] P^{|2|}_2(\cos\theta) \cos(2\phi) 
\end{eqnarray*}
This expression needs to be divided by $(1+\delta_2^2)$ for normalization to unity.

\section{Three-dimensional visualizations} \label{sec:results}
Here, we present three-dimensional visualizations of radiation patterns resulting from $\gamma$-ray scattering experiments. Transitions in even-mass and odd-mass nuclei will be considered separately. For even-mass nuclei, we will focus on an initial state spin-parity of $j_1^{\pi_1}$ $=$ $0^+$, since the large majority of stable even-mass nuclei have $0^+$ ground states. This implies that the first transition proceeds via a unique multipolarity, while the second one may be mixed, depending on the values of the final state spin and mixing ratio.

In odd-mass nuclei, multipolarity mixing can occur in both the first and the second transition. The only exception is for transitions, $j_1$ $\rightarrow$ $J$, with spins $1/2$ $\rightarrow$ $1/2$. However, sequences with $J$ $=$ $1/2$ will not be discussed further because they always result in an isotropic radiation pattern (Sect.~\ref{sec:formalism}). The angular correlation functions computed in the present work are summarized in Table~\ref{tab:summary}.
\begin{table*}
\begin{center}
\caption{Summary of polarization--direction angular correlation functions, $W(\theta\phi)$, and analyzing powers presented in this work.}\label{tab:summary}
\begin{tabular}{l c c c c}
\hline\noalign{\smallskip}
Spin sequence$^a$  		&     \multicolumn{2}{c}{Pattern}  &     \multicolumn{2}{c}{Analyzing Power}   \\
 \cmidrule(lr){2-3}\cmidrule(lr){4-5}
                        &     Sect./App.   	&     Fig.  &     Sect./App.   	&     Fig.   \\
\noalign{\smallskip}\hline\noalign{\smallskip}
$0 \rightarrow 1 \rightarrow 0$	   	        	       		&   \ref{sec:visual1}	    &	\ref{fig:patt0} & \\	
$0 \rightarrow 2 \rightarrow 0$	   	        	        	&   \ref{sec:visual1}	    &	\ref{fig:patt0}   \\	
$0 \rightarrow 1 \rightarrow 1$	   	        	        	&   \ref{sec:visual2}	    &	\ref{fig:patt011} & \ref{app:powers} & \ref{fig:ana_011} \\	
$0 \rightarrow 1 \rightarrow 2$     	   	        	    &   \ref{app:pattern}	    &	\ref{fig:patt012} & \ref{app:powers} & \ref{fig:ana_012} \\	
$0 \rightarrow 1 \rightarrow 3$	   	        	        	&   \ref{app:pattern}	    &	\ref{fig:patt013} & \ref{app:powers} & \ref{fig:ana_013} \\	
$0 \rightarrow 2 \rightarrow 1$	   	        	        	&   \ref{app:pattern}   	&	\ref{fig:patt021}  & \ref{app:powers} & \ref{fig:ana_021} \\	
$0 \rightarrow 2 \rightarrow 2$	   	        	        	&   \ref{app:pattern}   	&	\ref{fig:patt022}  & \ref{app:powers} & \ref{fig:ana_022}  \\	
$0 \rightarrow 2 \rightarrow 3$	   	        	        	&   \ref{app:pattern}   	&	\ref{fig:patt023}  & \ref{app:powers} & \ref{fig:ana_023} \\	
$0 \rightarrow 2 \rightarrow 4$	   	        	        	&   \ref{app:pattern}	    &	\ref{fig:patt024}  & \ref{app:powers} & \ref{fig:ana_024}  \\	
$0 \rightarrow 1 \xrightarrow{\text{U}} 1 \rightarrow 0$    &   \ref{sec:unobs}         &	\ref{fig:patt0110} & \ref{app:powers} & \ref{fig:ana_0110}  \\
$0 \rightarrow 1 \xrightarrow{\text{U}} 2 \rightarrow 0$    &   \ref{app:pattern}	    &	\ref{fig:patt0120} & \ref{app:powers} & \ref{fig:ana_0120}  \\
$0 \rightarrow 1 \xrightarrow{\text{U}} 3 \rightarrow 0$    &   \ref{app:pattern}	    &	\ref{fig:patt0130} & \ref{app:powers} & \ref{fig:ana_0130}  \\
$0 \rightarrow 2 \xrightarrow{\text{U}} 1 \rightarrow 0$    &   \ref{app:pattern}	    &	\ref{fig:patt0210} & \ref{app:powers} & \ref{fig:ana_0210}  \\
$0 \rightarrow 2 \xrightarrow{\text{U}} 2 \rightarrow 0$    &   \ref{app:pattern}	    &	\ref{fig:patt0220} & \ref{app:powers} & \ref{fig:ana_0220}  \\
$0 \rightarrow 2 \xrightarrow{\text{U}} 3 \rightarrow 0$    &   \ref{app:pattern}	    &	\ref{fig:patt0230} & \ref{app:powers} & \ref{fig:ana_0230}  \\
$1/2 \rightarrow 3/2 \rightarrow 1/2$  	                    &   \ref{sec:visual3}	    &	\ref{fig:patt051505} & \ref{app:powers} & \ref{fig:ana_051505} \\
$1/2 \rightarrow 5/2 \rightarrow 1/2$  	                    &   \ref{app:pattern}	    &	\ref{fig:patt052505} & \ref{app:powers} & \ref{fig:ana_052505} \\
$3/2 \rightarrow 3/2 \rightarrow 3/2$  	                    &   \ref{app:pattern}   	&	\ref{fig:patt151515} & \ref{app:powers} & \ref{fig:ana_151515} \\	
$3/2 \rightarrow 5/2 \rightarrow 3/2$  	                    &   \ref{app:pattern}   	&	\ref{fig:patt152515} & \ref{sec:single_mixing_ratio} & \ref{fig:analyzing_power_15_25_15} \\	
$3/2 \rightarrow 7/2 \rightarrow 3/2$  	                    &   \ref{app:pattern}   	&	\ref{fig:patt153515} & \ref{app:powers} & \ref{fig:ana_153515} \\	
$5/2 \rightarrow 3/2 \rightarrow 5/2$  	                    &   \ref{app:pattern}       &	\ref{fig:patt251525} & \ref{app:powers} & \ref{fig:ana_251525}   \\	
$5/2 \rightarrow 5/2 \rightarrow 5/2$  	                    &   \ref{app:pattern}   	&	\ref{fig:patt252525} & \ref{app:powers} & \ref{fig:ana_252525}   \\	
$5/2 \rightarrow 7/2 \rightarrow 5/2$  	                    &   \ref{app:pattern}	    &	\ref{fig:patt253525} & \ref{app:powers} & \ref{fig:ana_253525}   \\	
$5/2 \rightarrow 9/2 \rightarrow 5/2$  	                    &   \ref{app:pattern}	    &	\ref{fig:patt254525} & \ref{app:powers} & \ref{fig:ana_254525}   \\	
$3/2 \rightarrow 7/2 \rightarrow 5/2$   	                &   \ref{sec:visual3}	    &	\ref{fig:patt153525} & \ref{sec:double_mixing_ratio} & \ref{fig:ana_357555} \\	
\noalign{\smallskip}\hline
\end{tabular}
\end{center}
$^a$ The symbol ``U'' refers to an unobserved intermediate $\gamma$ ray (see Sect.~\ref{sec:unobs}).\\
\end{table*}

The visualizations presented here reveal a number of important general features. First, the polarization--direction correlations are independent of the final state parity, $\pi_2$, since the polarization of the second transition is not observed. For example, the angular correlation functions are the same for the sequences $0^+$ $\rightarrow$ $1^+$ $\rightarrow$ $0^+$ and 0$^+ $ $\rightarrow$ $1^+$ $\rightarrow$ $0^-$. Second, changing only the parity of the initial state or the intermediate state changes the sign of the function $E_n(1)$ according to Eq.~(\ref{eq:en1}). This sign change corresponds to a spatial rotation of the radiation pattern by $90^\circ$ about the incident beam direction ($x$-axis). Third, the polarization--direction correlation function is invariant under simultaneous changes in the parities of the initial and intermediate states, as can be seen from Eq.~(\ref{eq:en1}). This means that the radiation pattern, {\it e.g.}, of the sequence $0^+$ $\rightarrow$ $1^+$ $\rightarrow$ $0$ is the same as that of $0^-$ $\rightarrow$ $1^-$ $\rightarrow$ $0$.

\subsection{Pure transitions in even-mass nuclei} \label{sec:visual1}
\begin{figure*}
\includegraphics[width=2.0\columnwidth]{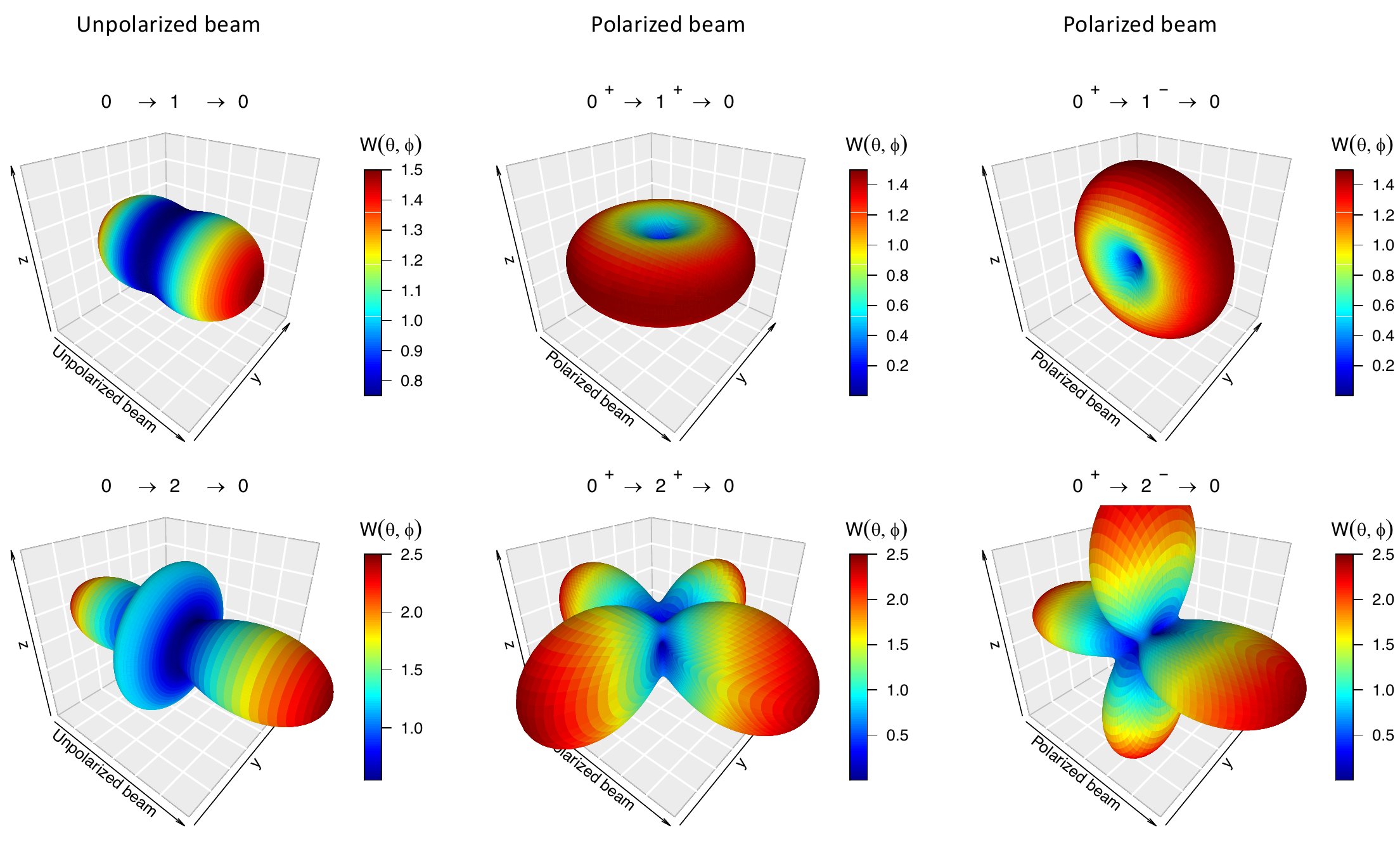} 
\caption{(Color online) Three-dimensional representations of radiation patterns obtained in $\gamma$-ray scattering experiments. The incident $\gamma$-ray beam moves along the positive $x$-axis. All presented sequences proceed via pure transitions, {\it i.e.}, they do not involve any mixing ratio. The colors in each panel (from red to blue) signify the deviation of the angular correlation function from unity (isotropy), with the color key given on the right hand side. (Left column) Direction--direction angular correlations, {\it i.e.}, measured with an incident {\it unpolarized} $\gamma$-ray beam. (Middle and right columns) Linear polarization--direction correlations, {\it i.e.}, measured with an incident {\it polarized} $\gamma$-ray beam. The plane of polarization coincides with the $x$--$y$ plane. The top and bottom row correspond to radiation patterns for dipole ($L_1$ $=$ $L_2$ $=$ $1$) and quadrupole ($L_1$ $=$ $L_2$ $=$ $2$) radiation, respectively. In each panel, the scattering target is located at the geometrical center of the pattern on display.
}
\label{fig:patt0}
\end{figure*}

Figure~\ref{fig:patt0} shows angular correlations for dipole ($L_1$ $=$ $L_2$ $=$ $1$; top row) and quadrupole ($L_1$ $=$ $L_2$ $=$ $2$; bottom row) radiations. The incident $\gamma$-ray beam moves along the positive $x$-axis. The colors, from red to blue, accentuate the magnitude of the angular correlation function. 

The images in the first column of Fig.~\ref{fig:patt0} are obtained for an unpolarized $\gamma$-ray beam and, thus, represent the direction--direction angular correlation, $W_{DD}(\theta)$. Since this function is independent of the parity of the radiation according to Eqs.~(\ref{eq:wdd}), (\ref{eq:an1}), and (\ref{eq:an2}), exactly the same patterns are obtained, {\it e.g.}, for the sequences $0^+$ $\rightarrow$ $1^+$ $\rightarrow$ $0^+$ and $0^+$ $\rightarrow$ $1^-$ $\rightarrow$ $0^+$ (top left), or for $0^+$ $\rightarrow$ $2^+$ $\rightarrow$ $0^+$ and $0^+$ $\rightarrow$ $2^-$ $ \rightarrow$ $0^+$ (bottom left). Furthermore, the function $W_{DD}(\theta)$ is rotationally symmetric about the incident beam direction (Sect.~\ref{sec:formalism}). Therefore, if several detectors are only arranged at a fixed angle of $\theta$ $=$ $90^\circ$ and different angles of $\phi$ (see Fig.~\ref{fig:geo2}), no information can be gathered regarding the intermediate-state spin, $J$.

The patterns in the second and third column are obtained with an incident linearly polarized $\gamma$-ray beam and, thus, represent the polarization--direction angular correlation, $W(\theta\phi)$, of Eq.~(\ref{eq:wtotal}). The $\gamma$-ray polarization direction ({\it i.e.}, the direction of the $\vec{E}$ vector) points parallel to the $y$-axis. The significant change in patterns, when using polarized compared to unpolarized incident $\gamma$ rays, is apparent. In addition, since the polarization--direction correlation function depends on the parity of the radiation according to Eq.~(\ref{eq:en1}), distinct patterns are obtained for the sequences $0^+$ $\rightarrow$ $1^+$ $\rightarrow$ $0$ and $0^+$ $\rightarrow$ $1^-$ $\rightarrow$ $0$, or for $0^+$ $\rightarrow$ $2^+$ $ \rightarrow$ $0$ and $0^+$ $\rightarrow$ $2^-$ $\rightarrow$ $0$. A number of detectors arranged at a fixed angle of $\theta$ $=$ $90^\circ$ and different angles $\phi$ (see Fig.~\ref{fig:geo2}) can easily distinguish between intermediate states of $J^\pi$ $=$ $1^+$ and $1^-$, or $2^+$ and $2^-$, or $1^-$ and $2^+$. A distinction between $J^\pi$ $=$ $1^+$ and $2^+$ is easily achieved by placing an additional detector at angles, {\it e.g.}, of $(\theta,\phi)$ $=$ $(45^\circ, 90^\circ)$. Detector configurations will be discussed in Sect.~\ref{sec:results2}.

As already mentioned above, the polarization--direction correlations are independent of the final state parity, $\pi_2$, and, thus, identical patterns are obtained, {\it e.g.}, for the sequences $0^+$ $\rightarrow$ $1^+$ $ \rightarrow$ $0^+$ and $0^+$ $\rightarrow$ $1^+$ $\rightarrow$ $0^-$. The patterns are also invariant under a simultaneous change in the parities of the initial and intermediate states, {\it i.e.}, they are the same, {\it e.g.}, for $0^+ $ $\rightarrow$ $1^+$ $\rightarrow$ $0$ and $0^-$ $\rightarrow$ $1^-$ $ \rightarrow$ $0$. In addition, by comparing the patterns in the second and third columns, it is apparent that changing only the parity of the intermediate state results in a rotation by $90^\circ$ about the incident beam direction ($x$-axis). 

\subsection{Mixed transitions in even-mass nuclei} \label{sec:visual2}
\begin{figure*}
\includegraphics[width=2.0\columnwidth]{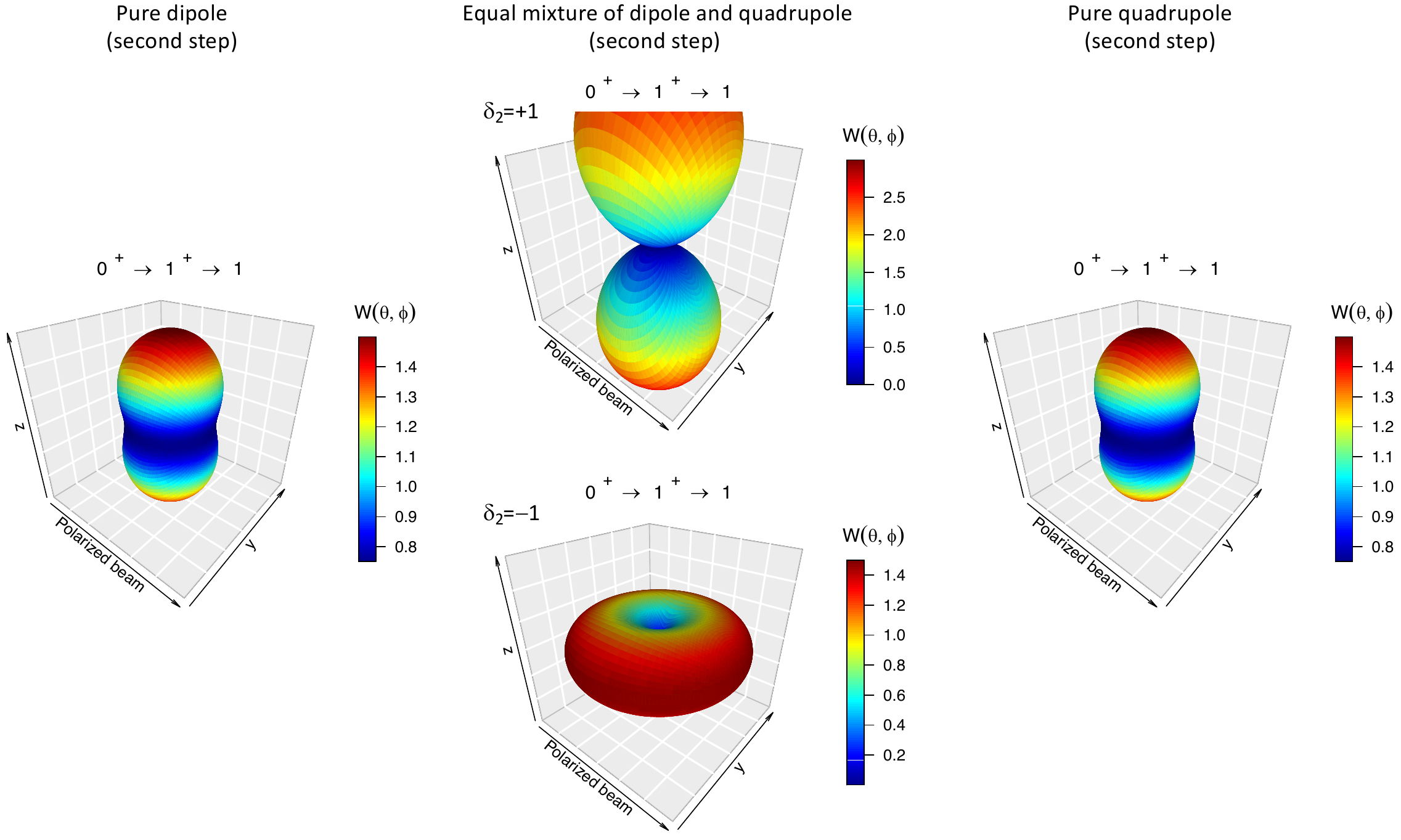} 
\caption{(Color online) Three-dimensional representations of linear polarization--direction correlations for the sequence $0^+$ $\rightarrow$ $1^+ $ $\rightarrow$ $1$, or, equivalently, $0^-$ $\rightarrow$ $1^-$ $\rightarrow $ $1$. For an explanation, see the caption of Fig.~\ref{fig:patt0}. The same correlation functions are obtained for positive or negative parity, $\pi_2$, of the final state, $j_2$. The images on the left and right correspond to pure dipole ($L_2$ $=$ $1$) and pure quadrupole ($L_2$ $=$ $2$) transitions, respectively, for the second step. The two middle panels correspond to an equal mixture of dipole and quadrupole amplitudes ($L_2$ $=$ $1$; $L_2^\prime$ $=$ $2$), but with opposite signs of the mixing ratio: (Middle top) $\delta_2$ $=$ $+1$; (Middle bottom) $\delta_2$ $=$ $-1$. To visualize the correlation functions for the sequences $0^+$ $\rightarrow 1^-$ $\rightarrow$ $1$ or $0^-$ $\rightarrow$ $1^+$ $\rightarrow$ $1$, rotate each of the displayed patterns by $90^\circ$ about the incident beam direction ($x$-axis).
}
\label{fig:patt011}
\end{figure*}

Figure~\ref{fig:patt011} illustrates the linear polarization--direction correlations for the sequence $0^+$ $\rightarrow$ $1^+$ $\rightarrow$ $1$. The first transition is of pure M1 character ($L_1$ $=$ $1$), but the second step can proceed via mixed multipolarities ($L_2$ $=$ $1$; $L_2^\prime$ $=$ $2$). As was the case above, the same correlation functions are obtained for a positive or negative parity, $\pi_2$, of the final state, $j_2$. The patterns on the left and right assume pure transitions for the second step. For this particular sequence, the correlation functions of these two pure transitions are the same because they can only differ by the $A_n(2)$ coefficient of Eq.~(\ref{eq:an2}). However, the $F$-coefficients are equal, {\it i.e.}, $F_2(1111)$ $=$ $F_2(2211)$, and, thus, the resulting correlation functions for pure dipole and pure quadrupole radiation in the second step are identical. 

The two images in the middle column of Fig.~\ref{fig:patt011} depict the results of $\gamma$-ray multipolarity mixing in the second step. The top and bottom panels are obtained for equal amplitudes of dipole and quadrupole radiations, but with a different sign of the mixing ratio (top: $\delta_2$ $=$ $+1$; bottom: $\delta_2$ $=$ $-1$). It can be seen that these shapes differ significantly from those of the pure radiations on the left and right. The sensitivity to the sign of the mixing ratio is also apparent. 

Correlation functions for other sequences of practical interest, starting from an initial (target nucleus ground) state of $j_1^{\pi_1}$ $=$ $0^+$, are presented in Figs.~\ref{fig:patt012}--\ref{fig:patt024}. The images demonstrate that the resulting radiation patterns for mixed transitions are not simply given by the incoherent sum of those for pure transitions. Instead, interference effects give rise to a large variety of shapes.

\subsection{Mixed transitions in odd-mass nuclei} \label{sec:visual3}
Among the many possible spin combinations of practical interest involving odd-mass nuclei, we will focus on initial state spins of $j_1$ $=$ $1/2$, $3/2$, and $5/2$. We shall mainly display results for equal spin values of the initial and final states, $j_1$ $=$ $j_2$. For radiation mixtures, we shall make the additional assumption of identical initial and final states, which requires $\delta_1$ $=$ $\delta_2$, as explained in Sect.~\ref{sec:phases}. 

\begin{figure*}
\includegraphics[width=2.0\columnwidth]{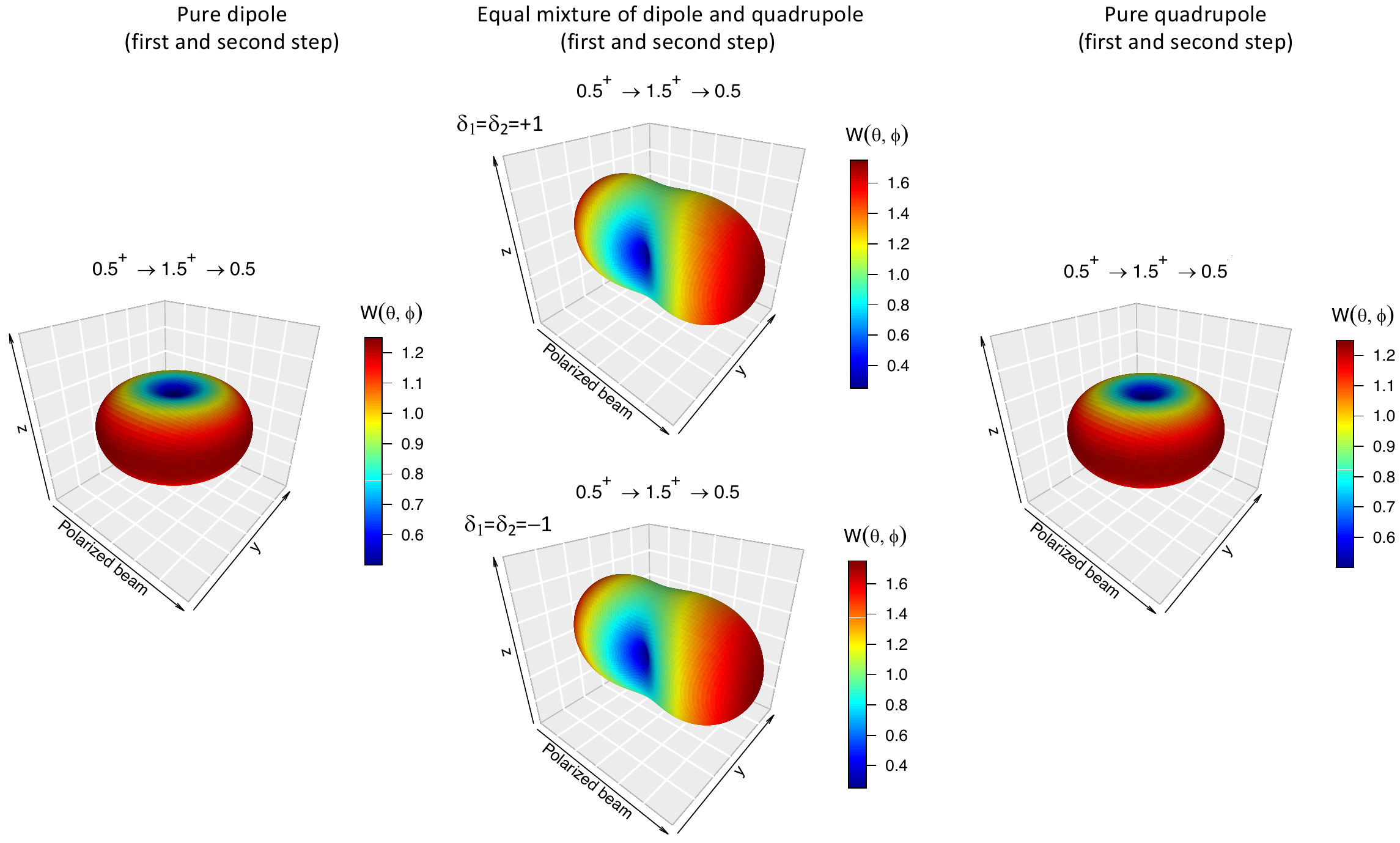} 
\caption{(Color online) Three-dimensional representations of linear polarization--direction correlations for the sequence $1/2^+$ $\rightarrow$ $ 3/2^+$ $\rightarrow$ $1/2$, or, equivalently, $1/2^-$ $\rightarrow$ $3/2^-$ $ \rightarrow$ $1/2$. For an explanation, see the caption of Fig.~\ref{fig:patt0}. The same correlation functions are obtained for positive or negative parity, $\pi_2$, of the final state, $j_2$. The images on the left and right correspond to pure transitions for both the first and second step (left: $L_1$ $=$ $L_2$ $=$ $1$; right: $L_1$ $=$ $L_2$ $=$ $2$). The two middle panels correspond to mixed transitions ($L_1$ $=$ $L_2$ $=$ $1$; $L_1^\prime$ $=$ $L_2^\prime$ $=$ $2$), where we assumed that the initial and final states are identical: (Middle top) $\delta_1$ $=$ $\delta_2$ $=$ $+1$; (Middle bottom) $\delta_1$ $=$ $\delta_2$ $=$ $-1$. To visualize the correlation functions for the sequences $1/2^+$ $\rightarrow$ $3/2^-$ $\rightarrow$ $1/2$ or $1/2^-$ $ \rightarrow$ $3/2^+$ $\rightarrow$ $1/2$, rotate each of the displayed patterns by $90^\circ$ about the incident beam direction ($x$-axis).
}
\label{fig:patt051505}
\end{figure*}

The angular correlations for the sequence $1/2^+$ $\rightarrow$ $3/2^+$ $\rightarrow$ $1/2$, or $1/2^-$ $\rightarrow$ $3/2^-$ $\rightarrow$ $1/2$, are found in Fig.~\ref{fig:patt051505}. In this particular case, the same patterns are obtained for the pure transitions, shown in the left and right panels. The reason for this behavior is that the coefficients $F_2(L_1L_1j_1J)$ and $F_2(L_1^\prime L_1^\prime j_1 J)$ in Eqs.~(\ref{eq:an1})--(\ref{eq:en1}) for $A_n(1)$, $A_n(2)$, and $E_n(1)$ are related by $F_2(11\frac{1}{2}\frac{3}{2})$ $=$ $-F_2(22\frac{1}{2}\frac{3}{2})$, which leaves $W_{DD}(\theta)$ and $W_{LP}(\theta\phi)$ in Eqs.~(\ref{eq:wdd}) and (\ref{eq:wlp}) unchanged. 

The two middle panels are obtained for equal amplitudes of the two multipolarities, but with opposite signs of the mixing ratio: $\delta_1$ $=$ $\delta_2$ $=$ $+1$ (top), and $\delta_1$ $=$ $\delta_2$ $=$ $-1$ (bottom). These two patterns are also identical for this particular spin sequence. Since $F_2(11\frac{1}{2}\frac{3}{2})$ $=$ $0.5$ and $F_2(22\frac{1}{2}\frac{3}{2})$ $=$ $-0.5$, the first and last terms in $A_n(1)$, $A_n(2)$, and $E_n(1)$ cancel, leaving only the interference (middle) term. A simultaneous change in the signs of $\delta_1$ and $\delta_2$ will leave the products $A_n(1)A_n(2)$ and $E_n(1)A_n(2)$, and, thus, $W_{DD}(\theta)$ and $W_{LP}(\theta\phi)$, unchanged.

Correlation functions for other sequences of practical interest starting from initial (target nucleus ground) states of $j_1$ $=$ $1/2$, $3/2$, and $5/2$ are presented in Figs.~\ref{fig:patt052505}--\ref{fig:patt254525}. Notice in Fig.~\ref{fig:patt151515} the isotropic emission pattern for the pure quadrupole case (right panel). The reason is that, although the angular momenta fulfill the condition of Eq.~(\ref{eq:selectionx}), the $F$ coefficients are zero because of the vanishing Racah coefficient, $W(\frac{3}{2} \frac{3}{2} 2 2 ; n \frac{3}{2})$, in Eq.~(\ref{eq:effcoeff}).

\begin{figure*}
\includegraphics[width=2.0\columnwidth]{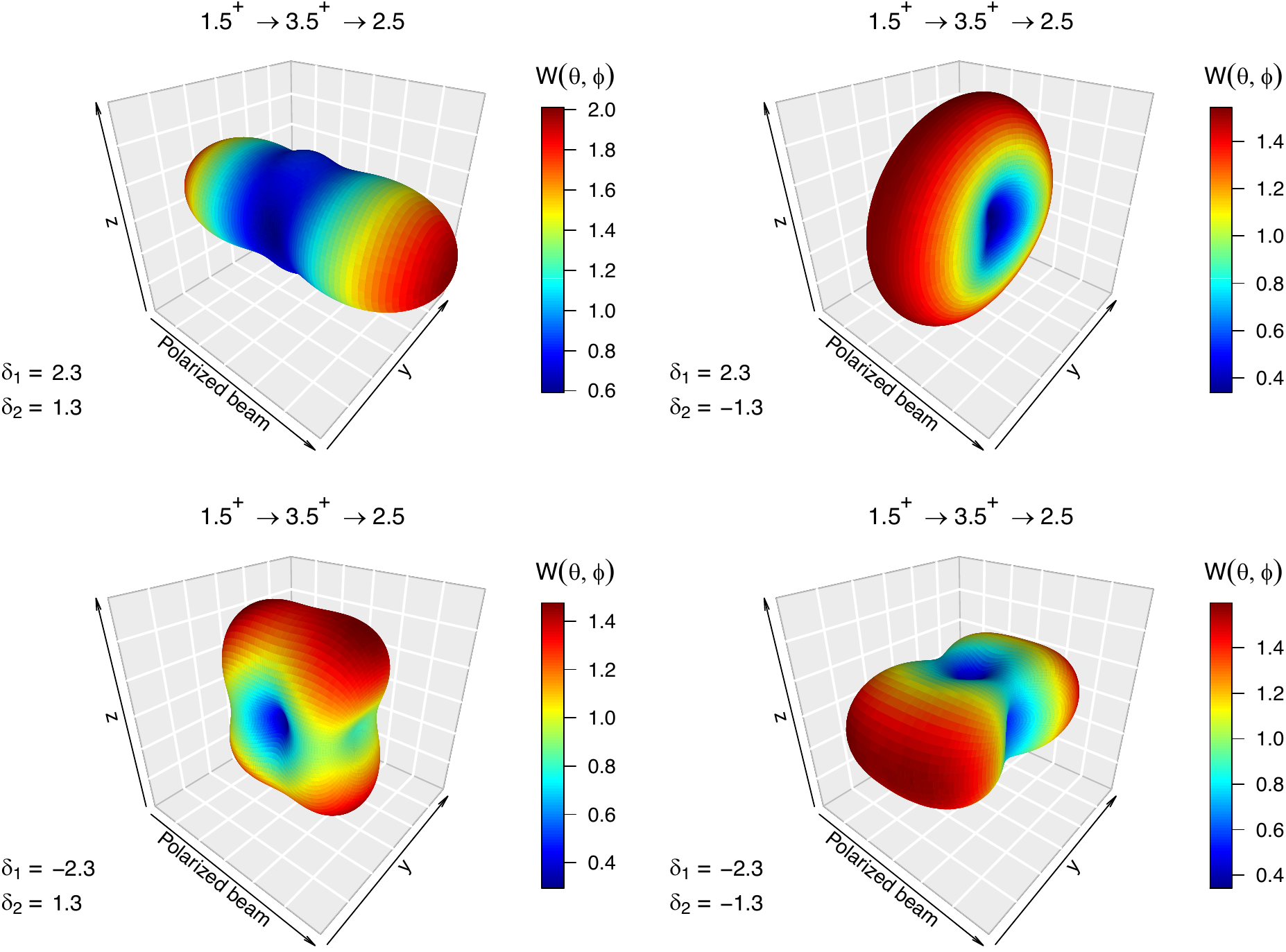} 
\caption{(Color online) Three-dimensional representations of linear polarization--direction correlations for the sequence $3/2^+$ $\rightarrow$ $ 7/2^+$ $\rightarrow$ $5/2$, or, equivalently, $3/2^-$ $\rightarrow$ $7/2^-$ $ \rightarrow$ $5/2$. For an explanation, see the caption of Fig.~\ref{fig:patt0}. All the patterns correspond to mixed transitions in the first and second step. The mixing ratios were kept at fixed magnitudes, $| \delta_1 |$ $=$ $2.3$ and $| \delta_2 |$ $=$ $1.3$, but the signs of the mixing ratios differ. The same correlation functions are obtained for positive or negative parity, $\pi_2$, of the final state, $j_2$. To visualize the correlation functions for the sequences $3/2^+$ $\rightarrow$ $7/2^-$ $ \rightarrow$ $5/2$ or $3/2^-$ $\rightarrow$ $7/2^+$ $\rightarrow$ $5/2$, rotate each of the displayed patterns by $90^\circ$ about the incident beam direction ($x$-axis).
}
\label{fig:patt153525}
\end{figure*}

Finally, three-dimensional visualizations for the sequence $3/2^+$ $\rightarrow$ $7/2^+$ $\rightarrow$ $5/2$, or $3/2^-$ $\rightarrow$ $7/2^-$ $\rightarrow$ $5/2$, are given in Fig.~\ref{fig:patt153525}. Since the initial and final states are not identical, the mixing ratios for the first and second transitions, $\delta_1$ and $\delta_2$, will, in general, differ in magnitude and sign. The correlation functions shown were generated with fixed values of $|\delta_1|$ $=$ $2.3$ and $| \delta_2 |$ $=$ 1.3, but for all four combinations of mixing ratio signs. It can be seen that changing the sign of either $\delta_1$ or $\delta_2$ gives rise to significantly different radiation patterns. As was the case before, the total correlation function is independent of the final state parity, $\pi_2$, and a change in the parity of only the initial or intermediate state will result in a rotation of the pattern by $90^\circ$ about the $x$-axis. 
%

\section{Formalism for the angular correlation involving an unobserved intermediate \texorpdfstring{$\gamma$}{g} ray}\label{sec:unobs}
\begin{figure}
\includegraphics[width=0.65\columnwidth]{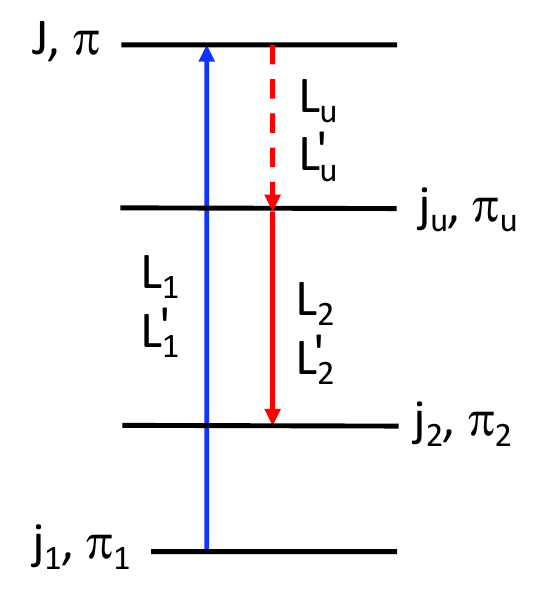} 
\caption{(Color online) Excitation and de-excitation involving three $\gamma$ rays: (i) absorption of incident $\gamma$ ray (blue arrow) and excitation of level $J^\pi$; (ii) $\gamma$-ray decay (dashed red arrow) to intermediate state, $j_u^{\pi_u}$; (iii) subsequent $\gamma$-ray decay (solid red arrow) to final level, $j_2^{\pi_2}$, which may or may not be identical with the initial state, $j_1^{\pi_1}$. Only the last $\gamma$ ray is detected, {\it i.e.}, the second one (dashed arrow) is unobserved.}
\label{fig:unobs}
\end{figure}

It is sometimes of interest to analyze the angular correlation when the state excited by the incident linearly polarized beam de-excites via a two-photon cascade, where only the last $\gamma$ ray is detected, {\it i.e.}, when the intermediate $\gamma$ ray is unobserved. The situation is shown in Fig.~\ref{fig:unobs}. We represent the correlation symbolically by
\begin{equation}
j_1 \left(   \begin{array}{c}
       \overrightarrow{L_1} \\
       L_1^\prime  
      \end{array}  \right) J   \left(   \begin{array}{c}
       L_u \\
       L_u^\prime  
      \end{array}  \right) j_u   \left(   \begin{array}{c}
       L_2 \\
       L_2^\prime  
      \end{array}  \right) j_2   \label{eq:stepsx}
\end{equation}
where all three steps may proceed via transitions of mixed multipolarities. The symbols $L_u$ and $L_u^\prime$ represent the unobserved intermediate $\gamma$ ray, while all other symbols have the same meaning as in Sect.~\ref{sec:formalism}.
The unobserved $\gamma$ ray mixes incoherently and the total correlation function is given by
\begin{equation}
W(\theta\phi) = W_{L_u}(\theta\phi) + \delta_u^2 W_{L_u^\prime}(\theta\phi)   \label{eq:wtotalu}
\end{equation}
where $\delta_u^2$ denotes the multipolarity mixing ratio for the unobserved photon. Equations~(\ref{eq:wdd}) and (\ref{eq:wlp}) need to be replaced by
\begin{equation}
W_{DD}(\theta) = \sum_{n=0,2,...} A_n(1) C_n A_n(2) P_n(\cos\theta) \label{eq:wddu}
\end{equation}
\begin{equation}
W_{LP}(\theta\phi) = \sum_{n=2,4,...} E_n(1) C_n A_n(2) P_n^{|2|}(\cos\theta) \cos(2\phi) \label{eq:wlpu}
\end{equation}
For $W_{L_u}(\theta\phi)$, the unobserved $\gamma$ ray introduces into both expressions an additional coefficient\footnote{The review of Fagg and Hanna \cite{Fagg:1959wo} contains a phase factor in the $C$ coefficient (which they call $U$; see their Eq.~(I-1$^\prime$)), leading sometimes to erroneous, and even negative, angular correlation functions for sequences involving unobserved intermediate $\gamma$ rays. The expression used in the present work is adopted from the review of Biedenharn \cite{biedenharn60}.}
\begin{equation}
C_n = \sqrt{(2J+1)(2j_u+1)} W(JnL_uj_u;Jj_u)     \label{eq:addu}
\end{equation}
where, for $W_{L_u^\prime}(\theta\phi)$, the sole change is to replace $L_u$ by $L_u^\prime$ in the Racah coefficient, $W$. The sums over $n$ are now restricted by
\begin{equation}
0 \leq n \leq \mathrm{min}(2J, 2j_u, 2L_{1,max}, 2L_{2,max}) \label{eq:selection2}
\end{equation}
Note that the multipolarities of the unobserved $\gamma$ ray, $L_u$ and $L_u^\prime$, do not limit the sums. For proper normalization, the total angular correlation, $W(\theta,\phi)$, in Eq.~(\ref{eq:wtotalu}) must be divided by $(1+\delta_1^2)(1+\delta_u^2)(1+\delta_2^2)$.  An isotropic radiation pattern, $W(\theta\phi)$ $=$ $1$, results if the spins of either intermediate state, $J$ or $j_u$, is $0$ or $1/2$.

\begin{figure*}
\includegraphics[width=2\columnwidth]{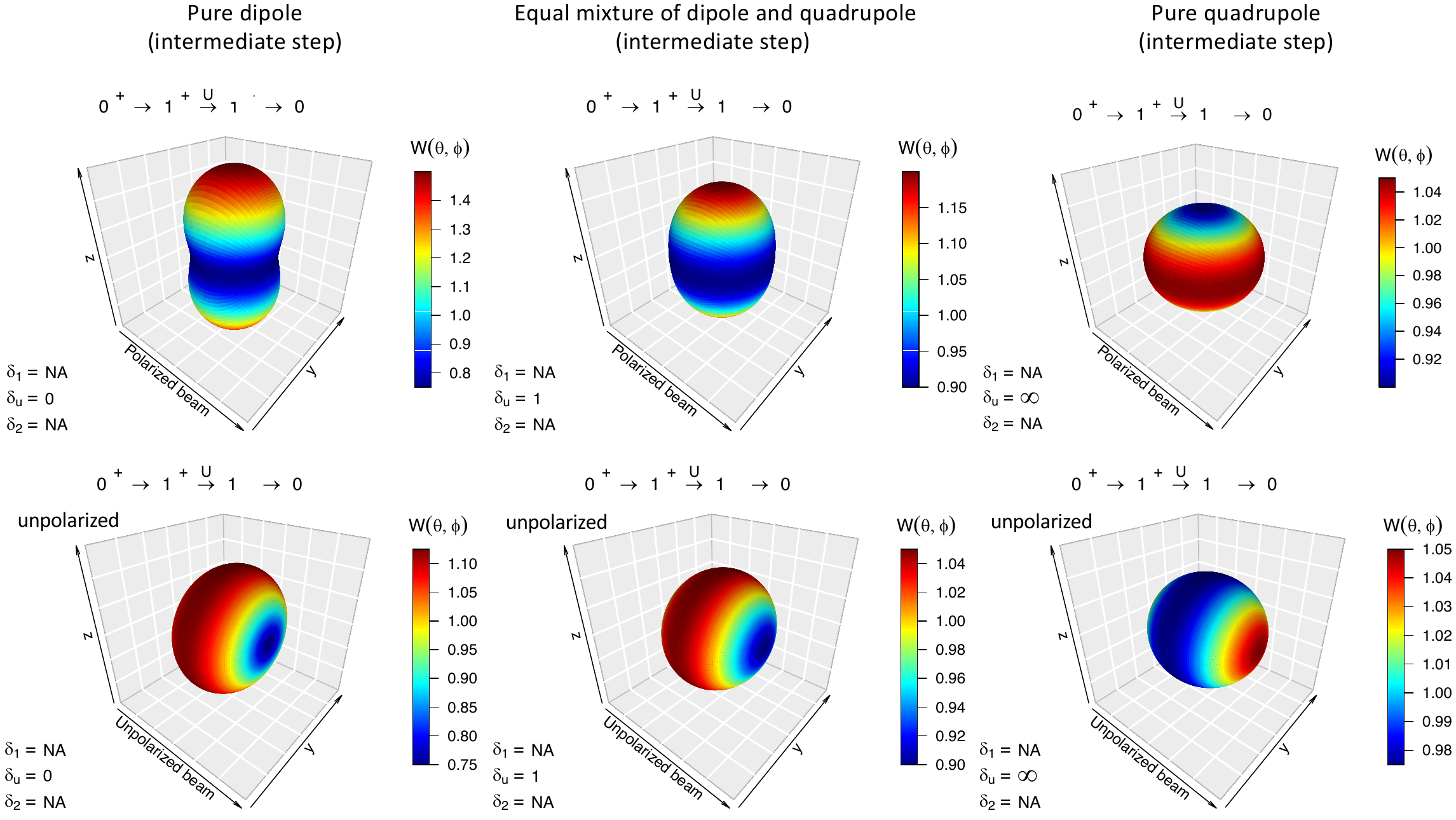} 
\caption{(Color online) Three-dimensional representations of angular correlations for the sequence $0^+$ $\rightarrow$ $1^+$ $\xrightarrow{\text{U}} $ $1$ $\rightarrow$ $0$. The symbol ``U'' indicates that the intermediate $\gamma$ ray in the sequence is unobserved. The panels in the left, middle, and right columns correspond to pure, equally mixed, and pure radiations, respectively, for the intermediate $\gamma$ ray. The top and bottom rows show the linear polarization--direction ({\it i.e.}, with polarized beam) and direction--direction ({\it i.e.}, with unpolarized beam) correlations, respectively. To visualize the correlation functions for the sequences $0^+$ $\rightarrow$ $1^-$ $\xrightarrow{\text{U}}$ $1$ $\rightarrow$ $0$ or $0^-$ $\rightarrow$ $1^+$ $\xrightarrow{\text{U}}$ $1$ $\rightarrow$ $0$, rotate each of the displayed patterns in the top row by $90^\circ$ about the incident beam direction ($x$-axis)}
\label{fig:patt0110}
\end{figure*}

An example is displayed in Fig.~\ref{fig:patt0110} for the sequence $0^+$ $\rightarrow$ $1^+$ $\xrightarrow{\text{U}} $ $1$ $\rightarrow$ $0$, where the symbol ``U'' denotes the unobserved transition. The top and bottom rows depict angular correlations for incident linearly polarized and unpolarized $\gamma$-ray beams, respectively. In this case, only the unobserved intermediate $\gamma$ ray can be of mixed multipolarity. The left, middle, and right columns present the radiation patterns for pure dipole, equally mixed  dipole and quadrupole, and pure quadrupole transitions, respectively. Although the correlation functions are not as pronounced as those for the two-step sequence $0^+$ $\rightarrow$ $1^+$ $\rightarrow$ $0$ (compare to Fig.~\ref{fig:patt0}), it can be seen that the anisotropies remain significant and can be useful for experimentally determining  spins, parities, and mixing ratios.

The symmetry properties of the correlation functions are similar to those discussed in earlier sections. The three-dimensional patterns are independent of the parities of either the intermediate level, $j_u^{\pi_u}$, or the final level, $j_2^{\pi_2}$. Changing the parities of either the initial state, $j_1^{\pi_1}$, or the intermediate level, $J^\pi$, results in a rotation of the linear polarization--direction correlation pattern by $90^\circ$ about the beam direction ($x$-axis). Changing both parities, $\pi_1$ and $\pi$, simultaneously will leave the pattern unchanged. Furthermore, the results in Fig.~\ref{fig:patt0110} are independent of the sign of the mixing ratio $\delta_u$ since the multipolarities of the unobserved intermediate radiation mix {\it incoherently} (see Eq.~(\ref{eq:wtotalu})).

Three--dimensional emission patterns for other transitions involving an unobserved intermediate transition can be found in Figs.~\ref{fig:patt0120} -- \ref{fig:patt0230}. See also Table~\ref{tab:summary}.

\section{Detector configurations and analyzing powers} \label{sec:results2}
The spatial placement of detectors is of central importance when measuring the radiation anisotropy in $\gamma$-ray scattering experiments. Consider again Fig.~\ref{fig:geo2}, showing different angle combinations, ($\theta$,$\phi$), on the surface of a sphere. The incident linearly polarized $\gamma$-ray beam moves along the positive $x$-axis. The two angle combinations that are most frequently used in measurements are ($\theta$,$\phi$) $=$ (90$^\circ$,0$^\circ$) and (90$^\circ$,90$^\circ$), indicated by the red circles at the top and left, respectively, in the figure. Since changing the parity of either the initial or the intermediate state results in a rotation of the radiation pattern by 90$^\circ$ about the beam direction, placing detectors at these two locations provides a useful probe to measure anisotropies and determine parities. This is demonstrated, for example, in Fig.~\ref{fig:patt0} (top middle and top right panels) for the sequences $0^+$ $\rightarrow$ $1^\pm$ $ \rightarrow$ $0$. However, measurements at just these two locations cannot distinguish between, {\it e.g.}, $0^+$ $\rightarrow$ $1^+$ $ \rightarrow$ $0$ and $0^+$ $\rightarrow$ $2^+$ $\rightarrow$ $0$ (bottom middle panel). Therefore, we propose a basic experimental geometry of four detectors, at angles ($\theta$,$\phi$) $=$ (90$^\circ$,0$^\circ$), (90$^\circ$,90$^\circ$), (45$^\circ$,0$^\circ$) and (45$^\circ$,90$^\circ$), as indicated by the four red circles in Fig.~\ref{fig:geo2}. Recall from Sect.~\ref{sec:formalism} that the angular correlation functions are reflection (mirror) symmetric about three orthogonal planes. For example, the same values of $W(\theta\phi)$ are obtained at (45$^\circ$,90$^\circ$) and (135$^\circ$,90$^\circ$), or at (90$^\circ$,90$^\circ$) and (90$^\circ$,270$^\circ$).

It is convenient in the data analysis to introduce the {\it analyzing power} (or {\it asymmetry}), defined\footnote{Equation~(15) in Ref.~\cite{kneissl96} shows the opposite order of terms in the numerator compared to our Eq.~(\ref{eq:analyzing}). But, since their definition of the angle $\phi$ differs from ours (see Sect.~\ref{sec:formalism}), both expressions yield the same sign of $A(\theta)$. On the other hand, Ref.~\cite{PhysRevLett.88.012502} defines $A(\theta)$ with the opposite sign compared to our Eq.~(\ref{eq:analyzing}).} by
\begin{equation}
A(\theta) = \frac{W(\theta, 0^\circ) - W(\theta, 90^\circ)}{W(\theta, 0^\circ) + W(\theta, 90^\circ)} \label{eq:analyzing}
\end{equation}
Because of the $\cos(2\phi)$ term in Eq.~(\ref{eq:wlp}), $W_{LP}(\theta, 0^\circ)$ and $W_{LP}(\theta, 90^\circ)$ have equal magnitudes but opposite signs. Therefore, the difference in the numerator of Eq.~(\ref{eq:analyzing}) is equal to $2W_{LP}(\theta, \phi=0^\circ)$ and the sum in the denominator corresponds to twice the intensity that would have been measured with an incident {\it unpolarized} $\gamma$-ray beam at the polar angle, $\theta$. The analyzing power can then be expressed in terms of $W_{DD}$ and $W_{LP}$ as
\begin{equation}
	A(\theta) = \frac{W_{LP} \left( \theta, \phi = 0^\circ\right)}{W_{DD} \left( 
	\theta \right)} \label{eq:analyzing_w}
\end{equation}
According to the properties of $W_{LP}$ (see Sect.~\ref{sec:formalism}), the analyzing power changes sign when the parity of either the initial or intermediate state is changed. Per definition, $A(\theta)$ $=$ $0$ for an unpolarized incident $\gamma$-ray beam.

With the four angle combinations suggested above, two analyzing powers can be determined, $A(\theta=90^\circ)$ and $A(\theta=45^\circ)$. Notice that if additional detectors are placed at angles of (90$^\circ$,45$^\circ$) or (45$^\circ$,45$^\circ$) (blue circles in Fig.~\ref{fig:geo2}), the intensities measured at these locations represent again those obtained with an unpolarized incident $\gamma$-ray beam because $W_{LP}(\theta, 45^\circ)$ $=$ $0$, according to Eq.~(\ref{eq:wlp}). 

\begin{figure}
\includegraphics[width=1.0\columnwidth]{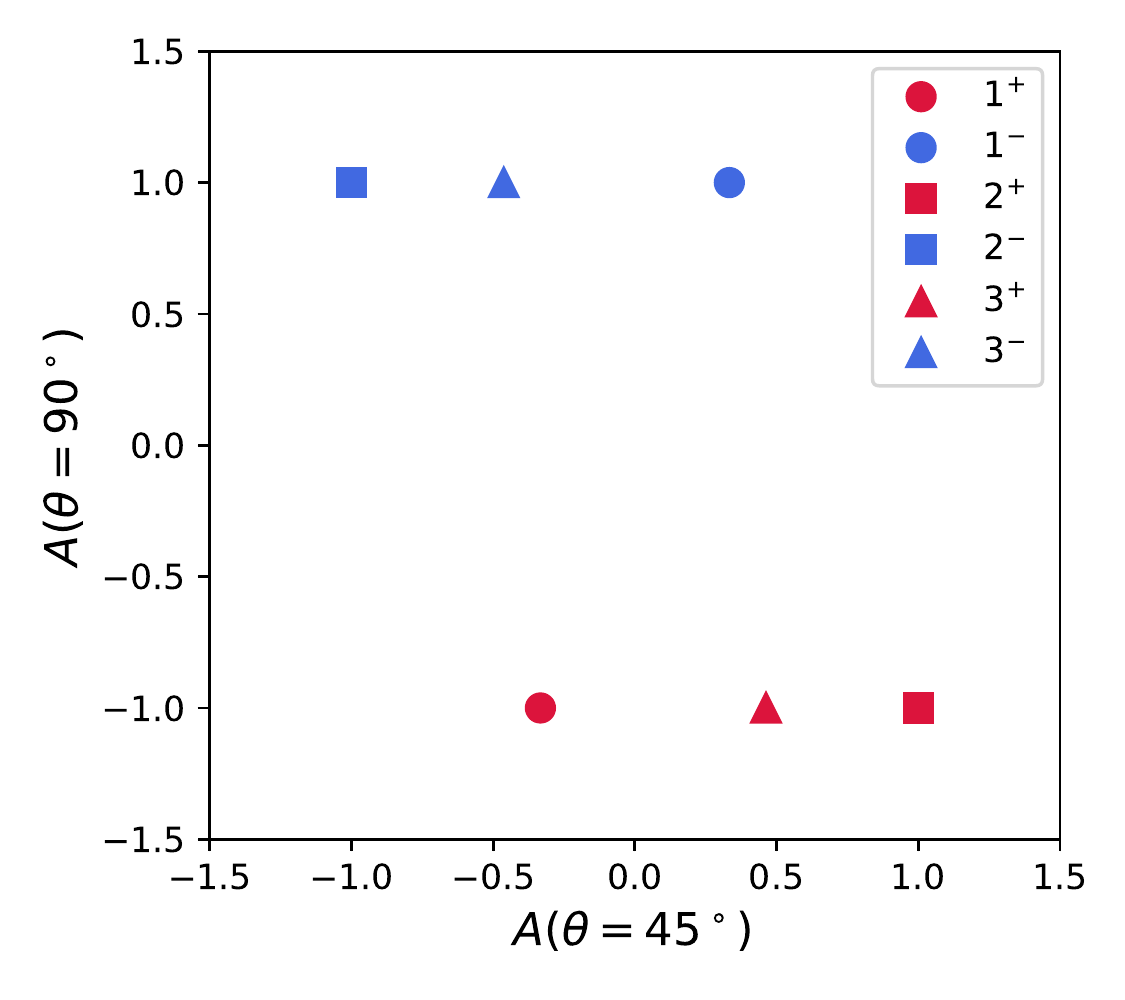} 
\caption{\label{fig:test} (Color online) Theoretical analyzing powers $A(\theta=90^\circ)$ {\it vs.} $A(\theta=45^\circ)$ for the pure transitions $0^+$~$\rightarrow$~$J^\pi$~$\rightarrow$~$0^+$, with $J^\pi$ $=$ $1^\pm$ (circles), $2^\pm$ (squares), and $3^\pm$ (triangles). The points are well separated and allow for a straightforward determination of unknown $J^\pi$ values by comparing the displayed values to the corresponding experimental results.}
\end{figure}

When only pure transitions are involved in the two-step spin sequence, the theoretical values of the two analyzing powers give rise to well-separated points in the $A(\theta=90^\circ)$ {\it vs.} $A(\theta=45^\circ)$ plane. These are shown in Fig.~\ref{fig:test} for the sequences $0^+$ $\rightarrow$ $J^\pi$ $\rightarrow$ $0^+$, with $J^\pi$ $=$ $1^\pm$, $2^\pm$, and $3^\pm$. In this case, it is straightforward to determine $J^\pi$ values by comparing measured analyzing powers to their theoretical values.

When a two-step spin sequence depends only on a single mixing ratio, either because only one transition is mixed (Sect.~\ref{sec:visual2}) or because the initial and final states are identical (Sect.~\ref{sec:visual3}), the dependence of $A(\theta=45^\circ)$ and $A(\theta=90^\circ)$ on this mixing ratio can be easily visualized. The situation is more complex if both steps proceed via mixed transitions and the initial and final states are different (see Sect.~\ref{sec:visual3}), {\it i.e.}, if two different mixing ratios, $\delta_1$ and $\delta_2$, need to be considered in the analysis. In that case, measuring the de-excitation to the ground state could be used to determine the $J^\pi$ value of the intermediate level and the mixing ratio, $\delta_1$, for the first step, while measuring the decay to the final excited state could be employed to find the mixing ratio, $\delta_2$, for the second step (see, {\it e.g.}, Ref.~\cite{Rusev:2009eo}). If the decay to the ground state cannot be observed, {\it e.g.}, because its branching ratio is too small, a measurement of $A(\theta=45^\circ)$ and $A(\theta=90^\circ)$ may still provide constraints on the values of $\delta_1$ and $\delta_2$.

The following subsections present examples of graphical representations for the dependence of the analyzing powers, $A(\theta=45^\circ)$ and $A(\theta=90^\circ)$, on a single (Sect.~\ref{sec:single_mixing_ratio}) or two (Sect.~\ref{sec:double_mixing_ratio}) multipolarity mixing ratios. We show how such graphs can be used to constrain allowed regions of $\delta_1$ or $\delta_2$ based on the measured analyzing powers. 

\subsection{Single multipolarity mixing ratio}
\label{sec:single_mixing_ratio}
\begin{figure*}
\centering
\includegraphics[width=1.75\columnwidth]{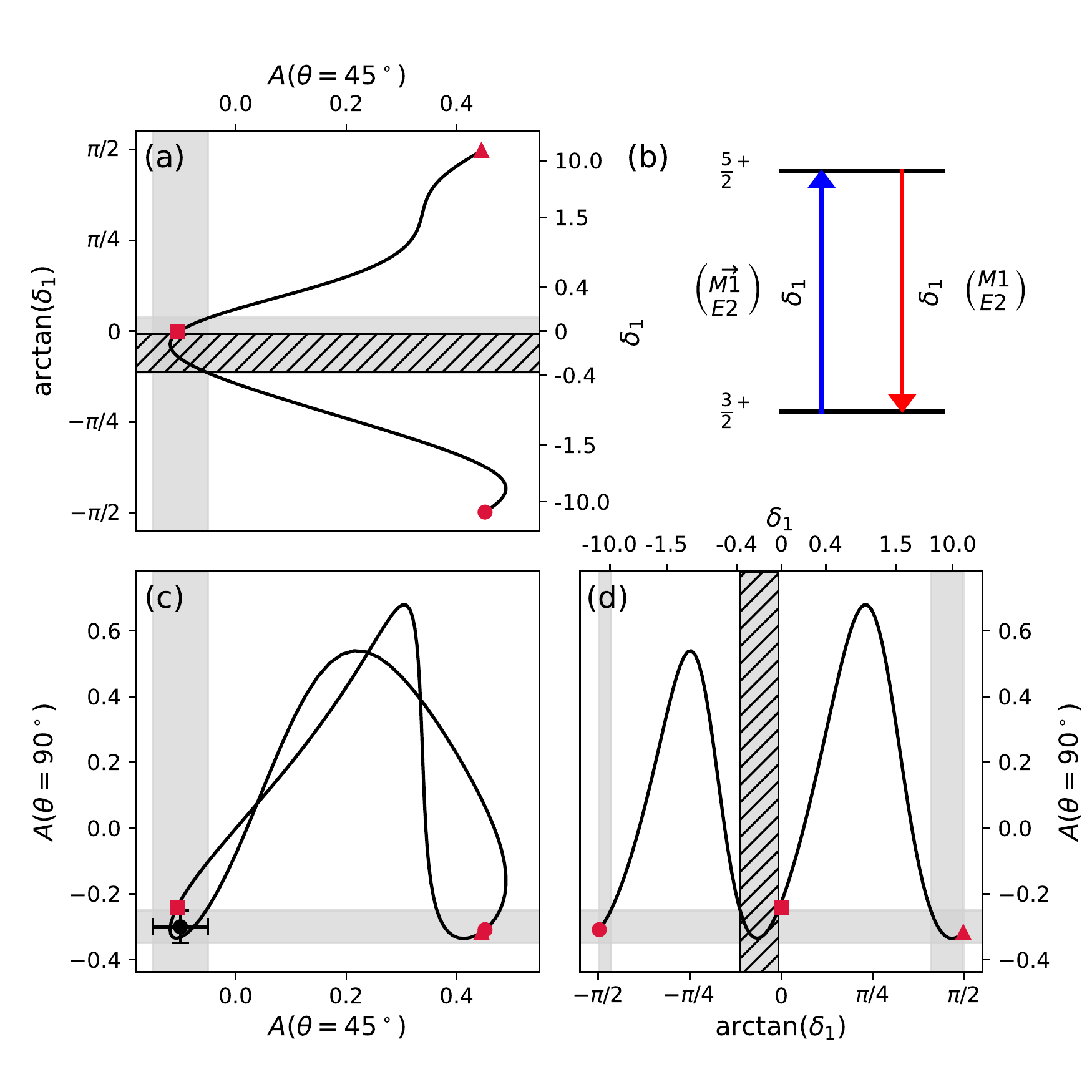}
\caption{\label{fig:analyzing_power_15_25_15} (Color online) Dependence of the analyzing powers, $A(\theta=45^\circ)$ and $A(\theta=90^\circ)$, on the multipolarity mixing ratio $\delta_1$ for the sequence $3/2^+$ $\rightarrow$ $5/2^+$ $ \rightarrow$ $3/2^+$ (see Fig.~\ref{fig:patt152515} for the corresponding radiation patterns). The level scheme in panel (b) shows the excitation to (blue arrow) and decay of (red arrow) the excited level back to the ground state. The lowest two possible multipolarities are M1 and E2, where the right arrow symbol for the first transition indicates a linearly polarized $\gamma$ ray. Since the initial and final states are identical, only a single mixing ratio, $\delta_1$, needs to be considered. Panels (a) and (d) illustrate the dependence of $A(\theta=45^\circ)$ and $A(\theta=90^\circ)$, respectively, on $\delta_1$. The red symbols signify mixing ratios of $\delta_1$ $\rightarrow$ $-\infty$ (circle), $\delta_1$ $=$ $0$ (square), and $\delta_1$ $\rightarrow$ $+\infty$ (triangle). Panel (c) shows all possible values of the analyzing powers (solid black line). A hypothetical data point covers an allowed region in the $A(\theta=90^\circ)$ {\it vs.} $A(\theta=45^\circ)$ plane, which corresponds to allowed ranges (gray shaded bands) for $\delta_1$ in panels (a) and (d). The hatched areas, showing the intersection ({\it i.e.}, conjunction) of these solutions, represent the best estimate for the mixing ratio, $\delta_1$, based on the data point.
}
\end{figure*}

Figure~\ref{fig:analyzing_power_15_25_15} presents an example for the dependence of the analyzing powers, $A(\theta=45^\circ)$ and $A(\theta=90^\circ)$, on a single multipolarity mixing ratio, $\delta_1$. The chosen sequence, $3/2^+$ $\rightarrow$ $5/2^+$ $\rightarrow$ $3/2^+$, which is depicted in panel (b), describes the elastic scattering of a photon. Since the initial and final states are identical, the mixing ratios for the two steps are the same ($\delta_1$ $=$ $\delta_2$; see Sect.~\ref{sec:phases}). Three-dimensional visualizations of the corresponding angular correlation functions for a few selected values of $\delta_1$ are given in Fig.~\ref{fig:patt152515}.

Panels (a) and (d) of Fig.~\ref{fig:analyzing_power_15_25_15} display the mixing ratio $\delta_1$ versus $A(\theta=45^\circ)$ and $A(\theta=90^\circ)$, respectively. The graphs were obtained by sampling each of the two analyzing powers on an equidistant grid in $\arctan(\delta_1)$ with $N_{\delta_1} = 101$ points. Using the sampled points, a plot of $A(\theta=90^\circ)$ {\it vs.} $A(\theta=45^\circ)$ was generated, see panel (c). Since both limits, $\delta_1 \to \infty$ and $\delta_1 \to -\infty$, correspond to a transition that is pure in the higher of the two multipolarities considered, they will yield the same values of $W_{LP}(\theta, \phi=0^\circ)$ or $A(\theta)$, according to Eqs.~(\ref{eq:wlp}), (\ref{eq:an2}), (\ref{eq:en1}). Therefore, the general form of such a graph is a closed loop. The red symbols in panels (a), (c), and (d) indicate analyzing power values obtained for mixing ratios of $\delta_1$ $\rightarrow$ $-\infty$ (circle), $\delta_1$ $=$ $0$ (square), and $\delta_1$ $\rightarrow$ $+\infty$ (triangle).

To illustrate how to use such a graph for interpreting experimental data, we superimposed in Fig.~\ref{fig:analyzing_power_15_25_15} a hypothetical data point at
\begin{align}
\label{eq:exp_single_mixing}
A \left( \theta = 45^\circ \right) &= -0.10\pm0.05 \\
A \left( \theta = 90^\circ \right) &= -0.30\pm0.05 \nonumber
\end{align}
A measurement of the asymmetry $A \left( \theta = 90^\circ \right)$ alone restricts $\delta_1$ to three possible ranges: $-\infty$ $<$ $\delta_1$ $\lesssim$ $-10.00$, $-0.36$ $\lesssim$ $\delta_1$ $\lesssim$ $-0.03$, and $3.80$ $\lesssim$ $\delta_1$ $\lesssim$ $\infty$ (gray bands in panel (d)). This ambiguous result does not allow to determine whether the transition is mainly of $M1$ or $E2$ character. The ambiguity is resolved by the simultaneous measurement of the asymmetry $A \left( \theta = 45^\circ \right)$. By itself, as shown in panel (a), it restricts the multipolarity mixing ratio to the range of $-0.36$ $\lesssim$ $\delta_1$ $\lesssim$ $0.10$, which favors a transition of predominant M1 character. The hatched areas depict the intersection of all solutions based on the measurements of both $A(\theta=90^\circ)$ and $A(\theta=45^\circ)$, which provides the best overall estimate, $-0.36$ $\lesssim$ $\delta_1$ $\lesssim$ $-0.03$. The angular correlation that gave rise to the hypothetical data point in Fig.~\ref{fig:analyzing_power_15_25_15} must have had a shape somewhere between the radiation patterns seen in the left and middle bottom panels of Fig.~\ref{fig:patt152515}.

Graphs similar to Fig.~\ref{fig:analyzing_power_15_25_15} are presented in App.~\ref{app:powers} for spin sequences of practical interest involving only a single multipolarity mixing ratio. We also included graphs for three--step processes when the intermediate transition is unobserved. The figure numbers are listed in Table~\ref{tab:summary}.

\subsection{Two multipolarity mixing ratios}
\label{sec:double_mixing_ratio}
When both the first and second $\gamma$-ray transitions are of mixed multipolarity, and the initial and final states are different, the values of $A(\theta=45^\circ)$ and $A(\theta=90^\circ)$ define a (not necessarily continuous) region of possible mixing ratios in the $\delta_1$ {\it vs.} $\delta_2$ plane. To visualize this four-dimensional problem, we will display the minimum and maximum values of both mixing ratios ($\delta_\mathrm{1,min}$, $\delta_\mathrm{1,max}$, $\delta_\mathrm{2,min}$, $\delta_\mathrm{2,max}$), if they exist, for all possible combinations of the two analyzing powers, $A(\theta=45^\circ)$ and $A(\theta=90^\circ)$. To this end, similar to the procedure outlined in Sect.~\ref{sec:single_mixing_ratio}, the analyzing powers were computed on an equidistant two-dimensional grid in $\arctan(\delta_1)$ and $\arctan(\delta_2)$ with $N_{\delta_1}$ $\times$ $N_{\delta_2} = 301 \times 301$ points. The resulting pairs of values were then sorted into a two-dimensional array with $N_A$ $\times$ $N_A$ equidistant bins between the respective minimum and maximum values of the analyzing powers. 

\begin{figure*}
\includegraphics[width=1.9\columnwidth]{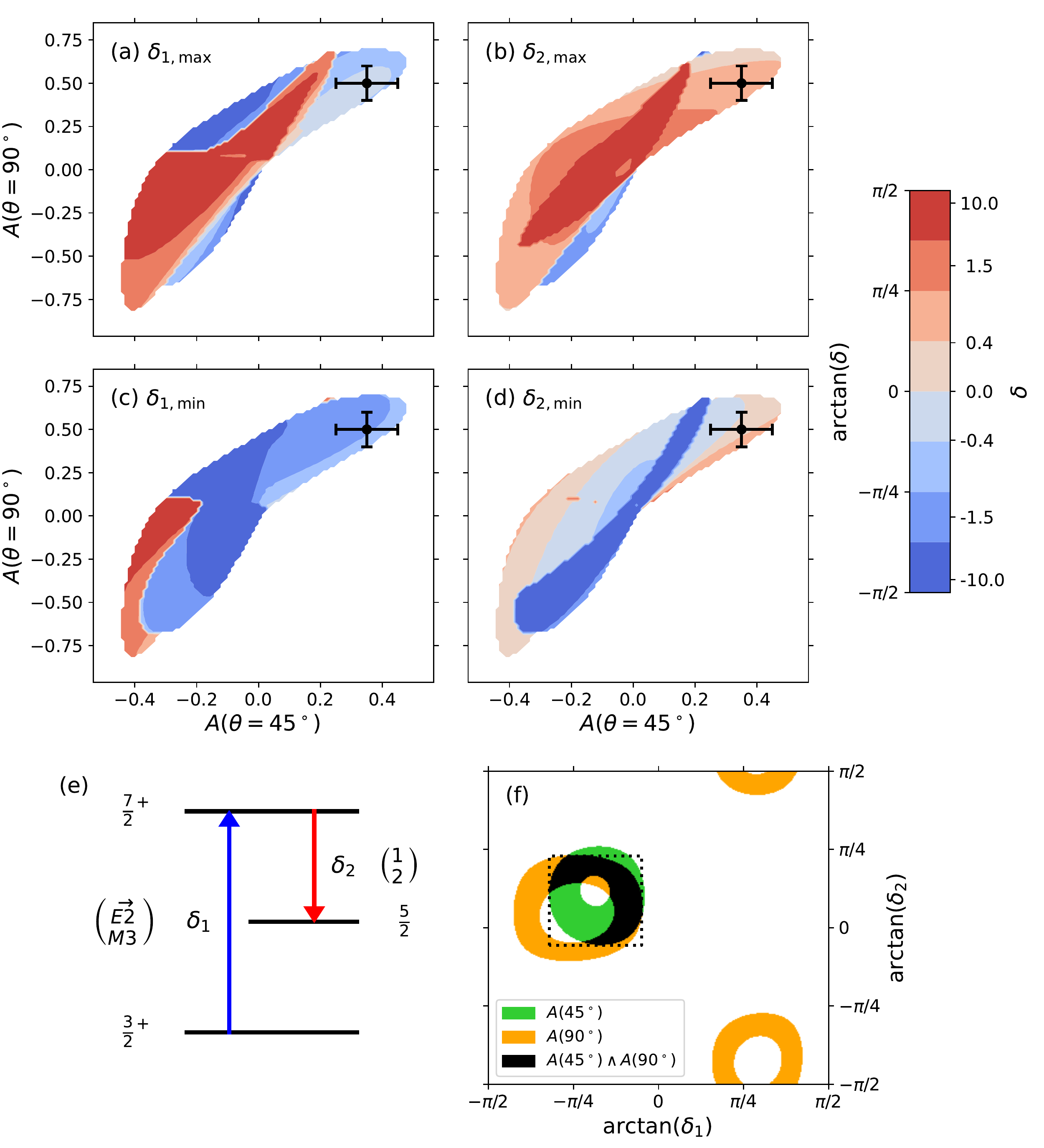} 
\caption{\label{fig:ana_357555} (Color online) (a), (b) Maximum ($\delta_\mathrm{max}$) and (c), (d) minimum ($\delta_\mathrm{min}$) possible values of $\delta_1$ (first column) and $\delta_2$ (second column) for given analyzing powers $A(\theta=45^\circ)$ and $A(\theta=90^\circ)$. The example is for the sequence $3/2^+$ $\rightarrow$ $7/2^+$ $\rightarrow$ $5/2$, which is depicted in panel (e) (see also Fig.~\ref{fig:patt153525}). For the upper four panels, the color key on the right-hand side indicates, for given regions, the values of the mixing ratios, with red and blue representing positive and negative values, respectively. The white areas indicate regions that cannot be reached by any combination of the two mixing ratios, $\delta_1$ and $\delta_2$. A data point, representing hypothetical values of $A(\theta=45^\circ)$ and $A(\theta=90^\circ)$, is superimposed in the top four panels. Panel (f) highlights regions in the $\delta_1$ {\it vs.} $\delta_2$ plane that are consistent with the measured $A(\theta=45^\circ)$ value alone (green area), the measured $A(\theta=90^\circ)$ value alone (orange areas), and the simultaneous measurement of $A(\theta=45^\circ)$ and $A(\theta=90^\circ)$ (black area). The dotted rectangle indicates the approximate solutions obtained from the $\delta_\mathrm{min}$--$\delta_\mathrm{max}$ method presented in the top four panels (see text).
}
\end{figure*}

Figure~\ref{fig:ana_357555} depicts an example of such a graph for the sequence $3/2^+$ $\rightarrow$ $7/2^+$ $\rightarrow$ $5/2$ (lower left panel). The upper four panels show $\delta_{min}$ and $\delta_{max}$ for the two transitions in the $A(90^\circ)$ $vs.$ $A(45^\circ)$ plane. Positive and negative values of the mixing ratios are displayed in shades of red and blue, respectively. Three-dimensional visualizations of the corresponding polarization-direction correlation functions for selected combinations of $\delta_1$ and $\delta_2$ are illustrated in Fig.~\ref{fig:patt153525}.

The white areas in the top four panels of Fig.~\ref{fig:ana_357555} indicate that many values of analyzing power values are forbidden for this spin sequence. Similar to Sect.~\ref{sec:single_mixing_ratio}, we add a hypothetical data point to the upper four panels at
\begin{align}
\label{eq:exp_double_mixing}
A \left( \theta = 45^\circ \right) &= 0.35\pm0.10 \\
A \left( \theta = 90^\circ \right) &= 0.50\pm0.10 \nonumber
\end{align}
In panel (a), which shows the maximum value of $\delta_1$, the mean value of the data point, $\left( 0.35, 0.5 \right)$, is located in a region with $-0.4$ $\lesssim$ $\delta_{1,\mathrm{max}}$ $\lesssim$ $0.0$. In panel (c), which depicts the minimum value of $\delta_1$, the mean value of the data point is located at the boundary between two contours, which provides a constraint of $-2.0$ $\lesssim$ $\delta_{1, \mathrm{min}}$ $\lesssim$ $-0.4$. Taken together, these results yield an allowed range of $-2.0$ $\lesssim$ $\delta_1$ $\lesssim$ $0.0$. Similarly, from panels (b) and (d) we find, for the mixing ratio of the second step, an allowed range of $-0.4$ $\lesssim$ $\delta_2$ $\lesssim$ $1.0$. 

The method discussed above provides only an approximate solution, representing a rectangular region in the $\delta_1$ {\it vs.} $\delta_2$ plane, which is depicted by the dotted rectangle in panel (f) of Fig.~\ref{fig:ana_357555}. The actual solutions, found numerically, that agree with the hypothetical data point are presented there as well. The green and orange areas correspond to solutions resulting from a measurement of either $A(\theta=45^\circ)$ or $A(\theta=90^\circ)$, respectively, while the black half-moon shaped area is consistent with the simultaneous measurement of both analyzing powers. It can be seen that the half-moon shape imposes tighter constraints than the rectangular area on the values of the two mixing ratios $\delta_1$ and $\delta_2$.

If a data point is located in certain regions of the $A(\theta=90^\circ)$ {\it vs.} $A(\theta=45^\circ)$ plane, the two mixing ratios will remain unconstrained. For example, this is the case for a data point near the center of the upper four panels, at about $\left( -0.1, -0.1 \right)$, where the magnitudes of the minimum and maximum mixing ratios range from large positive values (dark red) to large negative ones (dark blue). In such cases, the simple four-detector setup assumed here must be modified, either by adding more detectors at suitable angles or by measuring the polarization of the emitted $\gamma$ ray (see, {\it e.g.}, Ref.~\cite{Fagg:1959wo}).  The visual representations of the angular correlation patterns presented in this work will be helpful for designing such experiments.

\section{Comments on the data analysis}
\label{sec:analyzing_powers_use}

The expressions presented so far apply to ideal conditions since they disregard experimental artifacts. For example, in a real experiment, the polarization of the incoming $\gamma$-ray beam may not be perfectly linear, or the finite solid angle coverage of the detectors or the spatial extent of the radiation source may not be negligible.

For achieving adequate count rates, detectors must not be located too far from the sample. On the other hand, when the detector solid angle exceeds 10\% of $4\pi$, the measured angular correlation will be much closer to isotropy compared to a point detector (see, e.g., Fig.~14.1 in \cite{Hamilton:1975ug}). A compromise must then be reached between an acceptable detection efficiency and a distinct correlation pattern. In practice, this means one needs to accurately correct for the attenuation of the radiation pattern for a given detector. Rose \cite{Rose1953} showed that this correction can be performed by multiplying each coefficient in the Legendre polynomial expansion by a solid-angle attenuation factor. According to Eqs.~(\ref{eq:wdd}) and (\ref{eq:wlp}), the corrected polarization--direction correlation can then be written as
\begin{eqnarray}
W_{corr}(\theta\phi) = A_0(1)A_0(2) + \sum_{n=2,4,...} A_n(2)Q_n(k) \times \nonumber \\
\left[ A_n(1)P_n(\cos \theta) + E_n(1)P_n^{|2|}(\cos \theta)\cos(2\phi) \right] 
\label{eq:Wcorr}
\end{eqnarray}
where $Q_n(k)$ is the attenuation factor for detector $k$ located at a given distance and angles ($\theta$,$\phi$). Only if the measurement employs identical detectors at exactly the same distance under the same shielding conditions will the factors $Q_n$ be independent of location. In this case, the corrected analyzing power, $A_{corr}(\theta)$, has the same form as Eq.~(\ref{eq:analyzing_w}), but will contain factors $Q_n$ in the sums of both the numerator and denominator (see, {\it e.g.}, Eq.~(15.10) in Ref.~\cite{Twin:1975ug}). Otherwise, the corrected analyzing power should be obtained from Eqs.~(\ref{eq:analyzing}) and (\ref{eq:Wcorr}). 

In principle, $Q_n$ values can be determined by measuring well established correlations showing a strong anisotropy and corresponding approximately in energy to the actual case being studied. Calculated and measured attenuation factors for scintillation counters and germanium detectors can be found in Ref.~\cite{Rose1953} and Ref.~\cite{Bolotin1980}, respectively. Values of $Q_n$ computed using a Monte Carlo method are listed in Ref.~\cite{CAMP1969192}.

An imperfect linear polarization can be interpreted as a superposition of $N$ linearly polarized $\gamma$-ray beams with different orientations of the polarization vector. Formally, a rotation of the polarization vector is obtained by adding an offset, $\Delta \phi$, to the azimuthal angle. Therefore, we find for the absolute value of the numerator in Eq.~(\ref{eq:analyzing_w}) the inequality
\begin{align}
&\left| \frac{1}{N} \left[ W_{LP} \left( \theta, 0 \right) + \sum_i W_{LP}\left( \theta, \Delta \phi_i \right) \right] \right| \nonumber \\
=&\left| \frac{W_{LP} \left( \theta, 0 \right)}{N} \left[ 1 + \sum_i \cos \left( 2 \Delta \phi_i  \right) \right] \right| \leq \left| W_{LP} \left( \theta, 0 \right) \right|
\end{align}
In other words, an imperfect polarization always results in a reduction of the absolute value of the asymmetry, and the effect can be described by a multiplicative correction factor (see, {\it e.g.}, Ref.~\cite{kneissl96}).

Correction factors for the finite size of the sample, including the self-absorption of $\gamma$ rays, can be found in Ref.~\cite{Overwater1993}. A modern approach to considering all these effects in the data analysis involves the numerical simulation of the interactions of radiation with matter. These analysis techniques are based on implementing angular correlations, such as those presented here, into Monte Carlo radiation transport codes, {\it e.g.}, the toolkits Geant4 \cite{Agostinelli2003,Allison2006,Allison2016} or MCNP \cite{Werner2017}.

We have demonstrated above the usefulness of a minimal detection geometry consisting of four counters positioned at suitable spacial locations (see red dots in Fig.~\ref{fig:geo2}). But, sometimes, more complicated situations may be encountered. For example, a two-step cascade may involve two different mixing ratios, and at the same time, the spin and parity of the intermediate level may be unknown. Or, the measured angular correlation functions may exhibit only weak anisotropies. An example is depicted in Fig.~\ref{fig:patt251525}, demonstrating the difficulty of distinguishing between pure dipole radiation (left panel) and pure quadrupole radiation (right panel) if the angular correlation function cannot be measured with sufficient precision. In such cases, detectors may need to be positioned at additional angles, and the data analysis should employ likelihood-based statistical methods, such as those suggested by the Particle Data Group \cite{pdg20} or the Joint Committee for Guides in Metrology \cite{gum08}.

\section{Summary} \label{sec:summary}
We reviewed the general angular correlation formalism for scattering experiments using linearly polarized incident $\gamma$ ray beams. Such measurements are powerful probes for determining spins and parities of nuclear levels. When more than one $\gamma$-ray multipolarity can participate in the transitions, the measurement of the angular correlation also provides estimates of the multipolarity mixing ratios. These depend on the nuclear matrix elements and are, therefore, sensitive to the relative phases of the wave functions. Since considerable confusion still exists in the literature regarding these relative phases, we reviewed the phase relationships among different formalisms.

As an aid for designing nuclear resonance fluorescence (NRF) experiments, we computed the angular correlation functions and illustrated their graphical representations for spin sequences of practical interest. We proposed a minimal detection geometry consisting of four counters that are positioned at suitable spacial locations. Such a setup allows for measuring two independent analyzing powers and will facilitate the determination of spins, parities, and mixing ratios in many situations. The dependence of the analyzing powers on the multipolarity mixing ratios are presented graphically for spin sequences of practical interest.

Computer codes for calculating the angular correlation functions, written in R or Python, can be requested from the first or second author, respectively.

\begin{acknowledgement}
We would like to thank Robert Janssens, Gencho Rusev, and Filip Kondev for helpful comments. Supported by the DOE, Office of Science, Office of Nuclear Physics, under grants DE-FG02-97ER41041 (UNC) and DE-FG02-97ER41033 (TUNL).  
\end{acknowledgement}


\appendix\label{sec:app}

\section{Three-dimensional angular correlation patterns}\label{app:pattern}
In the following, we show three-dimensional representations of linear polarization-direction correlations for the spin-parity sequences listed in Table~\ref{tab:summary}. The incident linearly polarized $\gamma$-ray beam moves along the positive $x$-axis. The colors in each panel (from red to blue) signify the deviation from isotropy, with the color key given on the right-hand side. The plane of polarization coincides with the $x$--$y$ plane. In each panel, the scattering target is located at the geometrical center of the displayed pattern. The same linear polarization--direction correlation functions are obtained for positive or negative parity, $\pi_2$, of the final state, $j_2$. The same correlation functions are also obtained if the parities of the initial ($\pi_1$) and intermediate ($\pi$) states are flipped simultaneously. To visualize the linear polarization--direction correlation functions when the parity of either the initial or intermediate state is flipped, rotate each of the displayed patterns by $90^\circ$ about the incident beam direction ($x$-axis).

\begin{figure*}
\includegraphics[width=2\columnwidth]{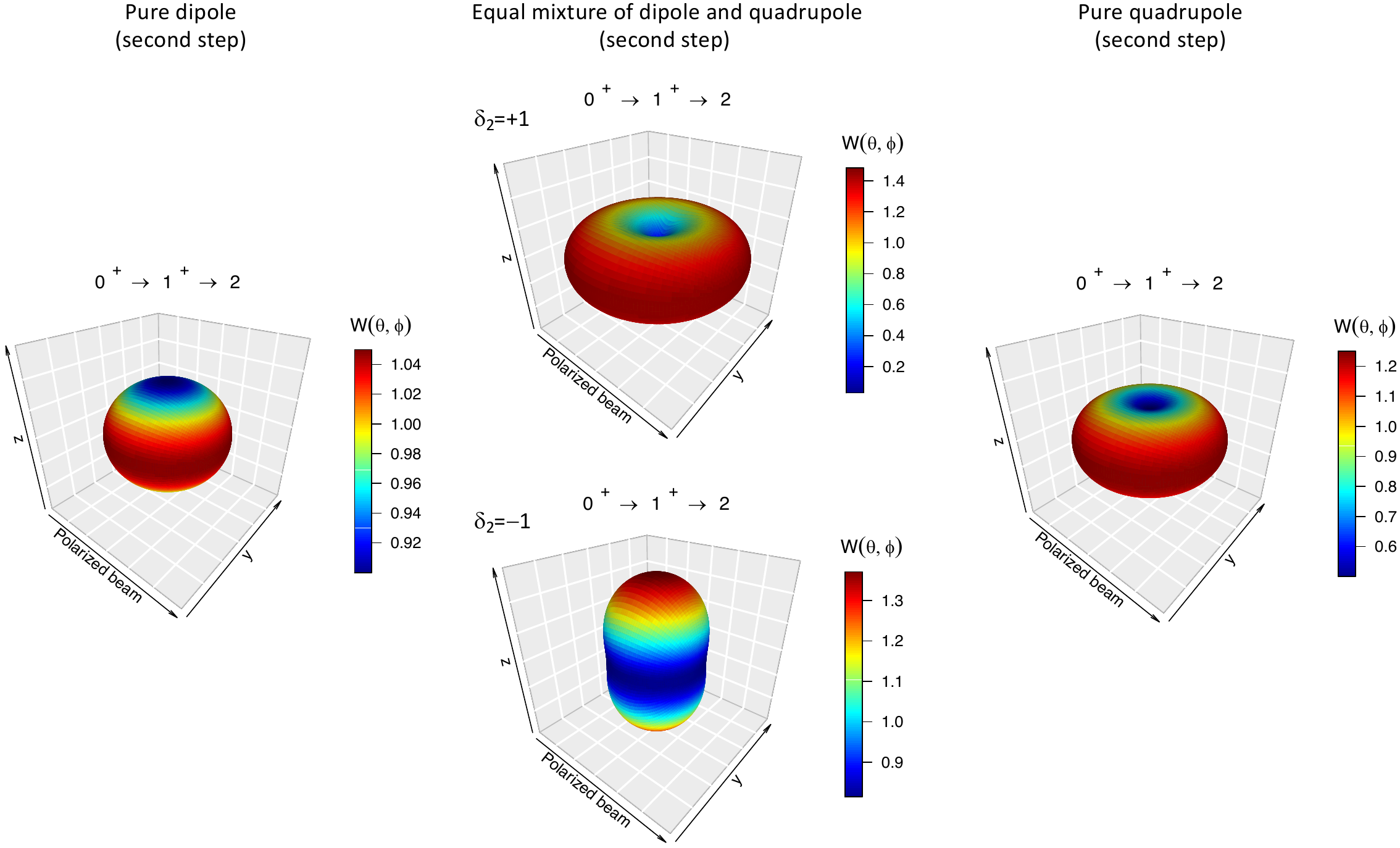} 
\caption{(Color online) Three-dimensional representations of linear polarization--direction correlations for the sequence $0^+$ $\rightarrow$ $1^+ $ $\rightarrow$ $2$.
}
\label{fig:patt012}
\end{figure*}

\begin{figure*}
\includegraphics[width=2\columnwidth]{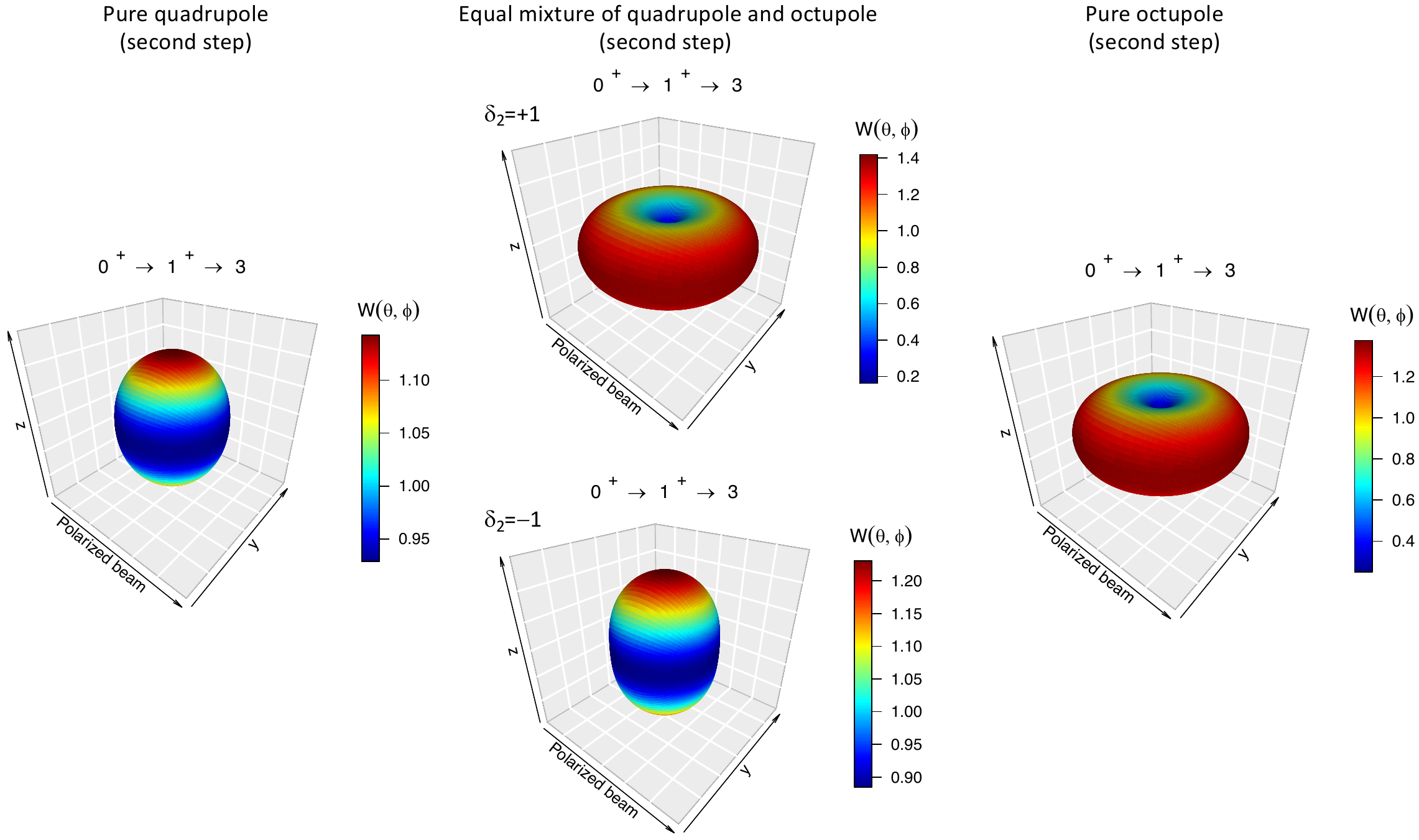}
\caption{(Color online) Three-dimensional representations of linear polarization--direction correlations for the sequence $0^+$ $\rightarrow$ $1^+$ $\rightarrow$ $3$.
}
\label{fig:patt013}
\end{figure*}

\begin{figure*}
\includegraphics[width=2\columnwidth]{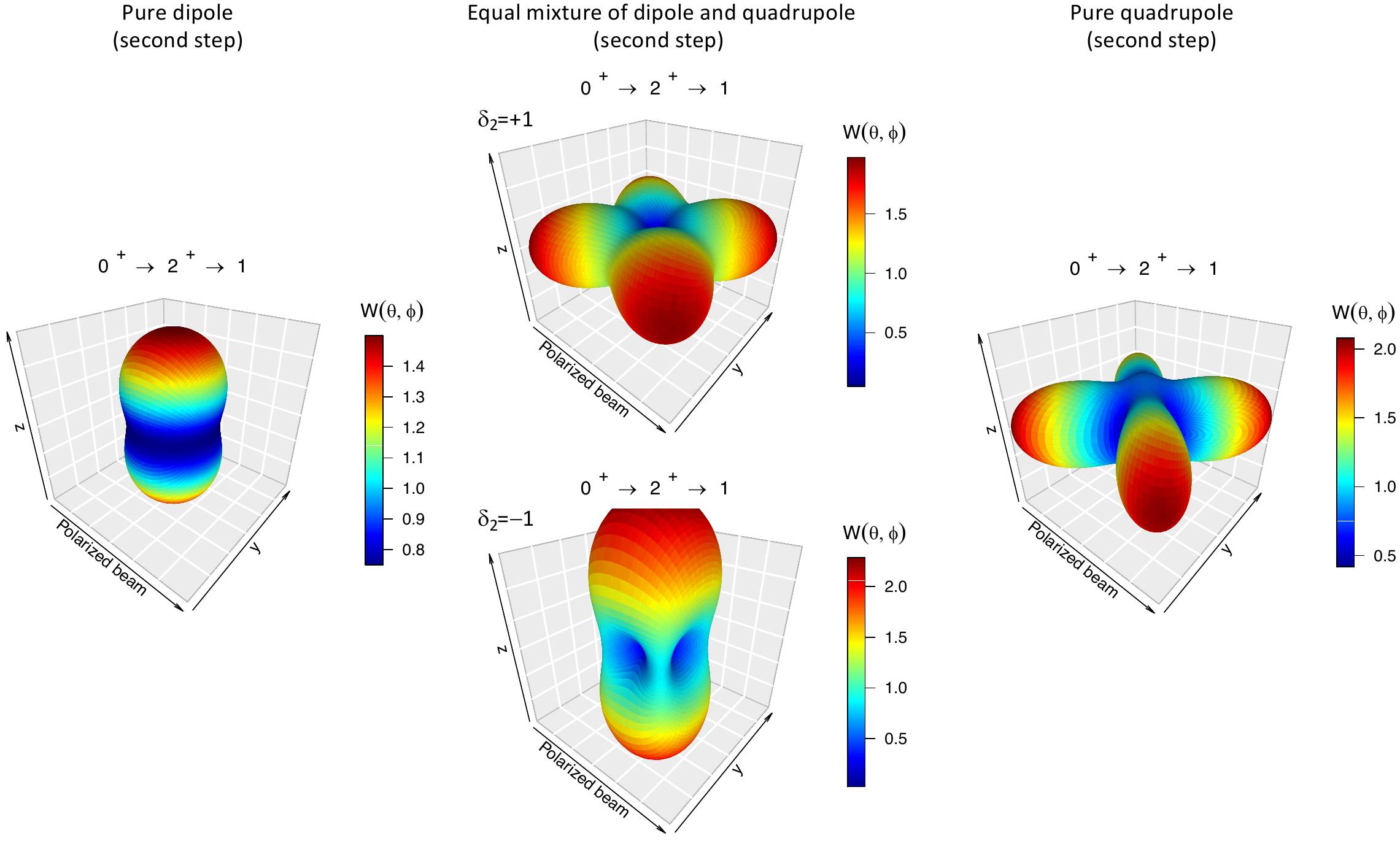} 
\caption{(Color online) Three-dimensional representations of linear polarization--direction correlations for the sequence $0^+$ $\rightarrow$ $2^+ $ $\rightarrow$ $1$.
}
\label{fig:patt021}
\end{figure*}

\begin{figure*}
\includegraphics[width=2\columnwidth]{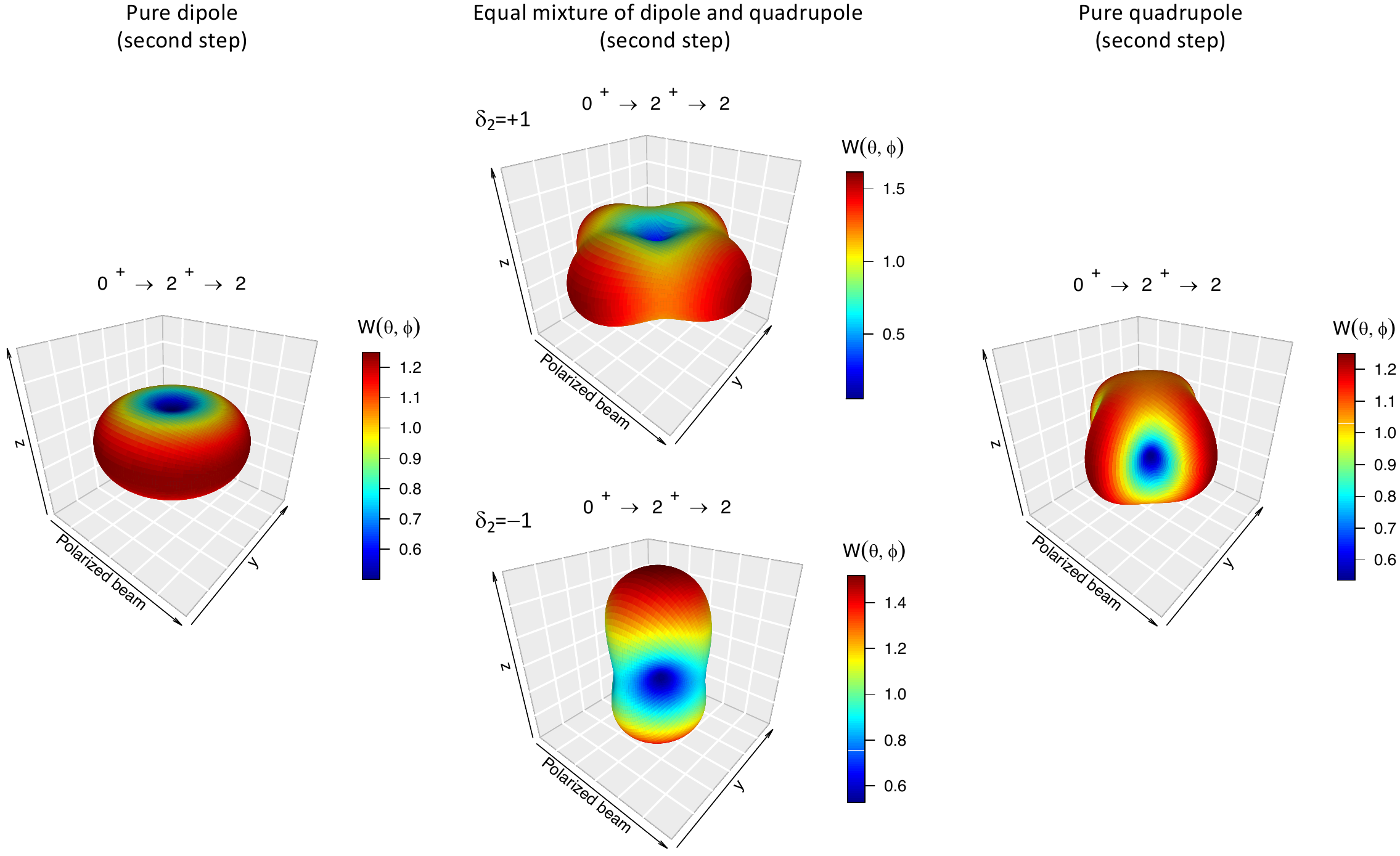} 
\caption{(Color online) Three-dimensional representations of linear polarization--direction correlations for the sequence $0^+$ $\rightarrow$ $2^+$ $\rightarrow$ $2$.
}
\label{fig:patt022}
\end{figure*}

\begin{figure*}
\includegraphics[width=2\columnwidth]{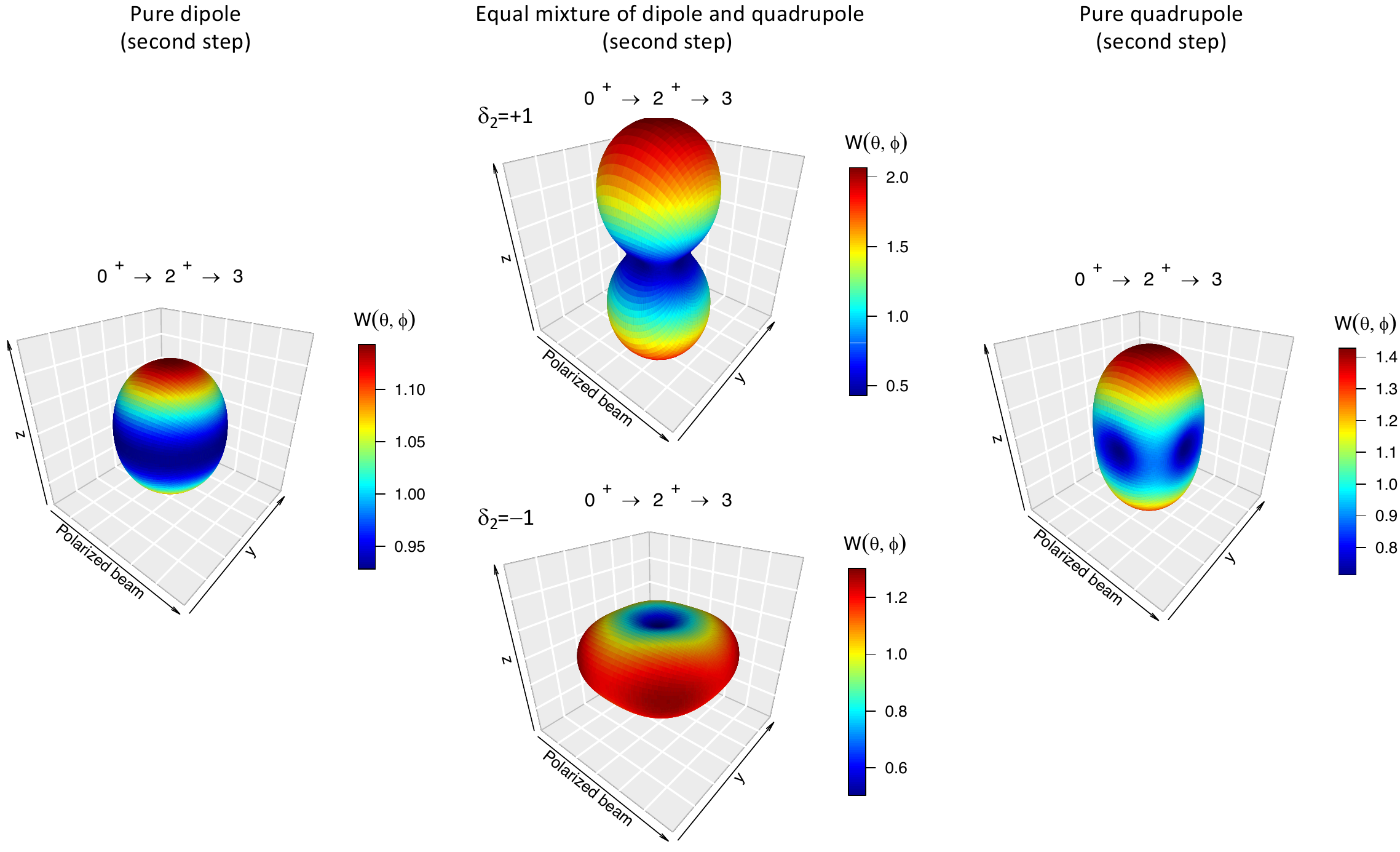} 
\caption{(Color online) Three-dimensional representations of linear polarization--direction correlations for the sequence $0^+$ $\rightarrow$ $2^+ $ $\rightarrow$ $3$.
}
\label{fig:patt023}
\end{figure*}

\begin{figure*}
\includegraphics[width=2\columnwidth]{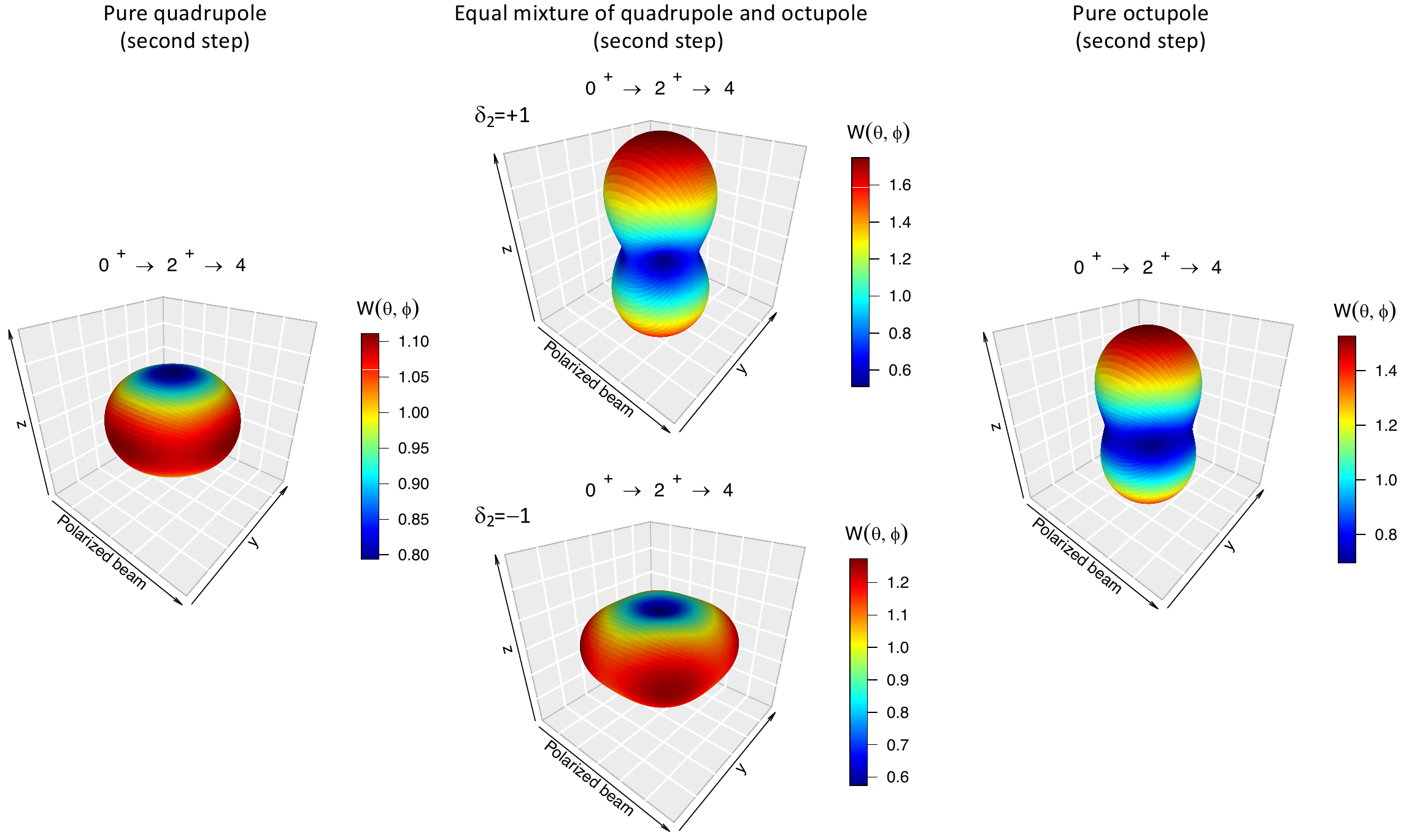} 
\caption{(Color online) Three-dimensional representations of linear polarization--direction correlations for the sequence $0^+$ $\rightarrow$ $2^+$ $\rightarrow$ $4$.
}
\label{fig:patt024}
\end{figure*}

For even-mass nuclei ({\it i.e.}, integer spin sequences), the panels on the left and right in Figs.~\ref{fig:patt012}--\ref{fig:patt024} correspond to pure transitions for the second step. The two middle panels in each figure correspond to an equal mixture of multipolarities ($| \delta |$ $=$ $1$), but with opposite signs of the mixing ratio. See Sect.~\ref{sec:visual2}.

\begin{figure*}
\includegraphics[width=2\columnwidth]{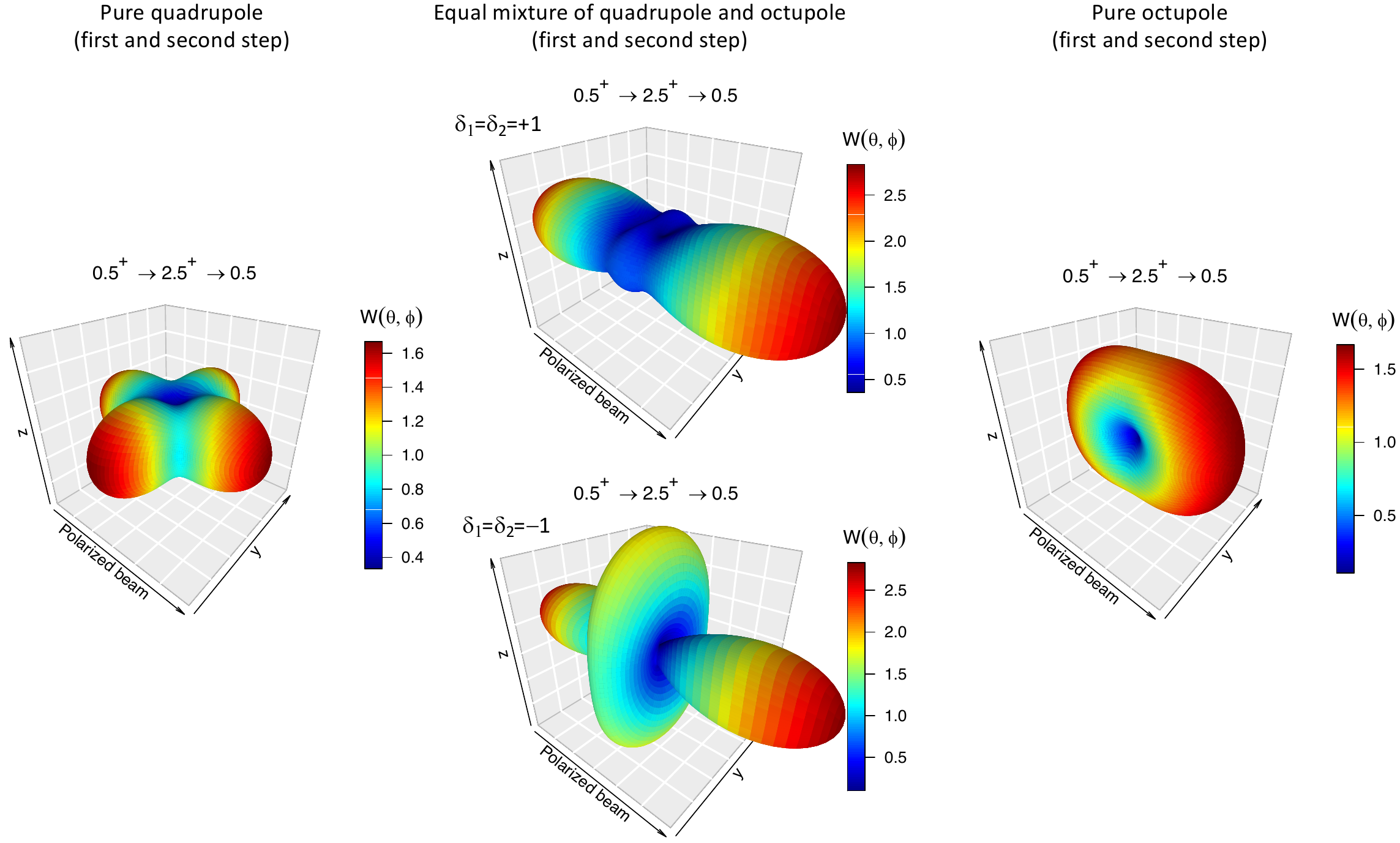} 
\caption{(Color online) Three-dimensional representations of linear polarization--direction correlations for the sequence $1/2^+$ $\rightarrow $ $5/2^+$ $\rightarrow$ $1/2$.
}
\label{fig:patt052505}
\end{figure*}

\begin{figure*}
\includegraphics[width=2\columnwidth]{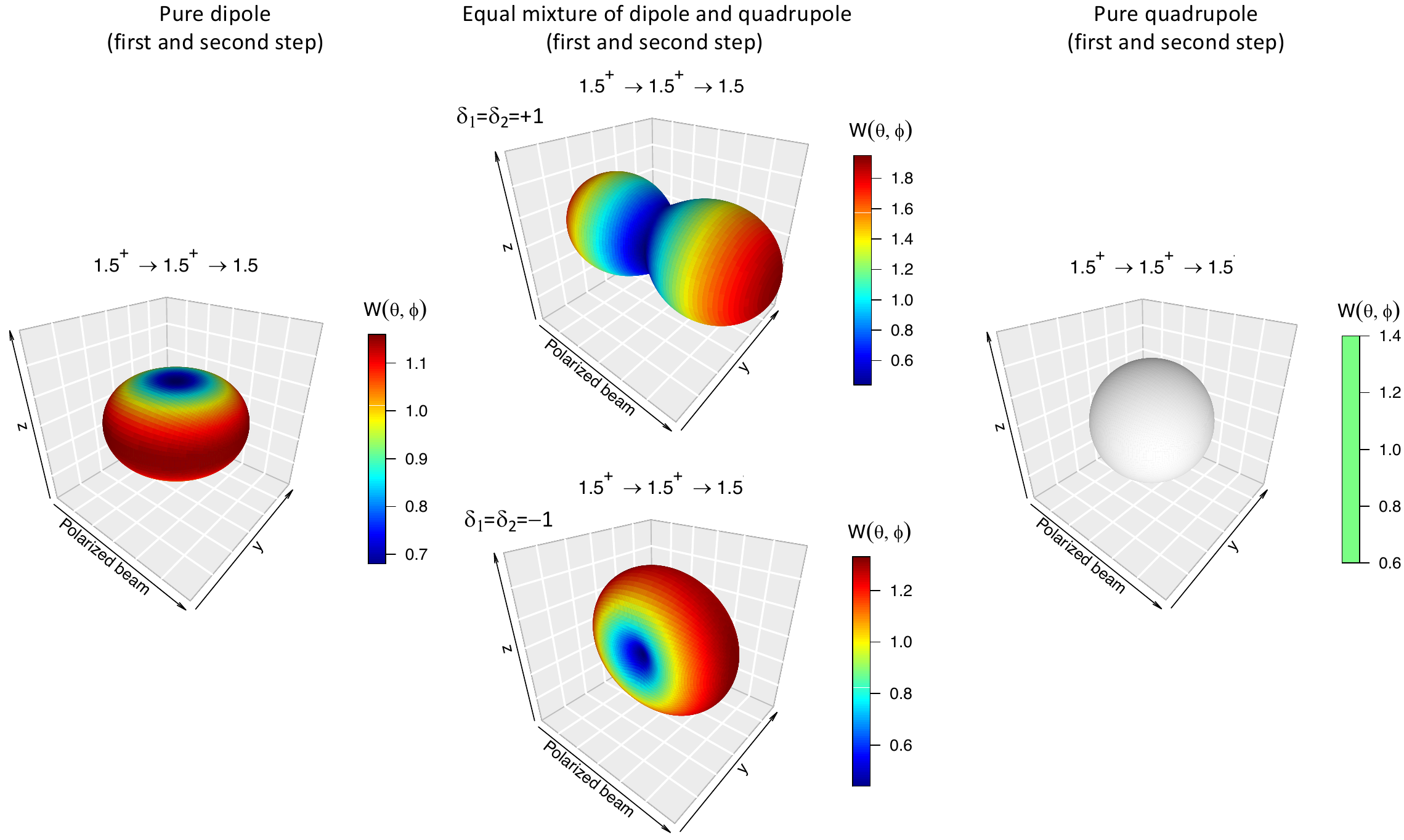} 
\caption{(Color online) Three-dimensional representations of linear polarization--direction correlations for the sequence $3/2^+$ $\rightarrow $ $3/2^+$ $\rightarrow$ $3/2$. Notice the isotropic emission pattern in the panel on the right.
}
\label{fig:patt151515}
\end{figure*}

\begin{figure*}
\includegraphics[width=2\columnwidth]{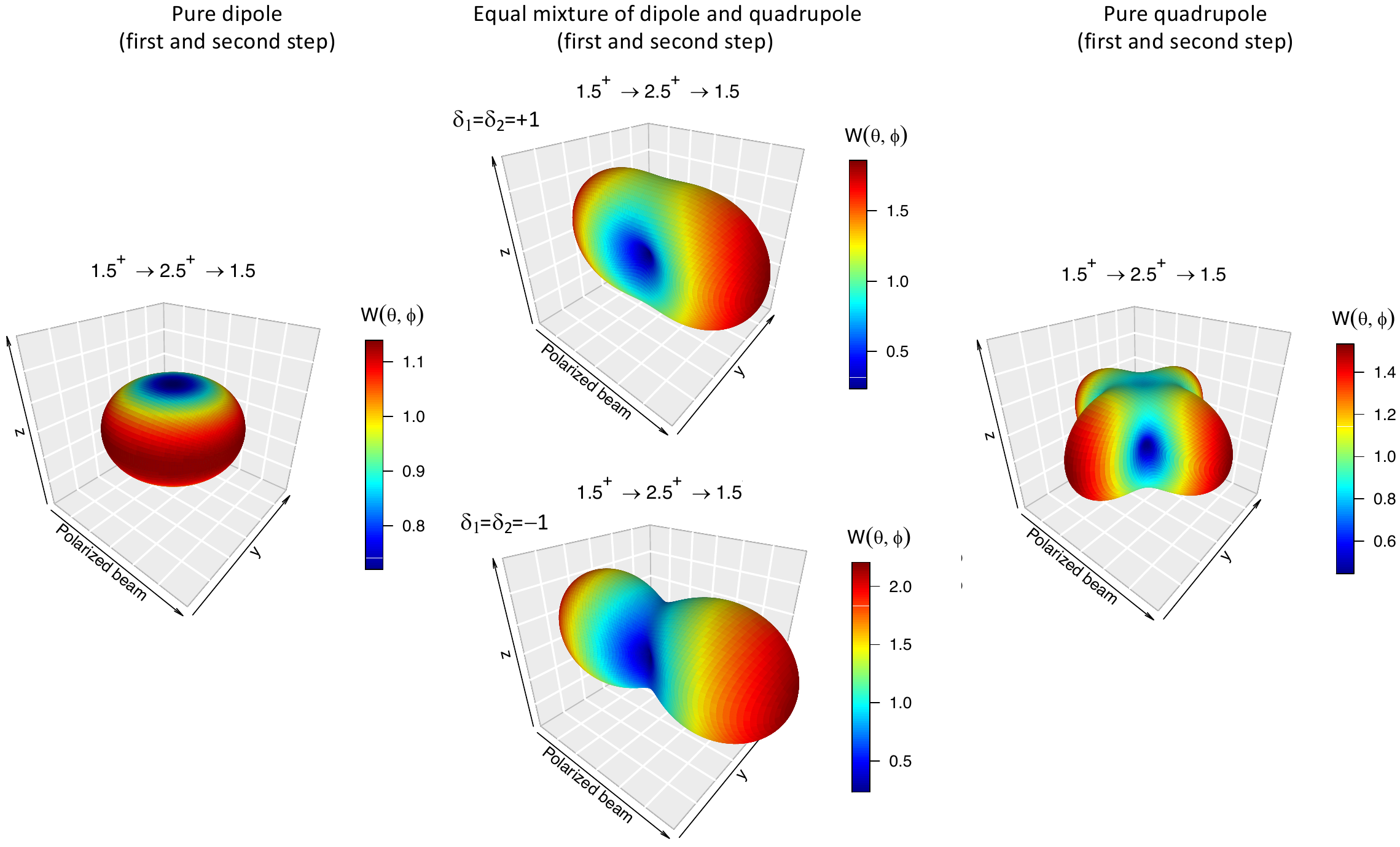} 
\caption{(Color online) Three-dimensional representations of linear polarization--direction correlations for the sequence $3/2^+$ $\rightarrow $ $5/2^+$ $\rightarrow$ $3/2$.
}
\label{fig:patt152515}
\end{figure*}

\begin{figure*}
\includegraphics[width=2\columnwidth]{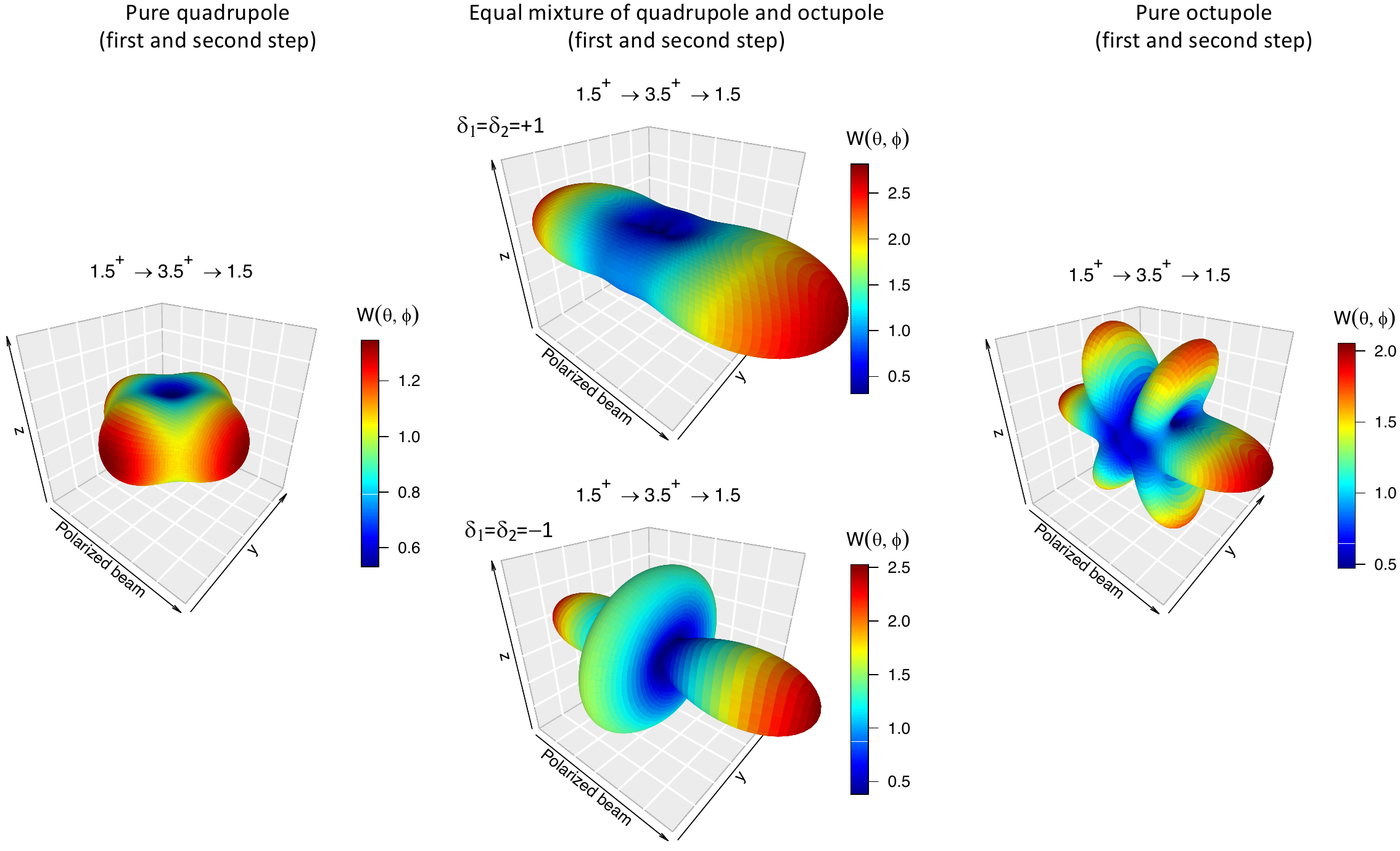} 
\caption{(Color online) Three-dimensional representations of linear polarization--direction correlations for the sequence $3/2^+$ $\rightarrow $ $7/2^+$ $\rightarrow$ $3/2$.
}
\label{fig:patt153515}
\end{figure*}

\begin{figure*}
\includegraphics[width=2\columnwidth]{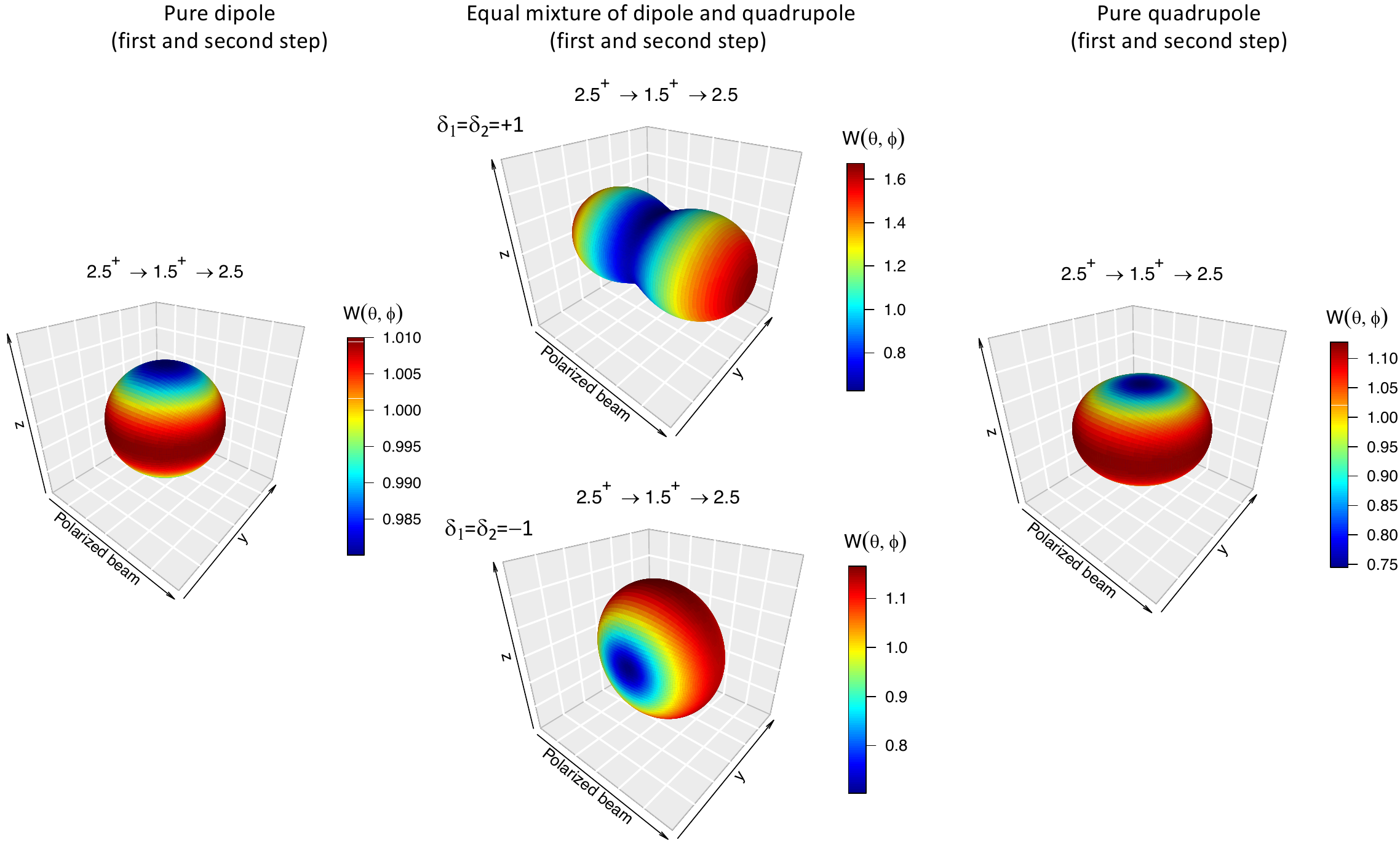} 
\caption{(Color online) Three-dimensional representations of linear polarization--direction correlations for the sequence $5/2^+$ $\rightarrow $ $3/2^+$ $\rightarrow$ $5/2$.
}
\label{fig:patt251525}
\end{figure*}

\begin{figure*}
\includegraphics[width=2\columnwidth]{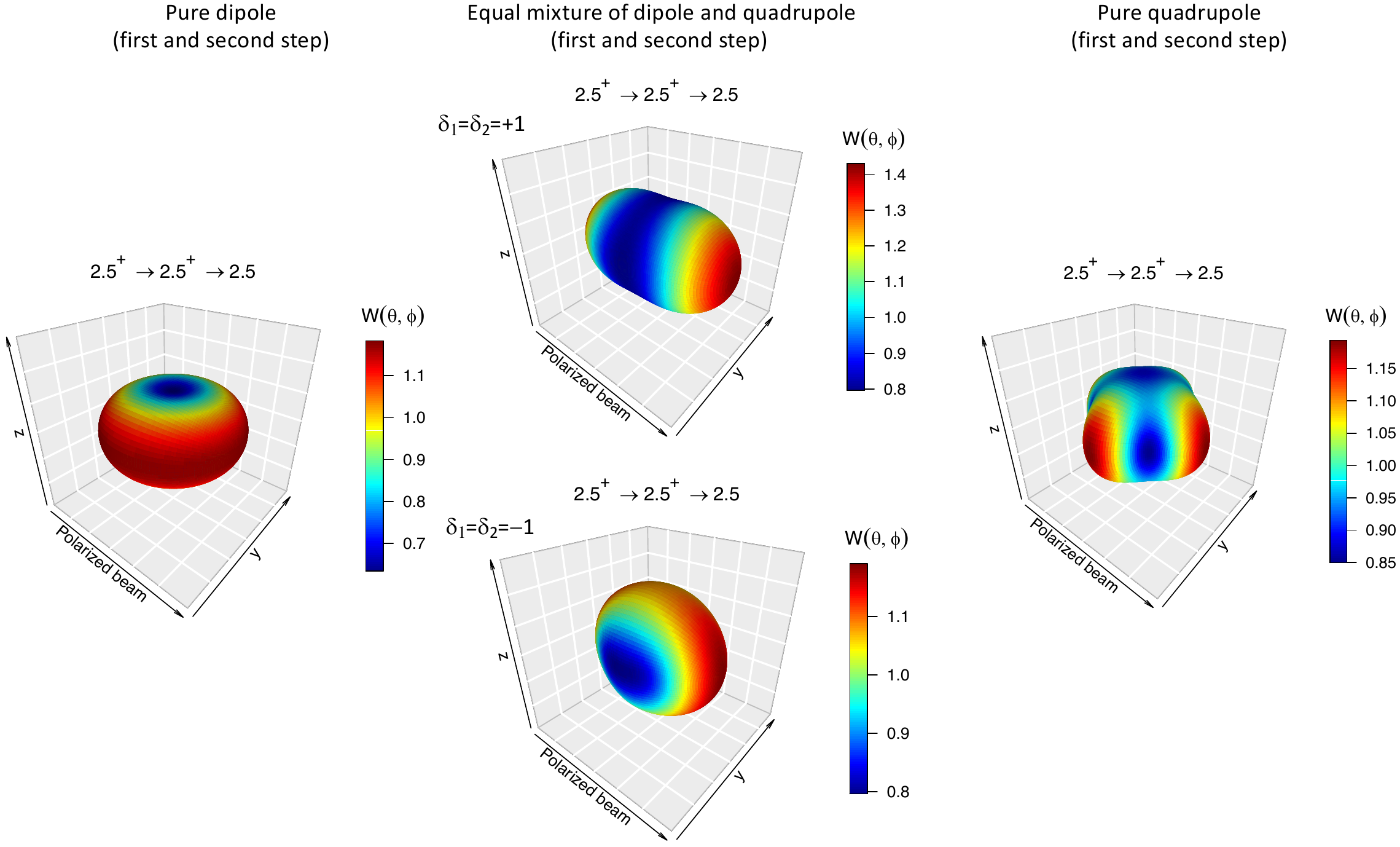} 
\caption{(Color online) Three-dimensional representations of linear polarization--direction correlations for the sequence $5/2^+$ $\rightarrow $ $5/2^+$ $\rightarrow$ $5/2$.
}
\label{fig:patt252525}
\end{figure*}

\begin{figure*}
\includegraphics[width=2\columnwidth]{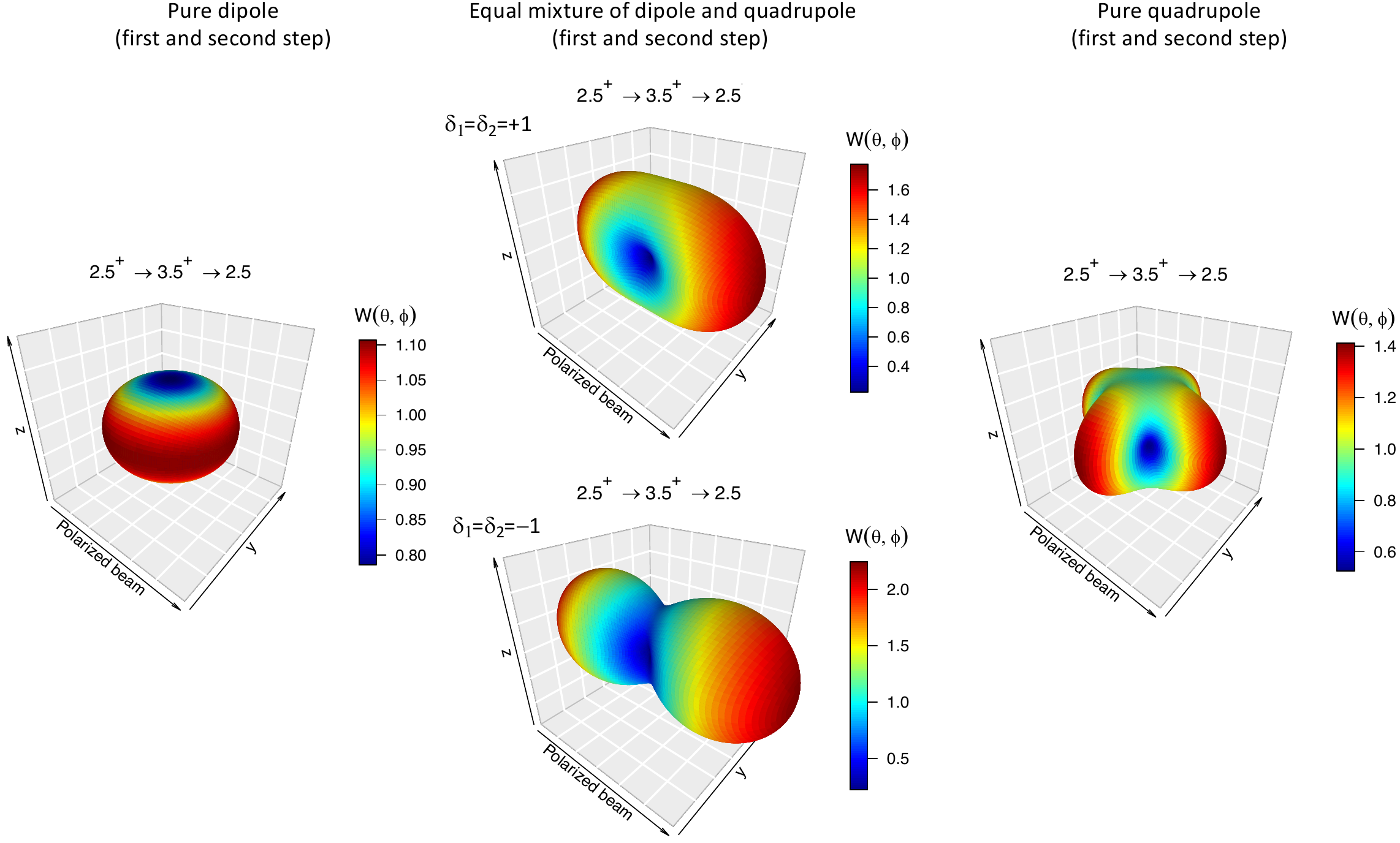} 
\caption{(Color online) Three-dimensional representations of linear polarization--direction correlations for the sequence $5/2^+$ $\rightarrow $ $7/2^+$ $\rightarrow$ $5/2$.
}
\label{fig:patt253525}
\end{figure*}

\begin{figure*}
\includegraphics[width=2\columnwidth]{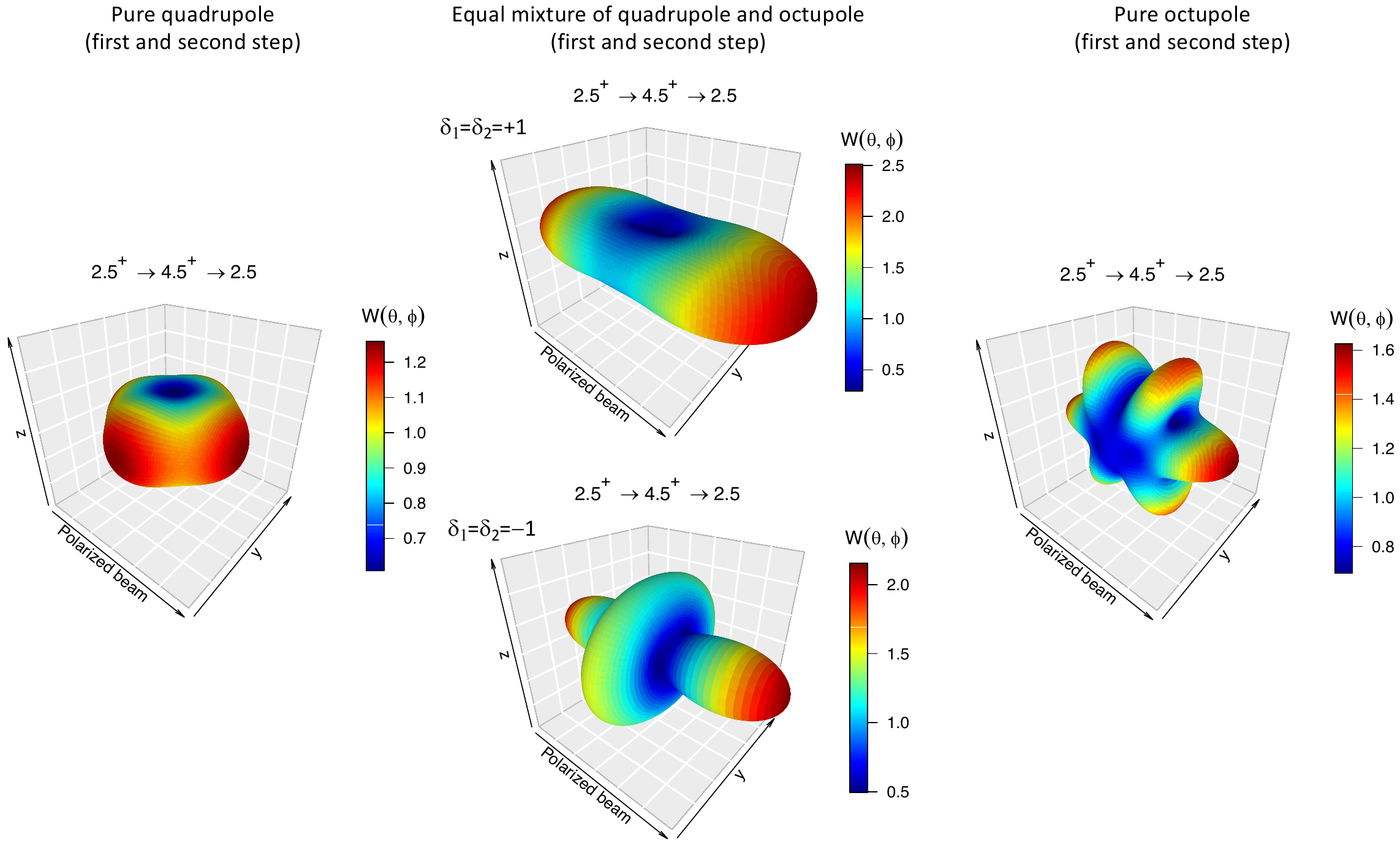} 
\caption{(Color online) Three-dimensional representations of linear polarization--direction correlations for the sequence $5/2^+$ $\rightarrow $ $9/2^+$ $\rightarrow$ $5/2$.
}
\label{fig:patt254525}
\end{figure*}

For odd-mass nuclei ({\it i.e.}, half-integer spin sequences), the panels on the left and right in Figs.~\ref{fig:patt052505}--\ref{fig:patt254525} correspond to pure transitions assuming the same multipolarity for the first and second step ($L_1$ $=$ $L_2$). The two middle panels in each figure correspond to an equal mixture of multipolarities ($| \delta |$ $=$ $1$) for opposite signs of the mixing ratio, with the additional assumption that the initial and final states are identical ({\it i.e.}, $\delta_1$ $=$ $\delta_2$). See Sect.~\ref{sec:visual3} and end of Sect.~\ref{sec:phases}.

\begin{figure*}
\includegraphics[width=2\columnwidth]{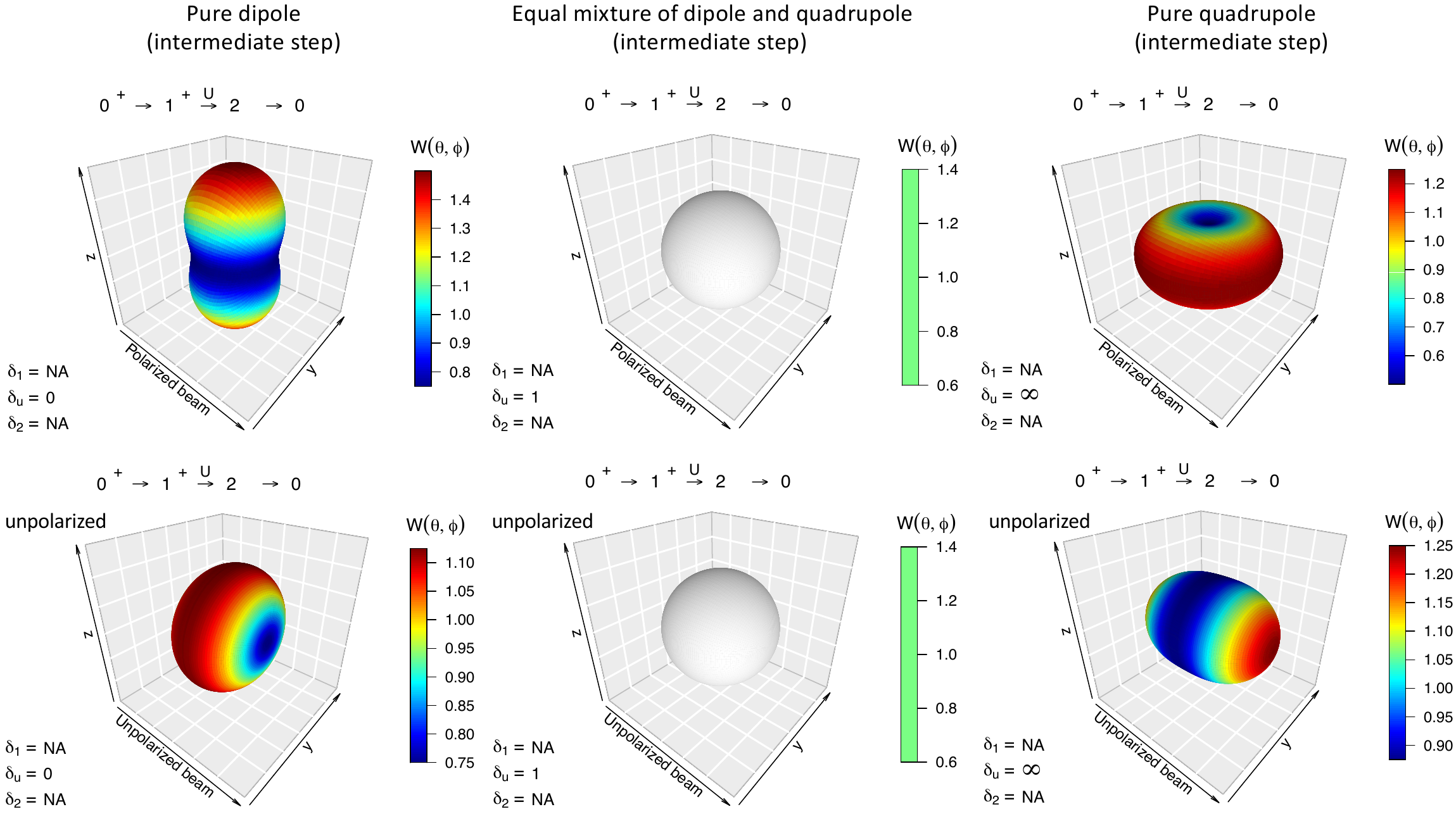} 
\caption{(Color online) Three-dimensional representations of angular correlations for the sequence $0^+$ $\rightarrow$ $1^+$ $\xrightarrow{\text{U}}$ $2$ $\rightarrow$ $0$. The symbol ``U'' indicates that the intermediate $\gamma$ ray in the sequence is unobserved. Notice the isotropic emission patterns in the middle row.
}
\label{fig:patt0120}
\end{figure*}

\begin{figure*}
\includegraphics[width=2\columnwidth]{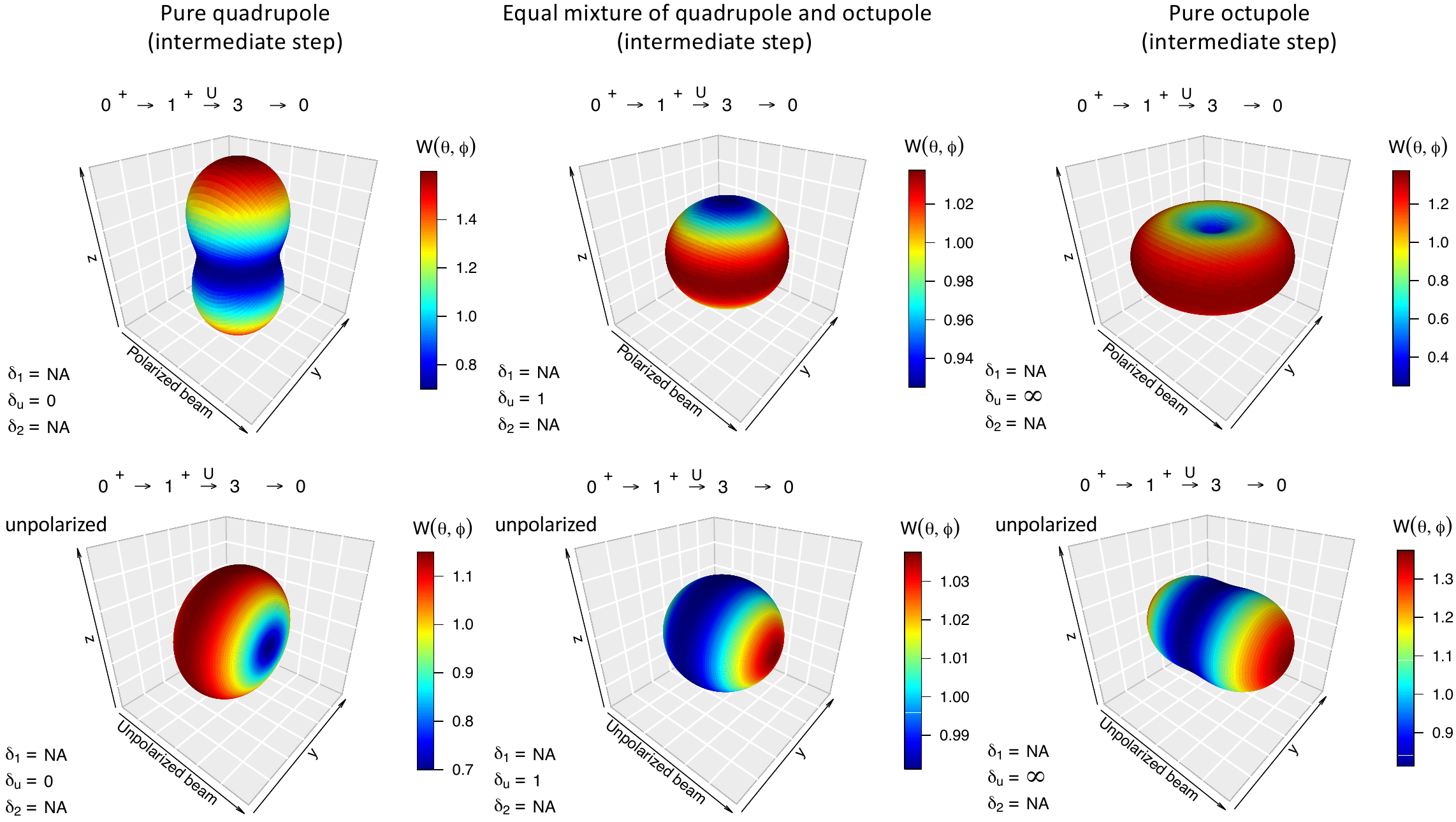} 
\caption{(Color online) Three-dimensional representations of angular correlations for the sequence $0^+$ $\rightarrow$ $1^+$ $\xrightarrow{\text{U}}$ $3$ $\rightarrow$ $0$. The symbol ``U'' indicates that the intermediate $\gamma$ ray in the sequence is unobserved.
}
\label{fig:patt0130}
\end{figure*}

\begin{figure*}
\includegraphics[width=2\columnwidth]{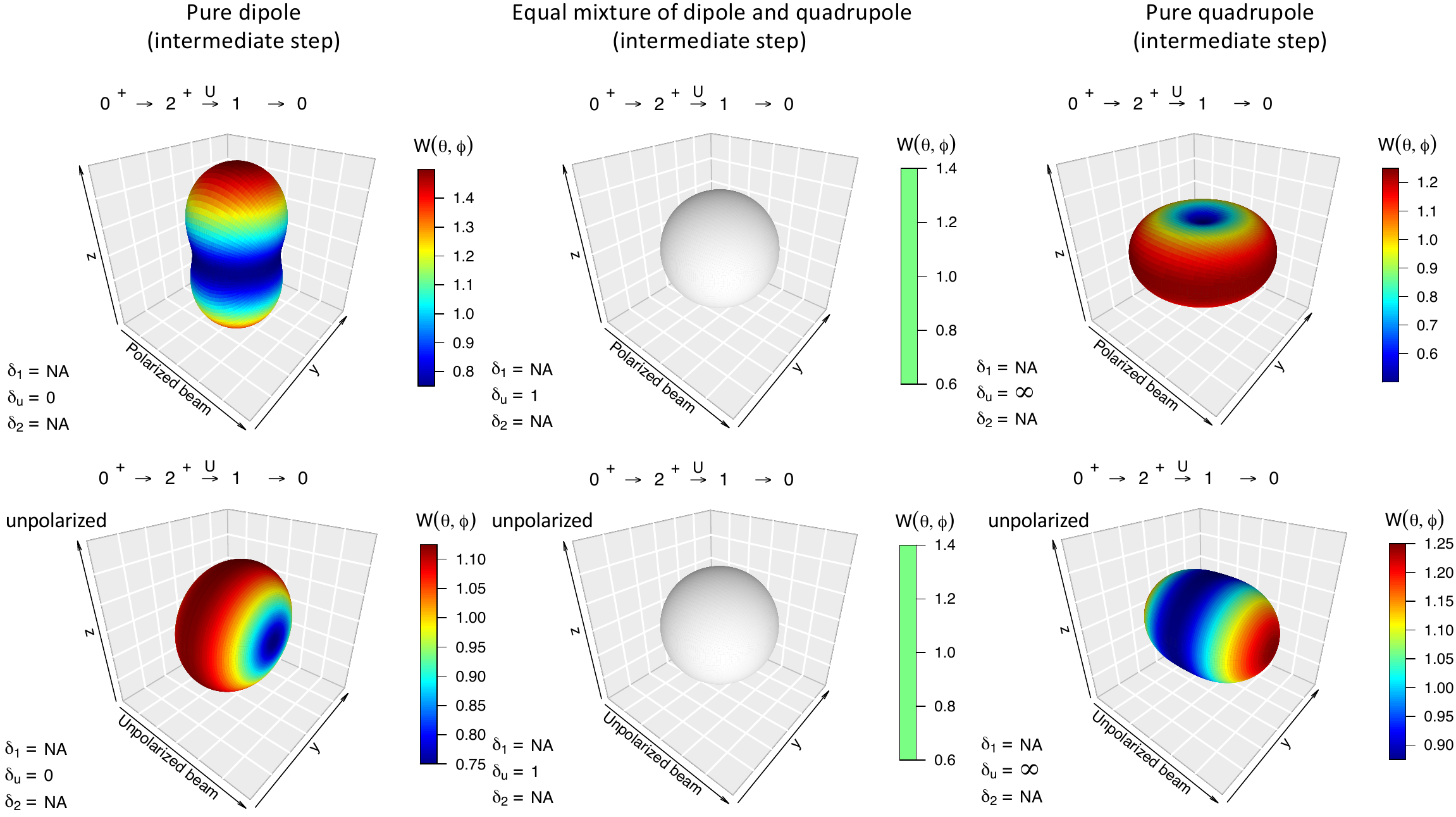} 
\caption{(Color online) Three-dimensional representations of angular correlations for the sequence $0^+$ $\rightarrow$ $2^+$ $\xrightarrow{\text{U}}$ $1$ $\rightarrow$ $0$. The symbol ``U'' indicates that the intermediate $\gamma$ ray in the sequence is unobserved. Notice the isotropic emission patterns in the middle row.
}
\label{fig:patt0210}
\end{figure*}

\begin{figure*}
\includegraphics[width=2\columnwidth]{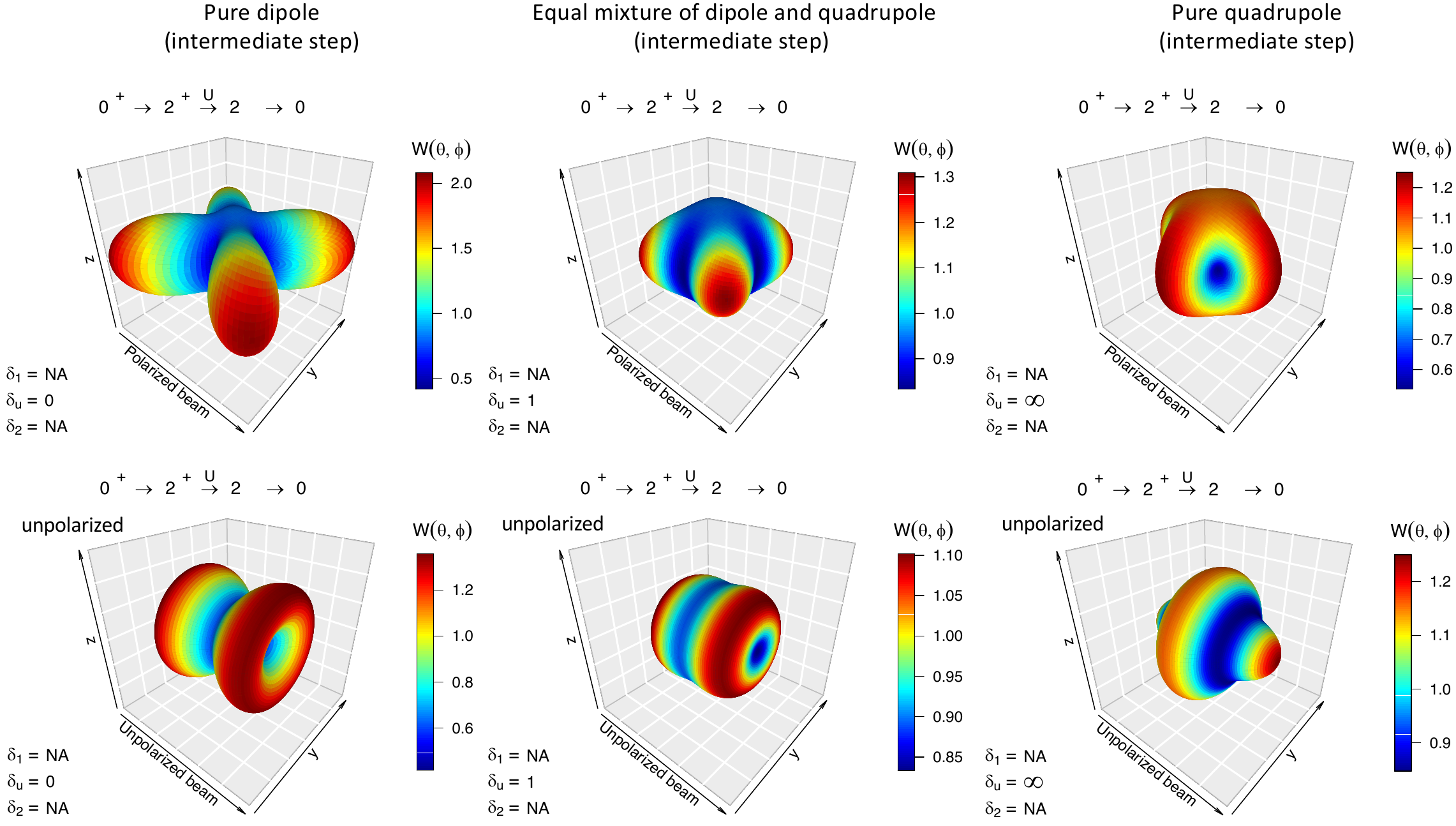} 
\caption{(Color online) Three-dimensional representations of angular correlations for the sequence $0^+$ $\rightarrow$ $2^+$ $\xrightarrow{\text{U}}$ $2$ $\rightarrow$ $0$. The symbol ``U'' indicates that the intermediate $\gamma$ ray in the sequence is unobserved.
}
\label{fig:patt0220}
\end{figure*}

\begin{figure*}
\includegraphics[width=2\columnwidth]{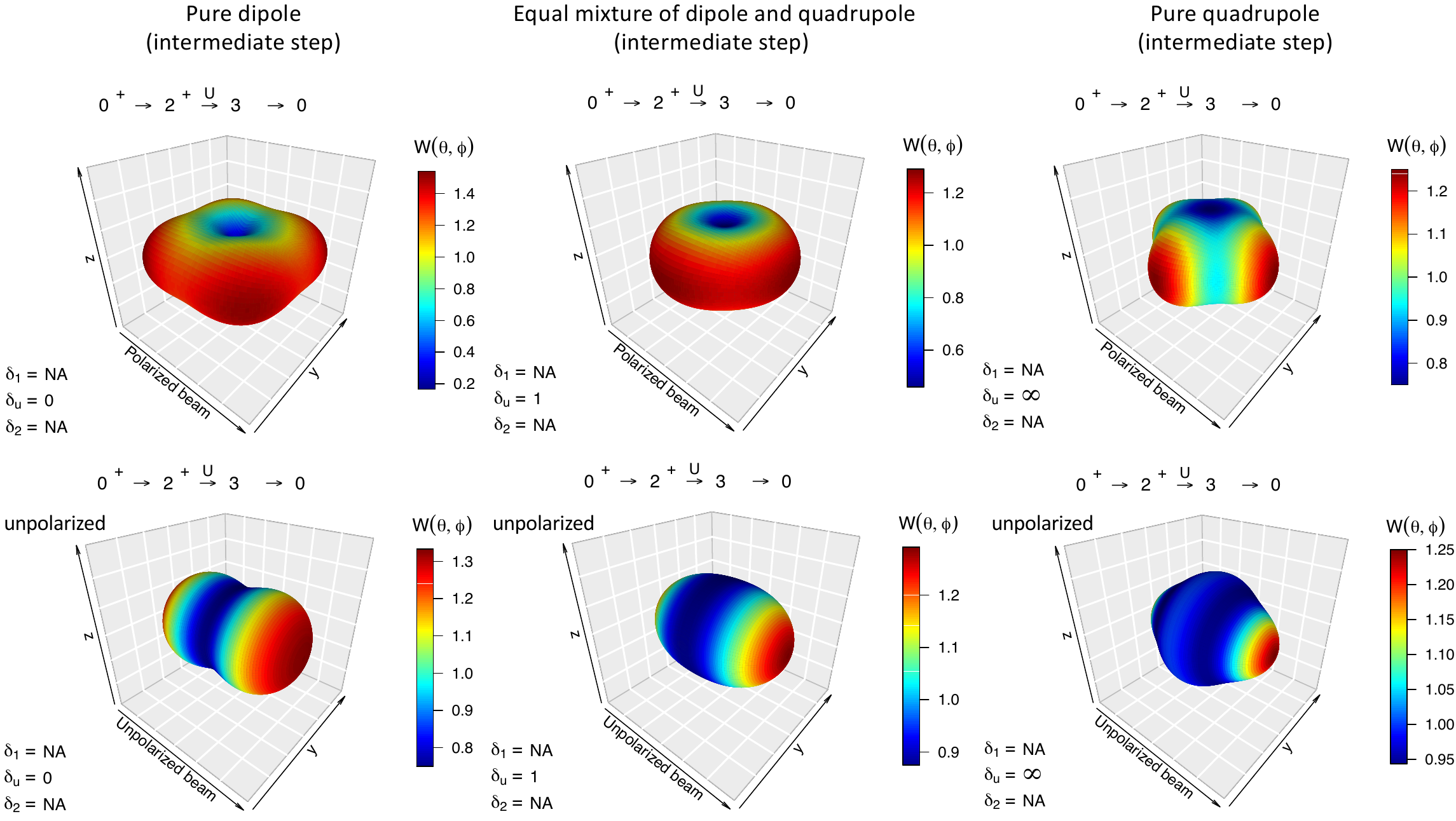} 
\caption{(Color online) Three-dimensional representations of angular correlations for the sequence $0^+$ $\rightarrow$ $2^+$ $\xrightarrow{\text{U}}$ $3$ $\rightarrow$ $0$. The symbol ``U'' indicates that the intermediate $\gamma$ ray in the sequence is unobserved.
}
\label{fig:patt0230}
\end{figure*}

Figures~\ref{fig:patt0120}--\ref{fig:patt0230} show three-dimensional emission patterns for the case that the intermediate $\gamma$ ray in the sequence is unobserved. The panels in the left, middle, and right columns correspond to pure, equally mixed, and pure radiations, respectively, for the intermediate $\gamma$ ray. The top and bottom rows show the linear polarization--direction ({\it i.e.}, with polarized beam) and direction--direction ({\it i.e.}, with unpolarized beam) correlations, respectively. For the symmetry properties of the patterns, see Sect.~\ref{sec:unobs}.


\section{Analyzing power graphs applicable to four--detector geometry}\label{app:powers}
To experimentally probe the three--dimensional angular correlation pattern, we suggested in Sect.~\ref{sec:results2} a detection geometry involving four counters at the locations indicated by the red circles in Fig.~\ref{fig:geo2}. This setup allows for the determination of two analyzing powers, $A(\theta = 90^\circ)$ and $A(\theta = 45^\circ)$, according to Eq.~(\ref{eq:analyzing}). The advantages of such a setup, including specific examples, have been discussed in Sects.~\ref{sec:single_mixing_ratio} and \ref{sec:double_mixing_ratio}. 

\begin{figure*}
\centering
\includegraphics[width=1.2\columnwidth]{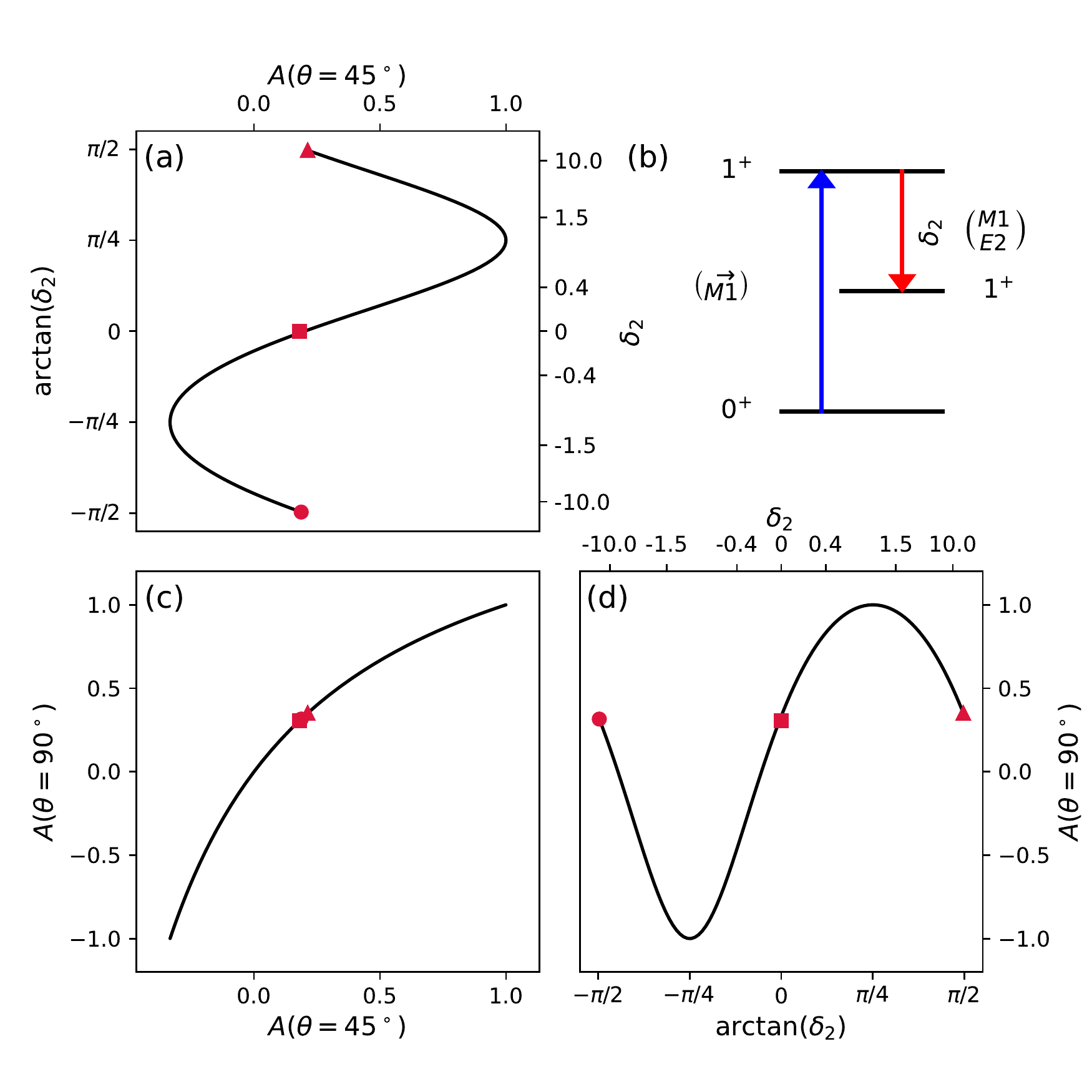} 
\caption{(Color online) Dependence of the analyzing powers, $A\left(\theta =  45^\circ \right)$ and $A \left(\theta = 90^\circ \right)$, on the multipolarity mixing ratio $\delta_2$ for the sequence $0^+$ $\rightarrow$ $1^+$ $\rightarrow$ $1$.
}
\label{fig:ana_011}
\end{figure*}

\begin{figure*}
\centering
\includegraphics[width=1.2\columnwidth]{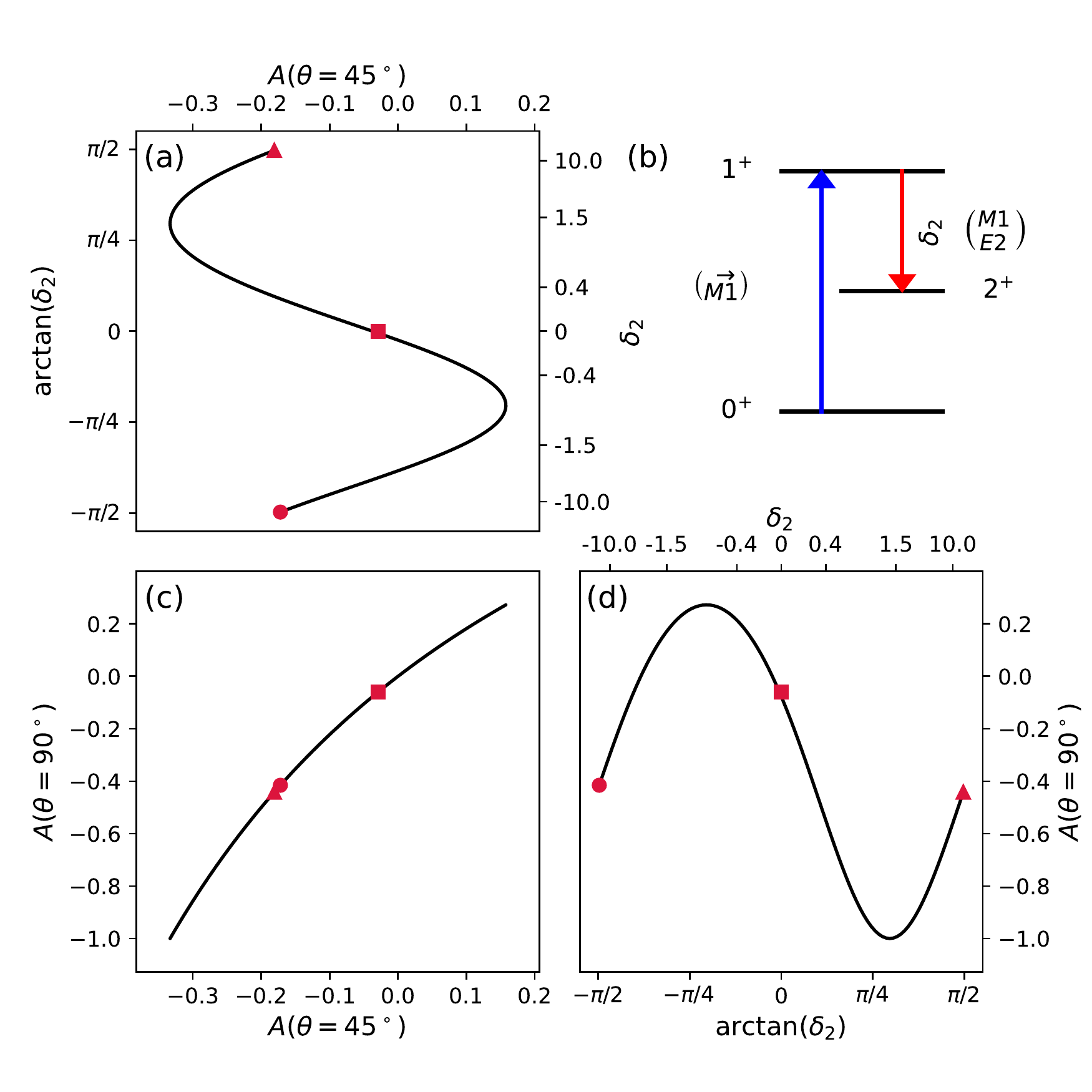} 
\caption{(Color online) Dependence of the analyzing powers, $A\left(\theta =  45^\circ \right)$ and $A \left(\theta = 90^\circ \right)$, on the multipolarity mixing ratio $\delta_2$ for the sequence $0^+$ $\rightarrow$ $1^+$ $\rightarrow$ $2$.
}
\label{fig:ana_012}
\end{figure*}

\begin{figure*}
\centering
\includegraphics[width=1.2\columnwidth]{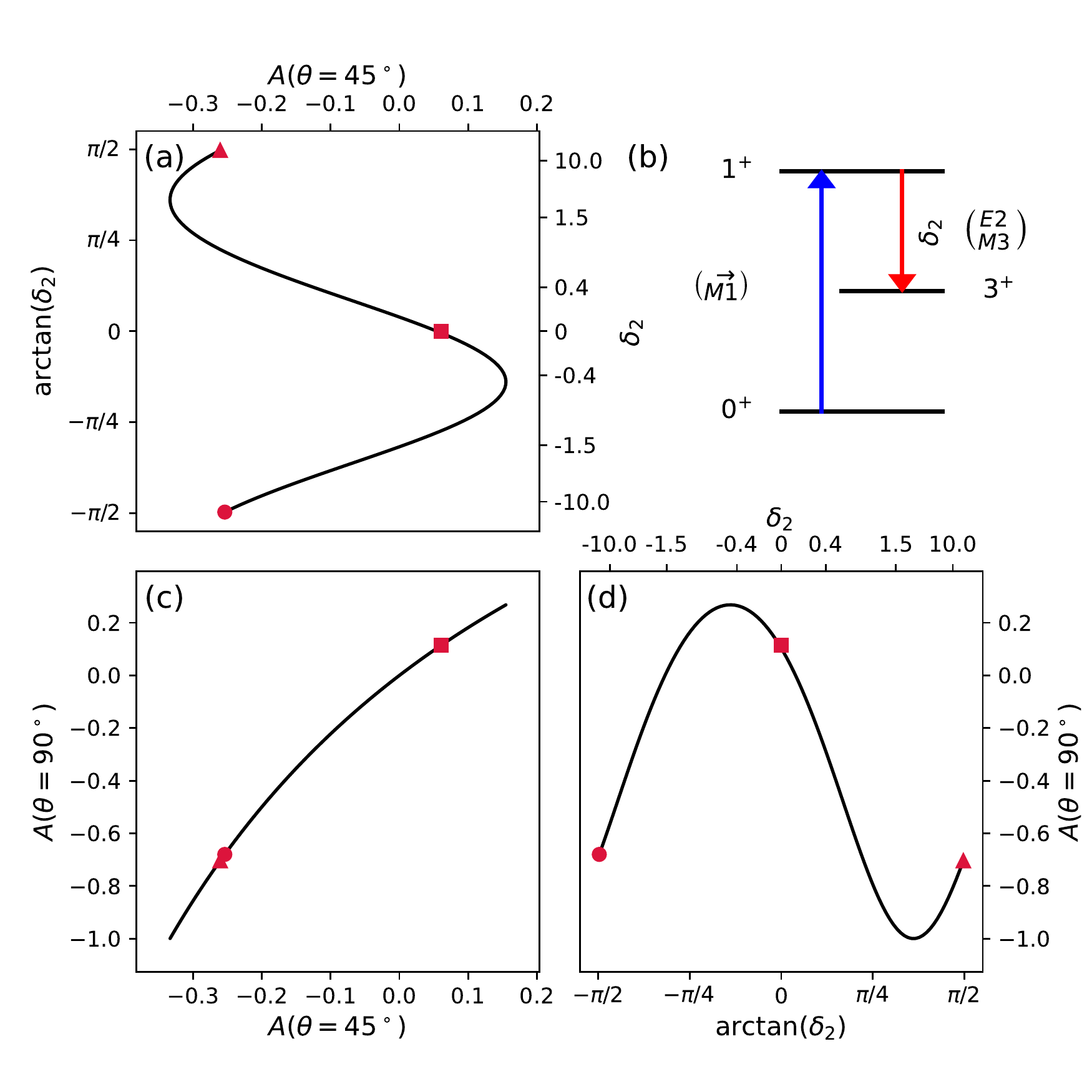} 
\caption{(Color online) Dependence of the analyzing powers, $A\left(\theta =  45^\circ \right)$ and $A \left(\theta = 90^\circ \right)$, on the multipolarity mixing ratio $\delta_2$ for the sequence $0^+$ $\rightarrow$ $1^+$ $\rightarrow$ $3$.
}
\label{fig:ana_013}
\end{figure*}

\begin{figure*}
\centering
\includegraphics[width=1.2\columnwidth]{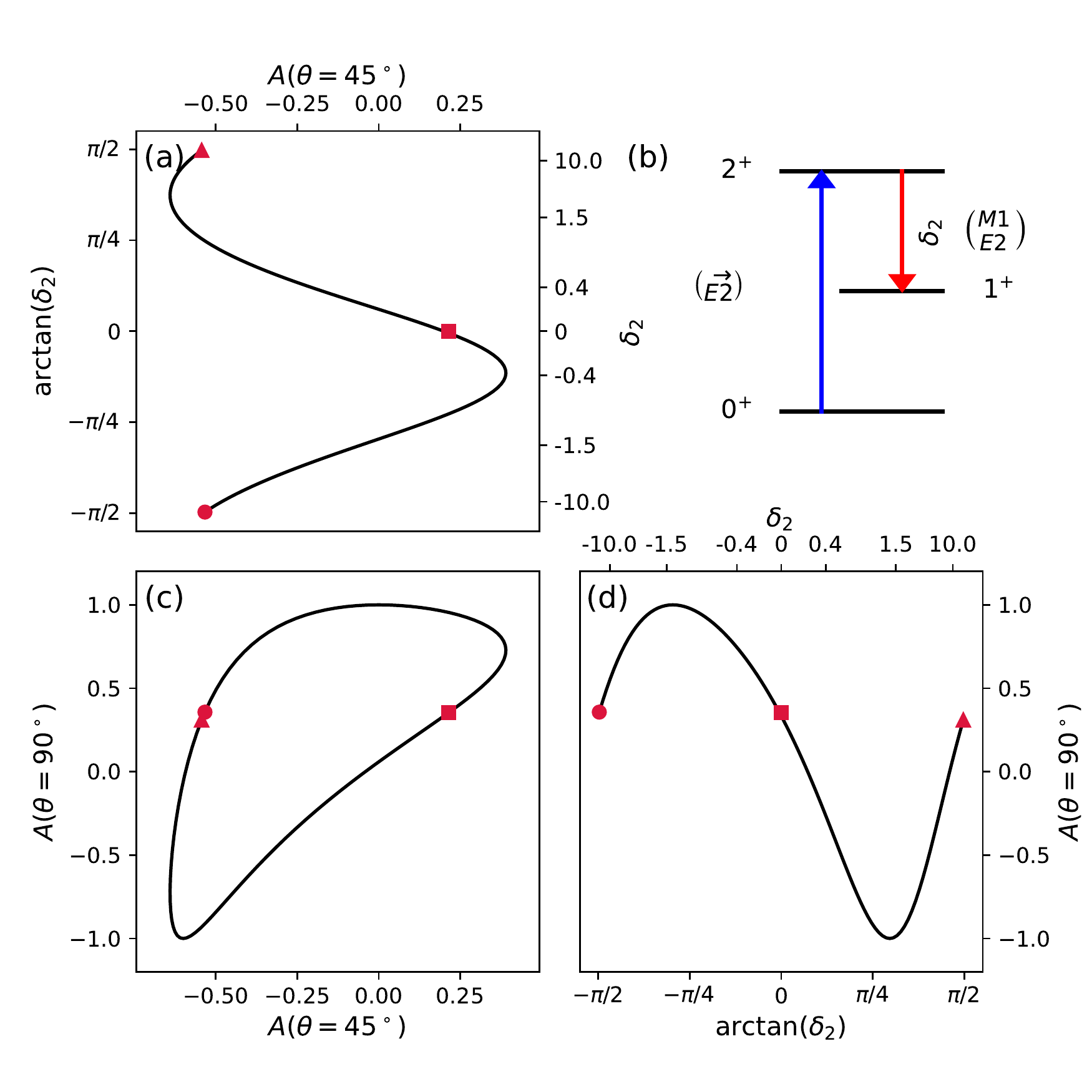} 
\caption{(Color online) Dependence of the analyzing powers, $A\left(\theta =  45^\circ \right)$ and $A \left(\theta = 90^\circ \right)$, on the multipolarity mixing ratio $\delta_2$ for the sequence $0^+$ $\rightarrow$ $2^+$ $\rightarrow$ $1$.
}
\label{fig:ana_021}
\end{figure*}

\begin{figure*}
\centering
\includegraphics[width=1.2\columnwidth]{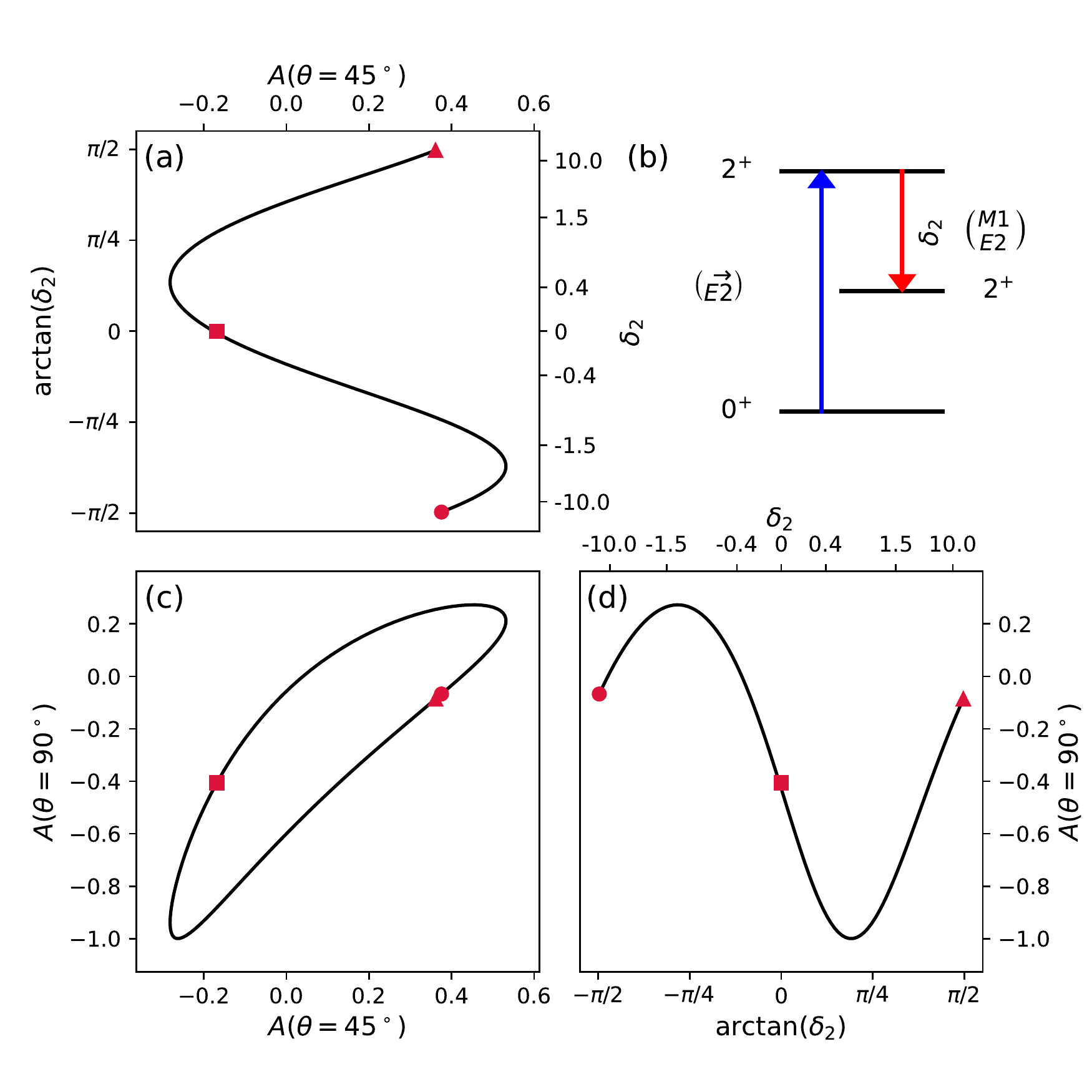} 
\caption{(Color online) Dependence of the analyzing powers, $A\left(\theta =  45^\circ \right)$ and $A \left(\theta = 90^\circ \right)$, on the multipolarity mixing ratio $\delta_2$ for the sequence $0^+$ $\rightarrow$ $2^+$ $\rightarrow$ $2$.
}
\label{fig:ana_022}
\end{figure*}

\begin{figure*}
\centering
\includegraphics[width=1.2\columnwidth]{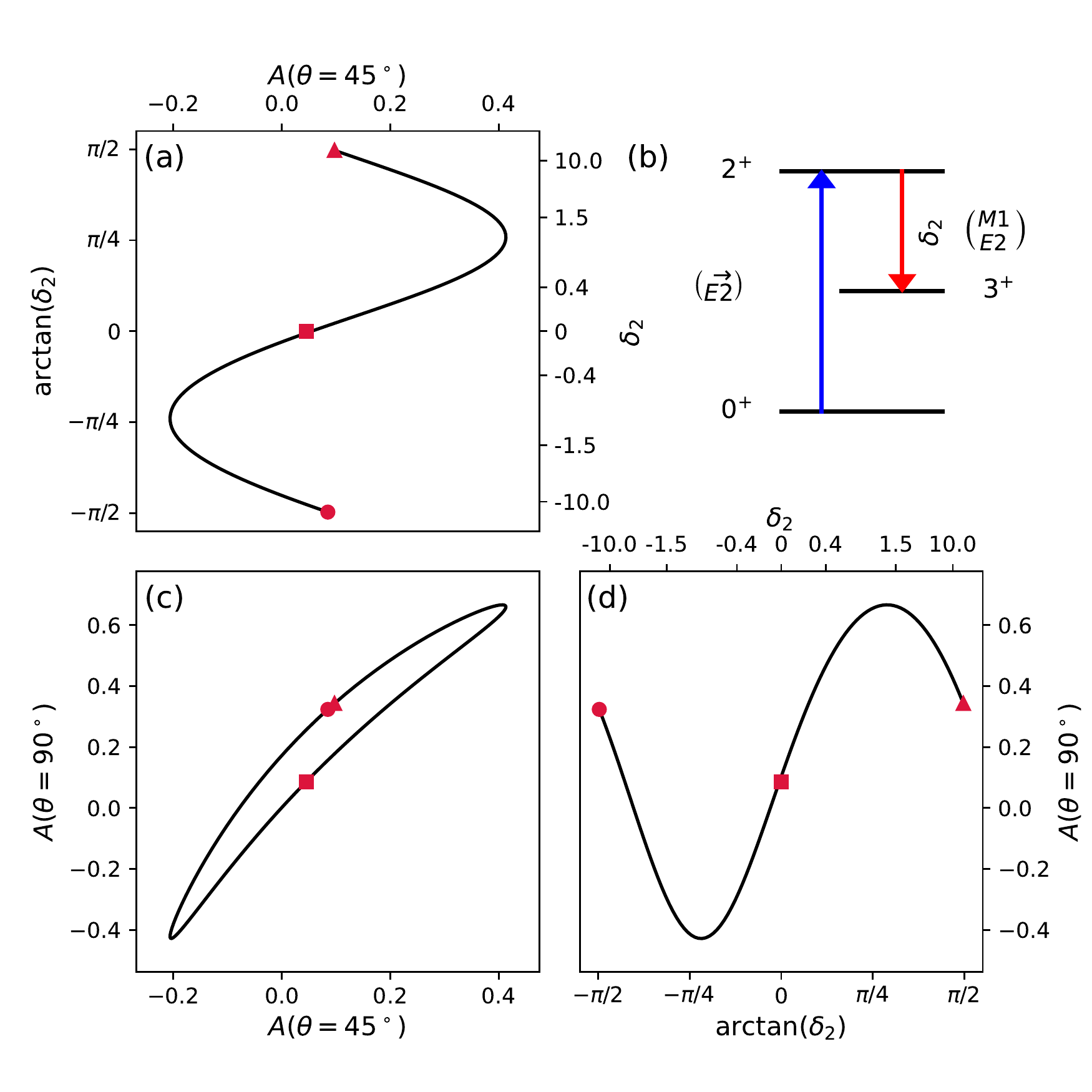} 
\caption{(Color online) Dependence of the analyzing powers, $A\left(\theta =  45^\circ \right)$ and $A \left(\theta = 90^\circ \right)$, on the multipolarity mixing ratio $\delta_2$ for the sequence $0^+$ $\rightarrow$ $2^+$ $\rightarrow$ $3$.
}
\label{fig:ana_023}
\end{figure*}

\begin{figure*}
\centering
\includegraphics[width=1.2\columnwidth]{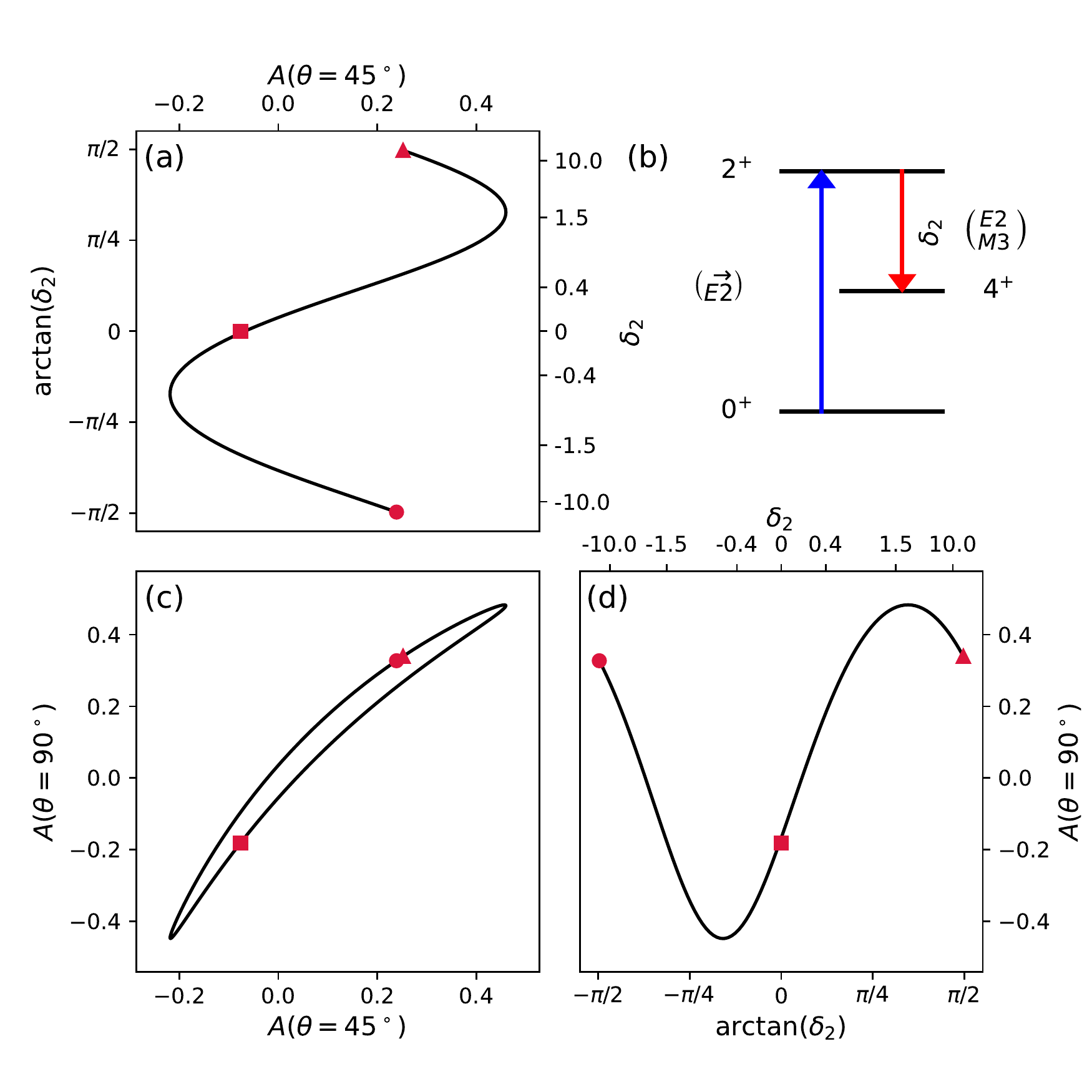} 
\caption{(Color online) Dependence of the analyzing powers, $A\left(\theta =  45^\circ \right)$ and $A \left(\theta = 90^\circ \right)$, on the multipolarity mixing ratio $\delta_2$ for the sequence $0^+$ $\rightarrow$ $2^+$ $\rightarrow$ $4$.
}
\label{fig:ana_024}
\end{figure*}

\begin{figure*}
\centering
\includegraphics[width=1.2\columnwidth]{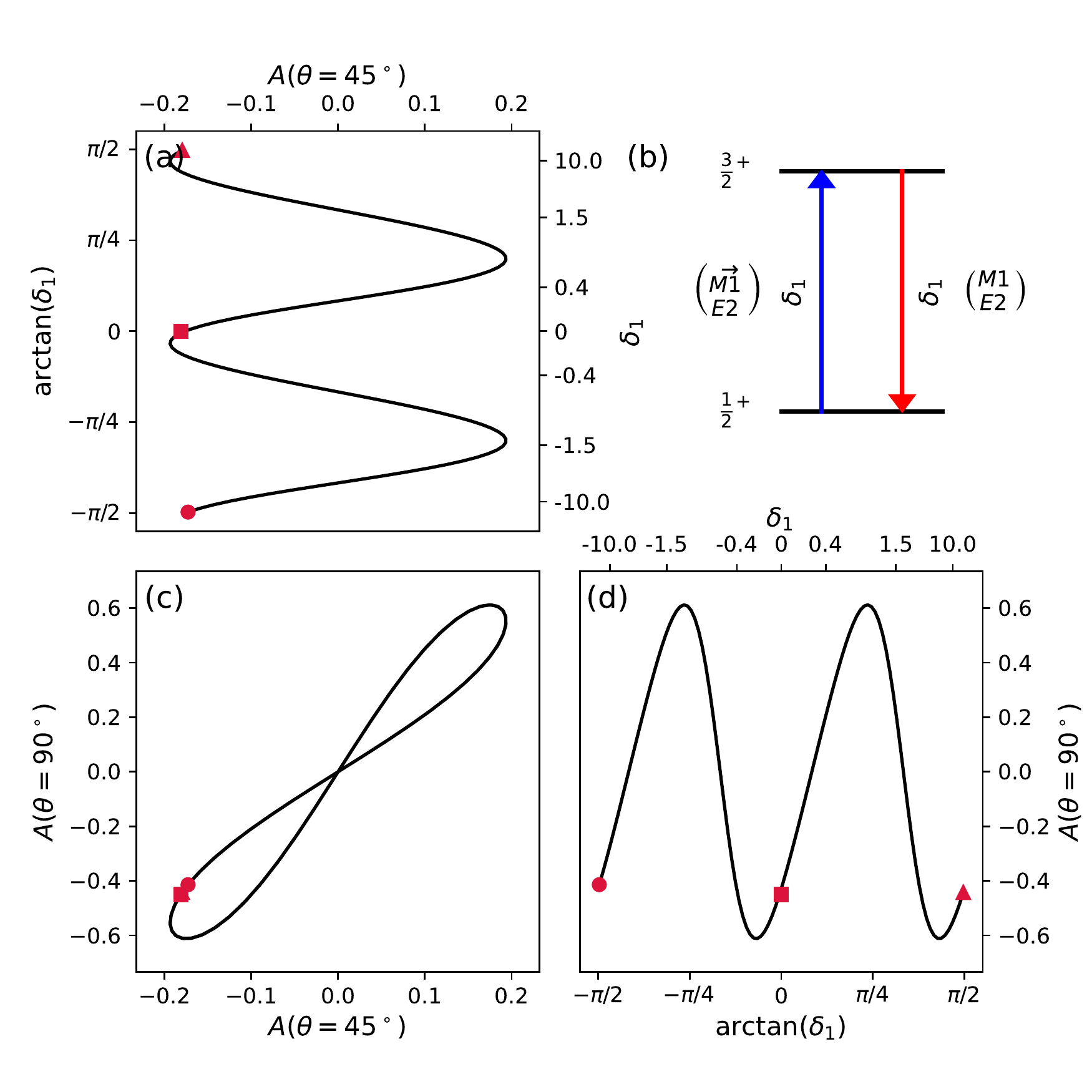} 
\caption{(Color online) Dependence of the analyzing powers, $A\left(\theta =  45^\circ \right)$ and $A \left(\theta = 90^\circ \right)$, on the multipolarity mixing ratio $\delta_1 = \delta_2$ for the elastic sequence $1/2^+$ $\rightarrow$ $3/2^+$ $\rightarrow$ $1/2$.
}
\label{fig:ana_051505}
\end{figure*}

\begin{figure*}
\centering
\includegraphics[width=1.2\columnwidth]{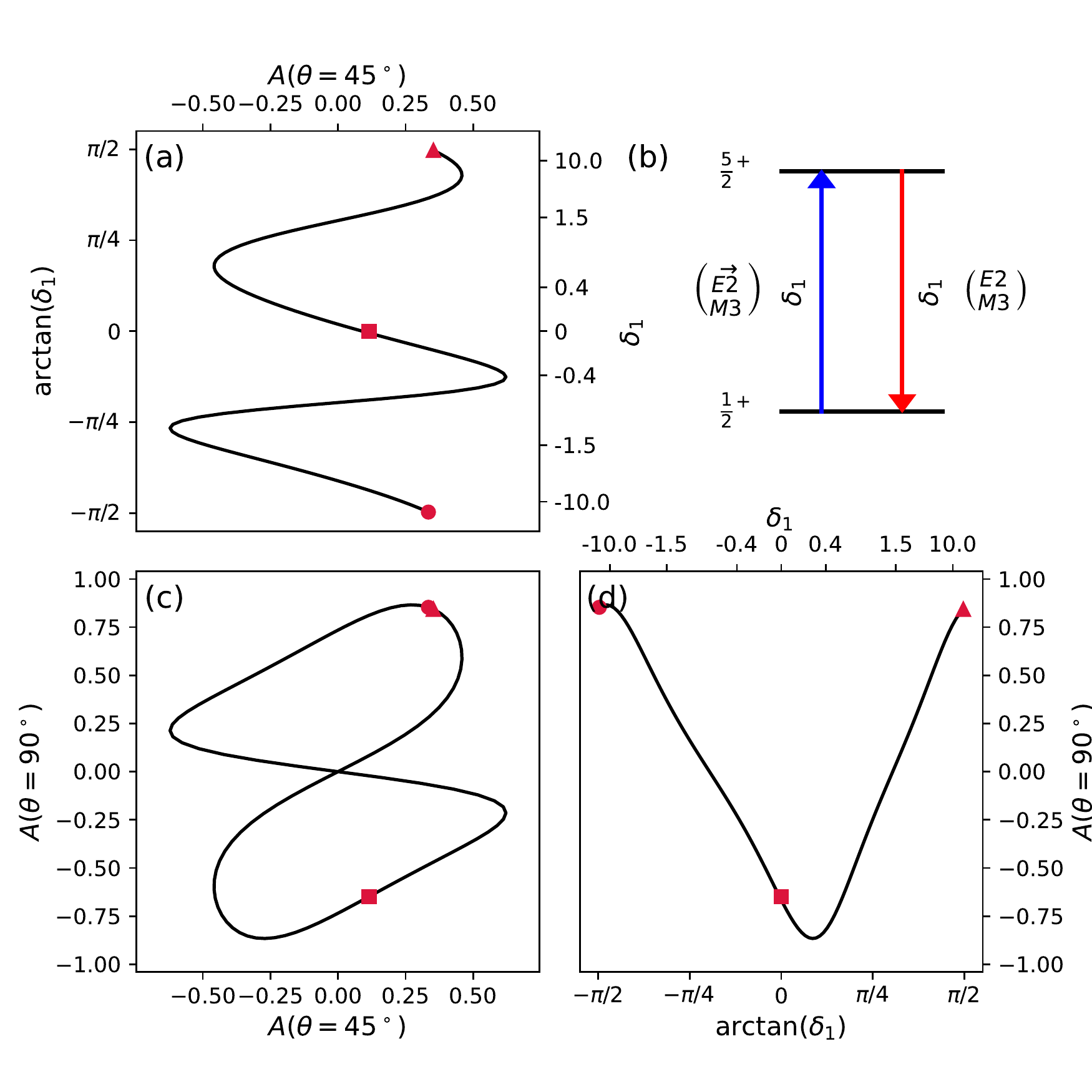} 
\caption{(Color online) Dependence of the analyzing powers, $A\left(\theta =  45^\circ \right)$ and $A \left(\theta = 90^\circ \right)$, on the multipolarity mixing ratio $\delta_1 = \delta_2$ for the elastic sequence $1/2^+$ $\rightarrow$ $5/2^+$ $\rightarrow$ $1/2$.
}
\label{fig:ana_052505}
\end{figure*}

\begin{figure*}
\centering
\includegraphics[width=1.2\columnwidth]{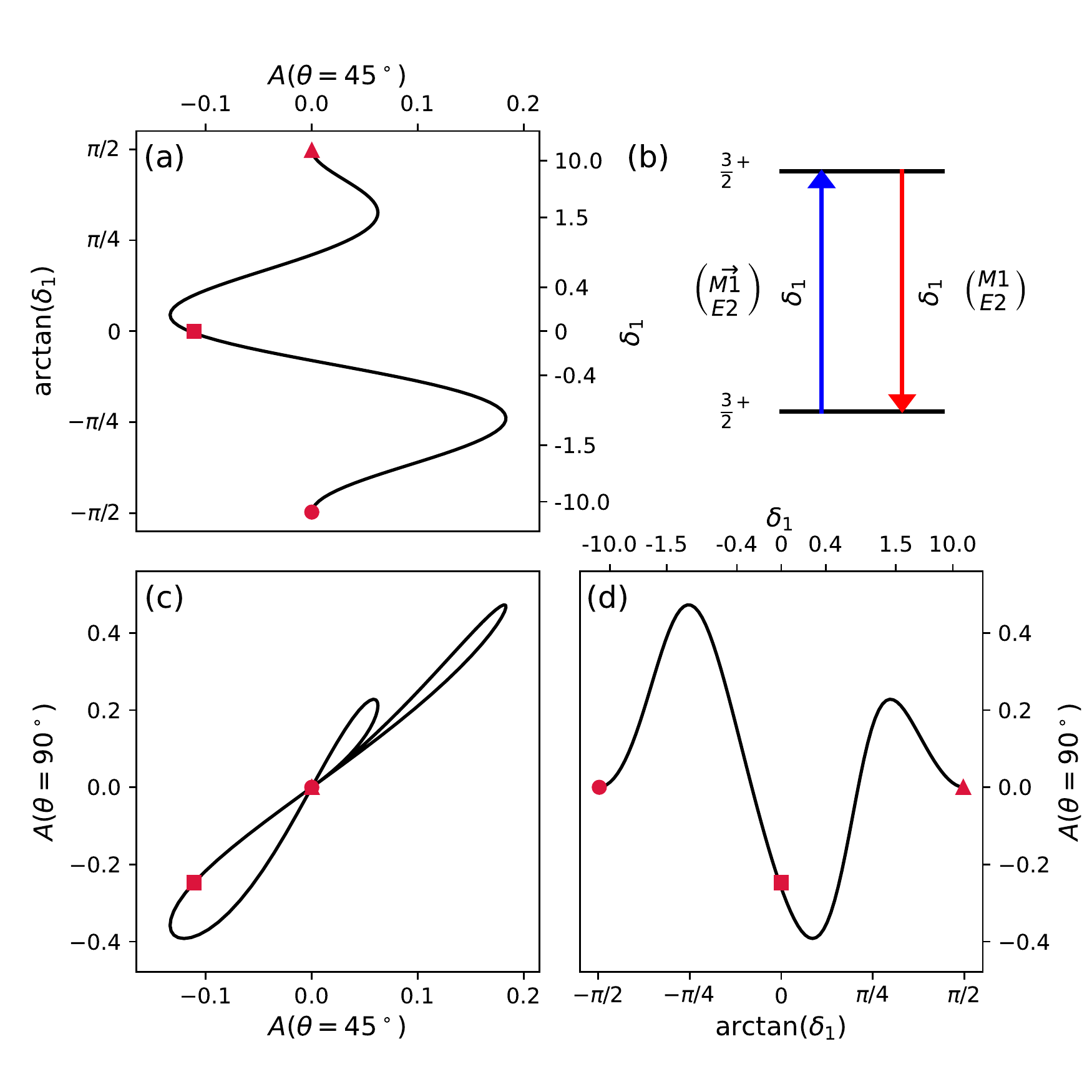} 
\caption{(Color online) Dependence of the analyzing powers, $A\left(\theta =  45^\circ \right)$ and $A \left(\theta = 90^\circ \right)$, on the multipolarity mixing ratio $\delta_1 = \delta_2$ for the elastic sequence $3/2^+$ $\rightarrow$ $3/2^+$ $\rightarrow $ $3/2$.
}
\label{fig:ana_151515}
\end{figure*}

\begin{figure*}
\centering
\includegraphics[width=1.2\columnwidth]{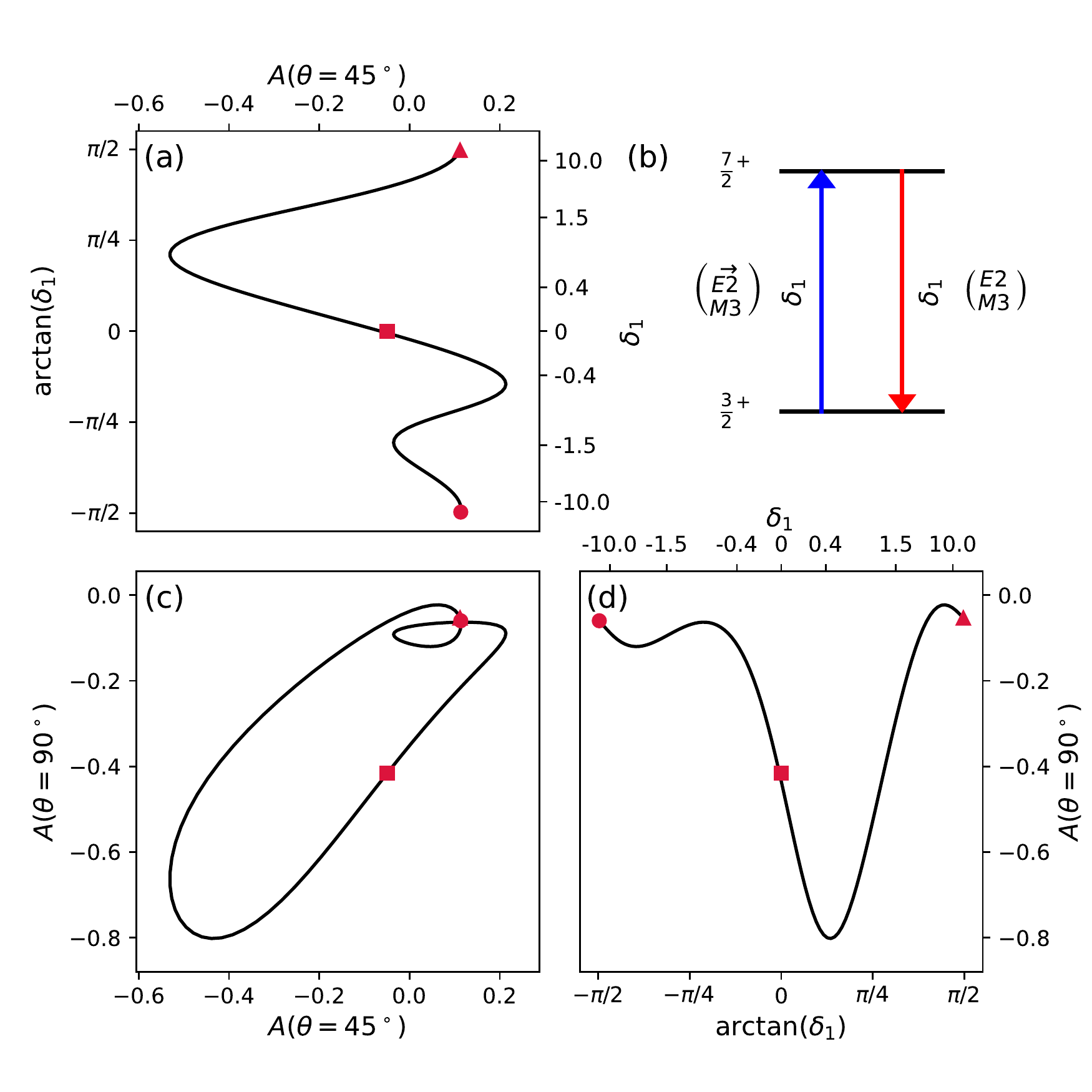} 
\caption{(Color online) Dependence of the analyzing powers, $A\left(\theta =  45^\circ \right)$ and $A \left(\theta = 90^\circ \right)$, on the multipolarity mixing ratio $\delta_1 = \delta_2$ for the elastic sequence $3/2^+$ $\rightarrow$ $7/2^+$ $\rightarrow $ $3/2$.
}
\label{fig:ana_153515}
\end{figure*}

\begin{figure*}
\centering
\includegraphics[width=1.2\columnwidth]{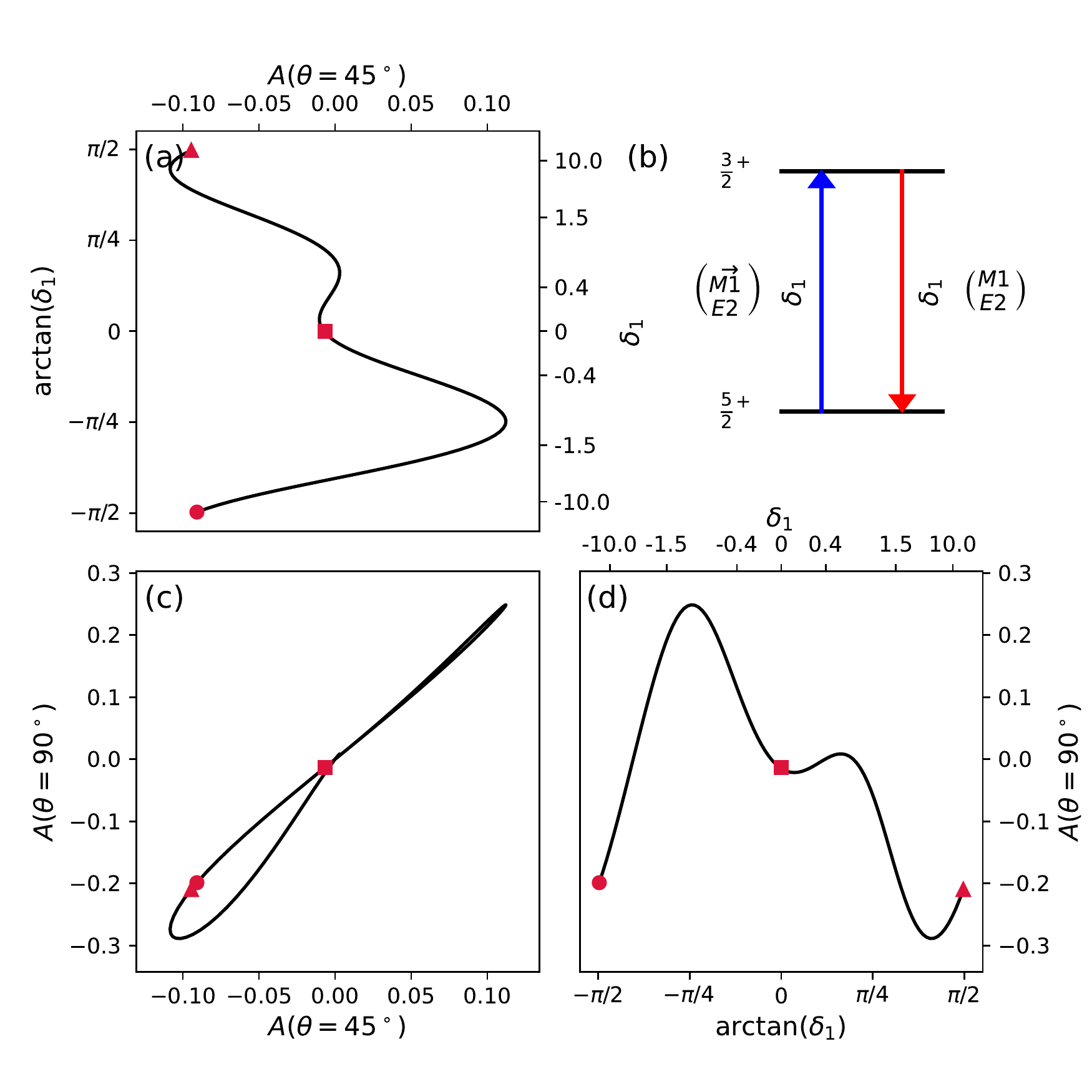} 
\caption{(Color online) Dependence of the analyzing powers, $A\left(\theta =  45^\circ \right)$ and $A \left(\theta = 90^\circ \right)$, on the multipolarity mixing ratio $\delta_1 = \delta_2$ for the elastic sequence $5/2^+$ $\rightarrow$ $3/2^+$ $\rightarrow $ $5/2$.
}
\label{fig:ana_251525}
\end{figure*}

\begin{figure*}
\centering
\includegraphics[width=1.2\columnwidth]{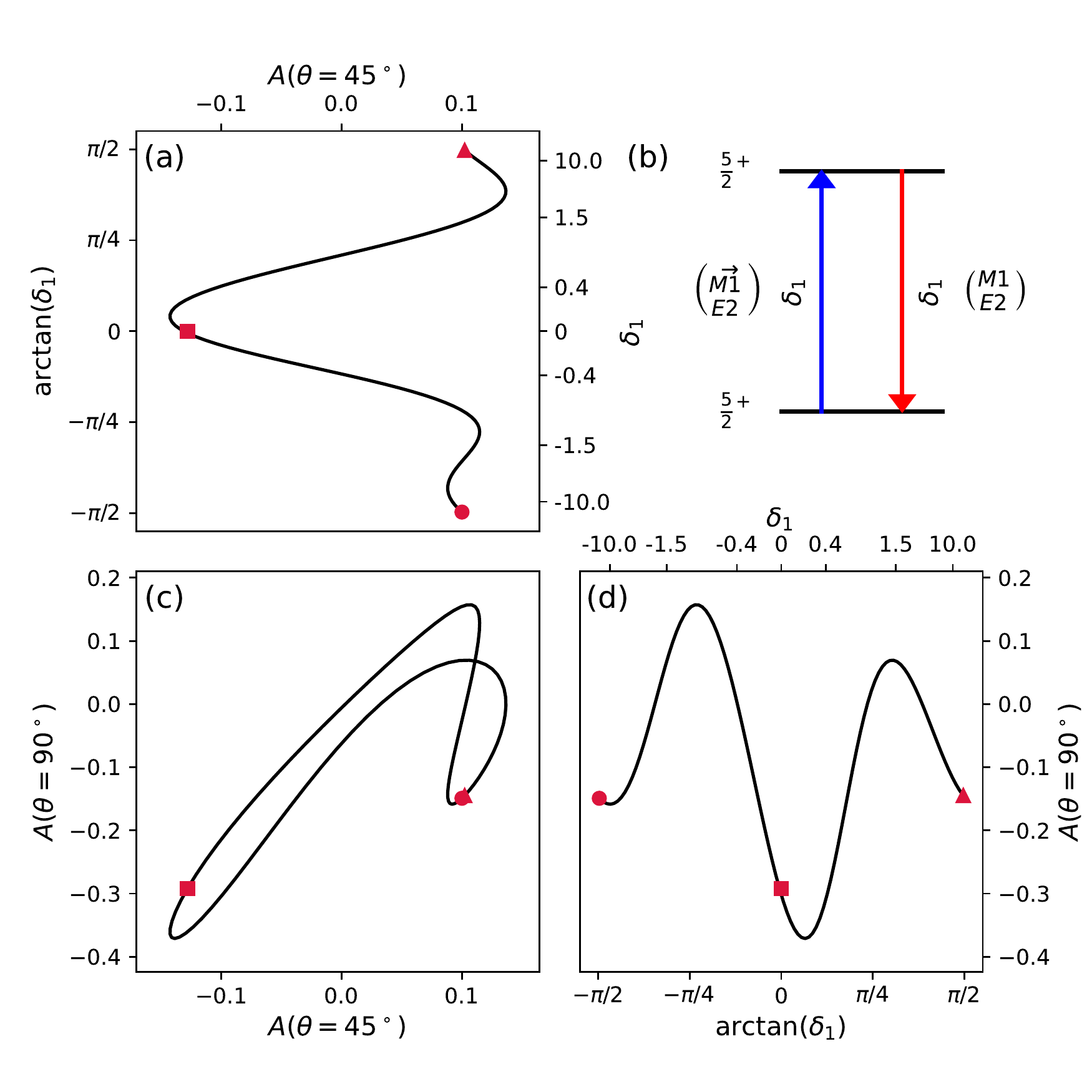} 
\caption{(Color online) Dependence of the analyzing powers, $A\left(\theta =  45^\circ \right)$ and $A \left(\theta = 90^\circ \right)$, on the multipolarity mixing ratio $\delta_1 = \delta_2$ for the elastic sequence $5/2^+$ $\rightarrow$ $5/2^+$ $\rightarrow $ $5/2$.
}
\label{fig:ana_252525}
\end{figure*}

\begin{figure*}
\centering
\includegraphics[width=1.2\columnwidth]{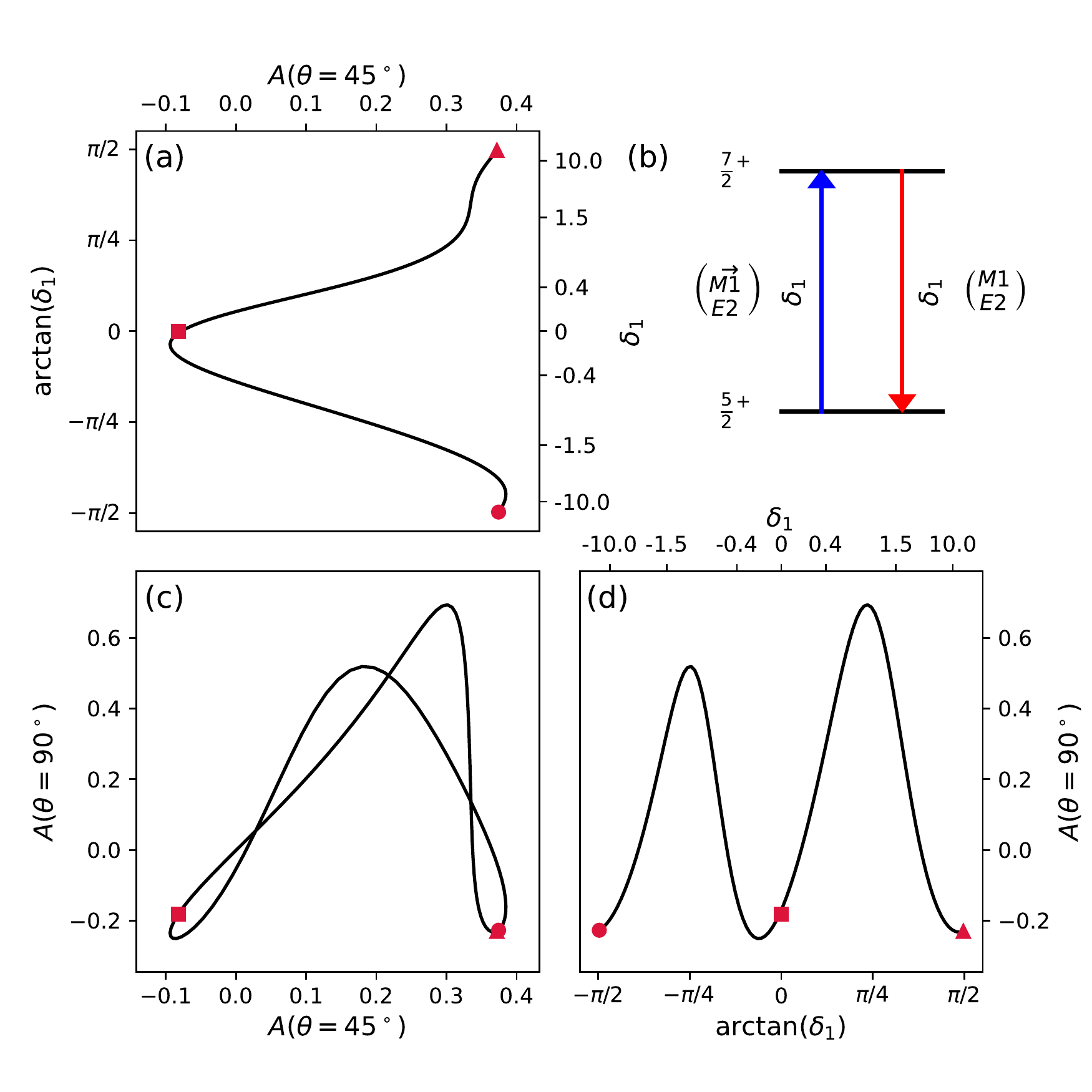} 
\caption{(Color online) Dependence of the analyzing powers, $A\left(\theta =  45^\circ \right)$ and $A \left(\theta = 90^\circ \right)$, on the multipolarity mixing ratio $\delta_1 = \delta_2$ for the elastic sequence $5/2^+$ $\rightarrow$ $7/2^+$ $\rightarrow $ $5/2$.
}
\label{fig:ana_253525}
\end{figure*}

\begin{figure*}
\centering
\includegraphics[width=1.2\columnwidth]{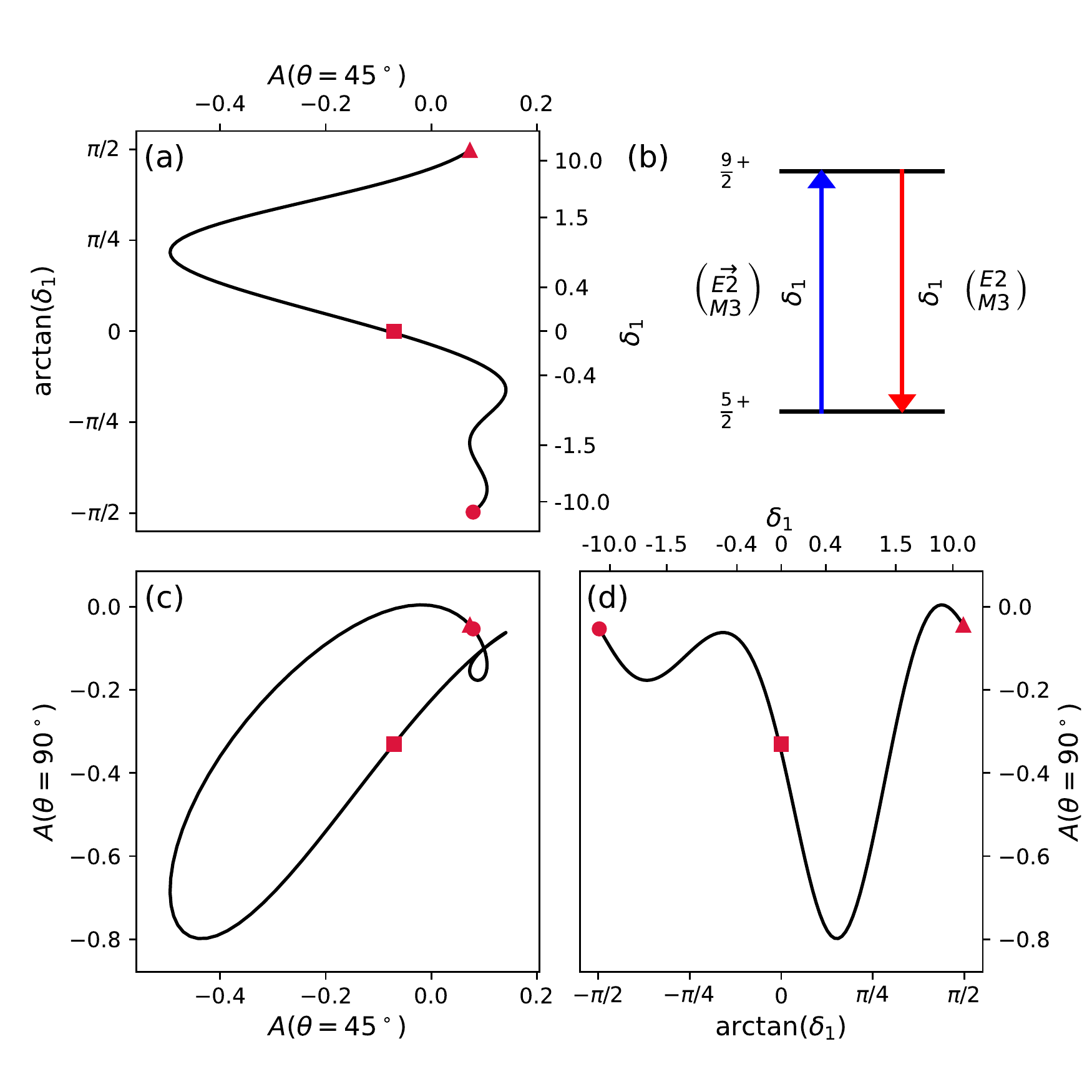} 
\caption{(Color online) Dependence of the analyzing powers, $A\left(\theta =  45^\circ \right)$ and $A \left(\theta = 90^\circ \right)$, on the multipolarity mixing ratio $\delta_1 = \delta_2$ for the elastic sequence $5/2^+$ $\rightarrow$ $9/2^+$ $\rightarrow $ $5/2$.
}
\label{fig:ana_254525}
\end{figure*}

In the following, we present graphs similar to Fig.~\ref{fig:analyzing_power_15_25_15} that will be helpful in the data analysis. All figures depict cases involving a single mixing ratio only, either because the initial state is $0^+$ (for even--mass nuclei), or because it is assumed that the initial and final states are identical (odd--mass nuclei). Figures~\ref{fig:ana_011}--\ref{fig:ana_254525} show analyzing powers for the spin-parity sequences listed in Table~\ref{tab:summary}. For an explanation of the figures, see the caption of Fig.~\ref{fig:analyzing_power_15_25_15}. The same figures are obtained for positive or negative parity, $\pi_2$, of the final state, $j_2$. The same figures are also obtained if the parities of the initial ($\pi_1$) and intermediate ($\pi$) states are flipped simultaneously.

In Sect.~\ref{sec:single_mixing_ratio}, we pointed out that both limits, $\delta$ $\to$ $+\infty$ and $\delta$ $\to$ $-\infty$, correspond to a transition which is pure in the higher of the two multipolarities. Since both limits of the mixing ratio give the same angular correlation pattern, the general form of a $A(\theta = 90^\circ)$ {\it vs.} $A(\theta = 45^\circ)$ graph is a closed loop. As an aid for the visualization, we added three red symbols to each figure. The circles, squares, and triangles indicate the values of $A(\theta = 90^\circ)$ and $A(\theta = 45^\circ)$ when $\delta$ $\to$ $-\infty$, $\delta$ $=$ $0$, and $\delta$ $\to$ $+\infty$, respectively.

For example, Fig.~\ref{fig:ana_021} depicts the analyzing powers for the sequence $0^+$ $\rightarrow$ $2^+$ $\rightarrow$ $1$. When varying the mixing ratio from $\delta_2$ $\to$ $-\infty$ to $\delta_2$ $\to$ $+\infty$, we start at the position of the red circle in panel (c), moving up and to the right along the loop, until reaching the red square ($\delta_2$ $=$ $0$), and continue towards the red triangle ($\delta_2$ $\to$ $+\infty$).  

Consider now Fig.~\ref{fig:patt011}, presenting angular correlation patterns for the sequence $0^+$ $\rightarrow$ $1^+$ $\rightarrow$ $1$. Notice that the patterns on the left ($\delta_2$ $=$ $0$) and right ($\delta_2$ $\to$ $\pm\infty$) are identical (see Sect.~\ref{sec:visual2} for an explanation). Adding more counters to the four-detector setup will not help to distinguish $\delta_2$ $=$ $0$ from $\delta_2$ $\to$ $\pm\infty$. This ambiguity causes the general shape of a two--dimensional loop to become a (one--dimensional) line in the $A(\theta = 90^\circ)$ {\it vs.} $A(\theta = 45^\circ)$ graph, as can be seen in Fig.~\ref{fig:ana_011}. 

In Fig.~\ref{fig:patt024}, the angular correlation patterns of the sequence $0^+$ $\rightarrow$ $2^+$ $\rightarrow$ $4$ for $\delta_2$ $=$ $+1$ (middle top panel) and $\delta_2$ $=$ $\pm\infty$ (right panel) are similar, but not identical. This causes the narrow shape of the two--dimensional loop in panel (c) of Fig.~\ref{fig:ana_024}. If those two angular correlation patterns would have been identical, the narrow loop would become a line in the $A(\theta = 90^\circ)$ {\it vs.} $A(\theta = 45^\circ)$ plane.

\begin{figure*}
\centering
\includegraphics[width=1.2\columnwidth]{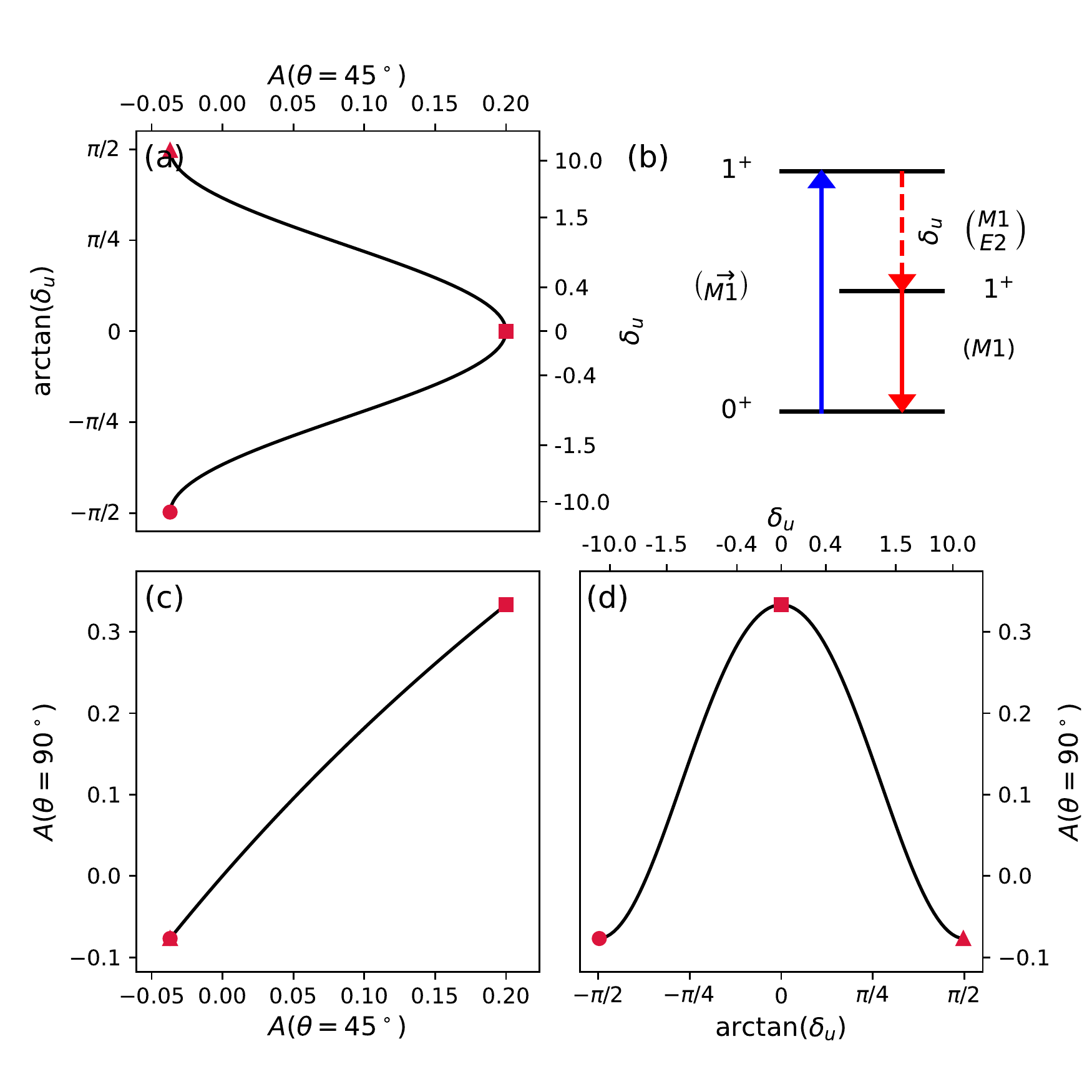} 
\caption{(Color online) Dependence of the analyzing powers, $A\left(\theta =  45^\circ \right)$ and $A \left(\theta = 90^\circ \right)$, on the multipolarity mixing ratio $\delta_u$ for the sequence $0^+$ $\rightarrow$ $1^+$ $\xrightarrow{\text{U}}$ $1$ $\rightarrow$ $0$. The symbol ``U'' indicates that the intermediate $\gamma$ ray in the sequence is unobserved.
}
\label{fig:ana_0110}
\end{figure*}

\begin{figure*}
\centering
\includegraphics[width=1.2\columnwidth]{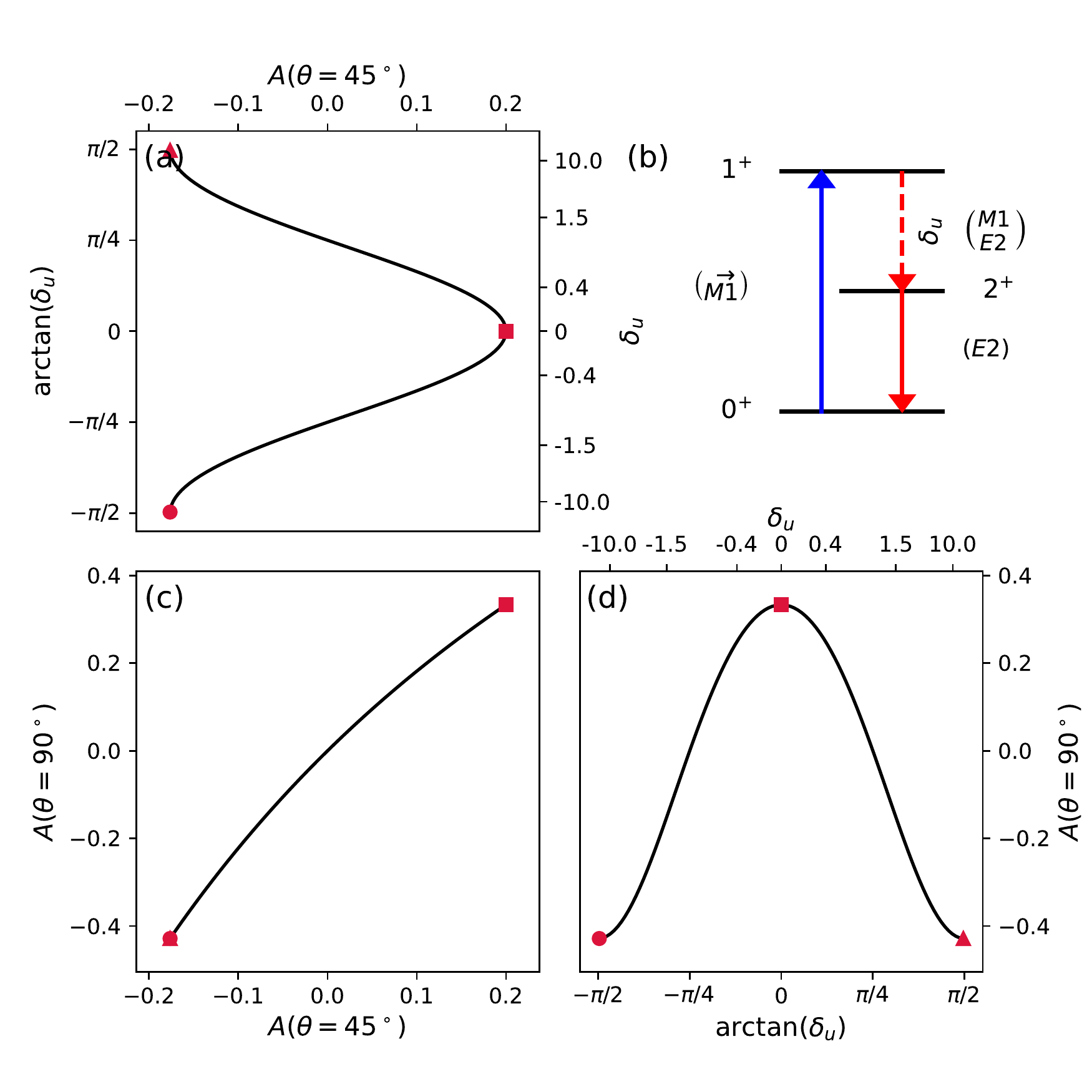} 
\caption{(Color online) Dependence of the analyzing powers, $A\left(\theta =  45^\circ \right)$ and $A \left(\theta = 90^\circ \right)$, on the multipolarity mixing ratio $\delta_u$ for the sequence $0^+$ $\rightarrow$ $1^+$ $\xrightarrow{\text{U}}$ $2$ $\rightarrow$ $0$. The symbol ``U'' indicates that the intermediate $\gamma$ ray in the sequence is unobserved.
}
\label{fig:ana_0120}
\end{figure*}

\begin{figure*}
\centering
\includegraphics[width=1.2\columnwidth]{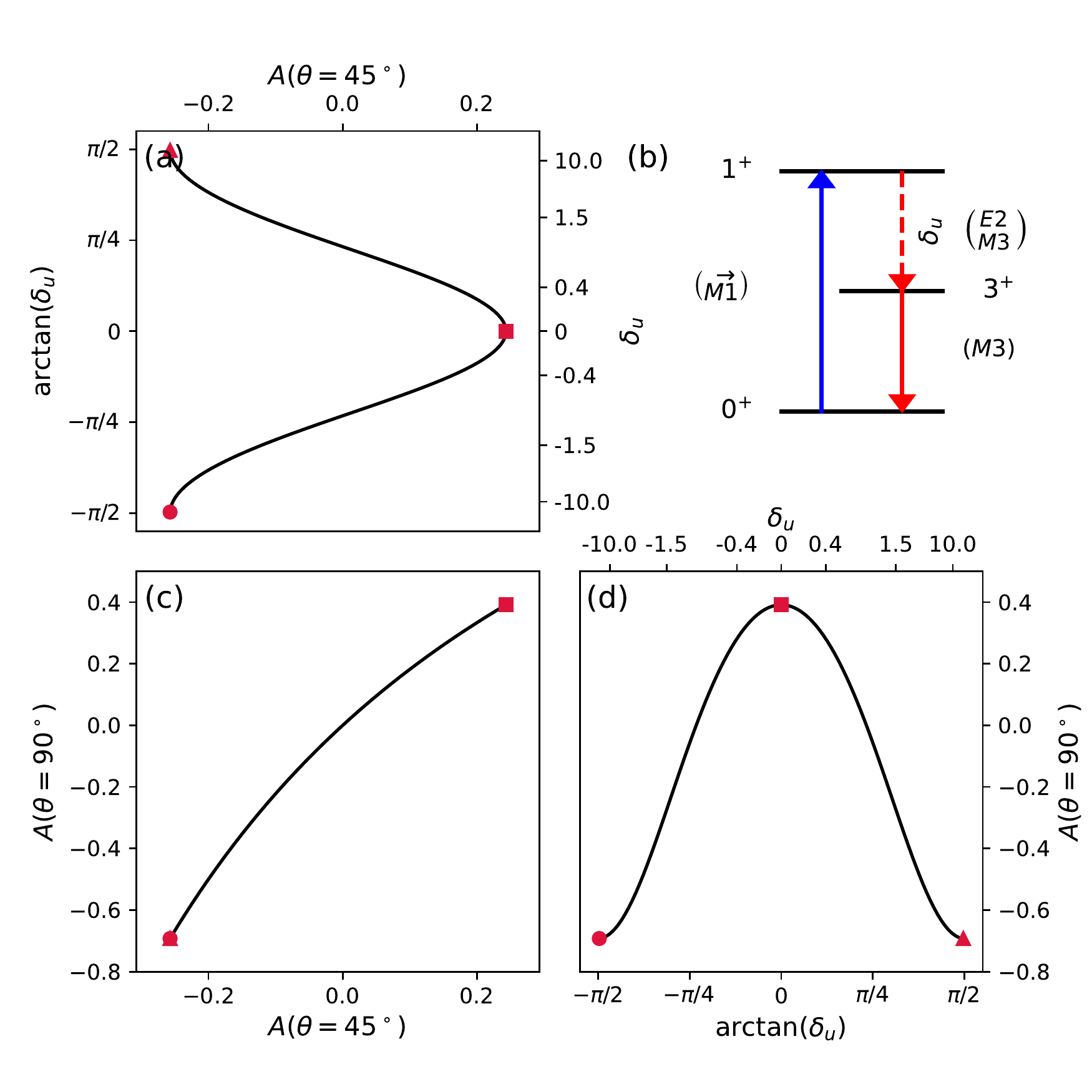} 
\caption{(Color online) Dependence of the analyzing powers, $A\left(\theta =  45^\circ \right)$ and $A \left(\theta = 90^\circ \right)$, on the multipolarity mixing ratio $\delta_u$ for the sequence $0^+$ $\rightarrow$ $1^+$ $\xrightarrow{\text{U}}$ $3$ $\rightarrow$ $0$. The symbol ``U'' indicates that the intermediate $\gamma$ ray in the sequence is unobserved.
}
\label{fig:ana_0130}
\end{figure*}

\begin{figure*}
\centering
\includegraphics[width=1.2\columnwidth]{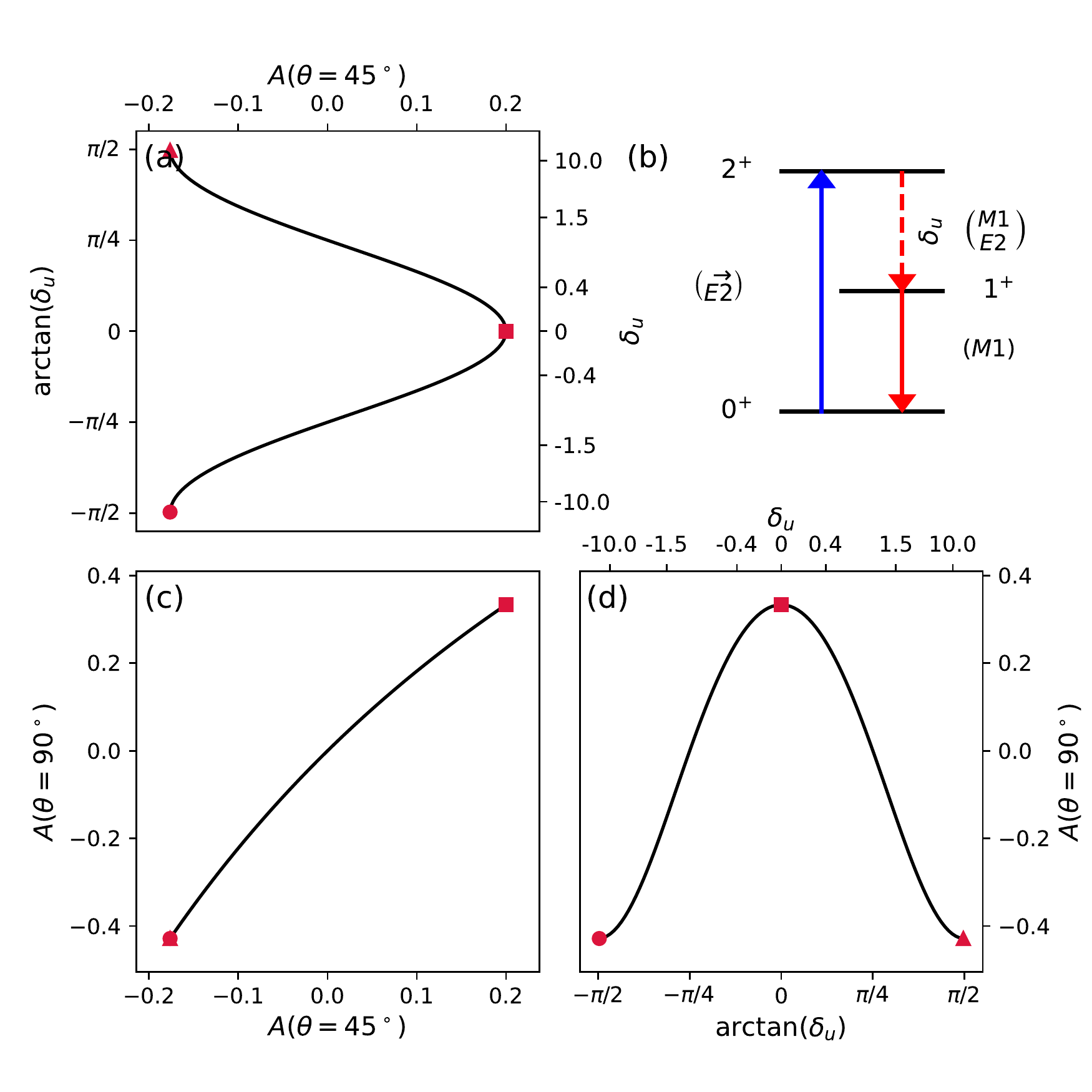} 
\caption{(Color online) Dependence of the analyzing powers, $A\left(\theta =  45^\circ \right)$ and $A \left(\theta = 90^\circ \right)$, on the multipolarity mixing ratio $\delta_u$ for the sequence $0^+$ $\rightarrow$ $2^+$ $\xrightarrow{\text{U}}$ $1$ $\rightarrow$ $0$. The symbol ``U'' indicates that the intermediate $\gamma$ ray in the sequence is unobserved.
}
\label{fig:ana_0210}
\end{figure*}

\begin{figure*}
\centering
\includegraphics[width=1.2\columnwidth]{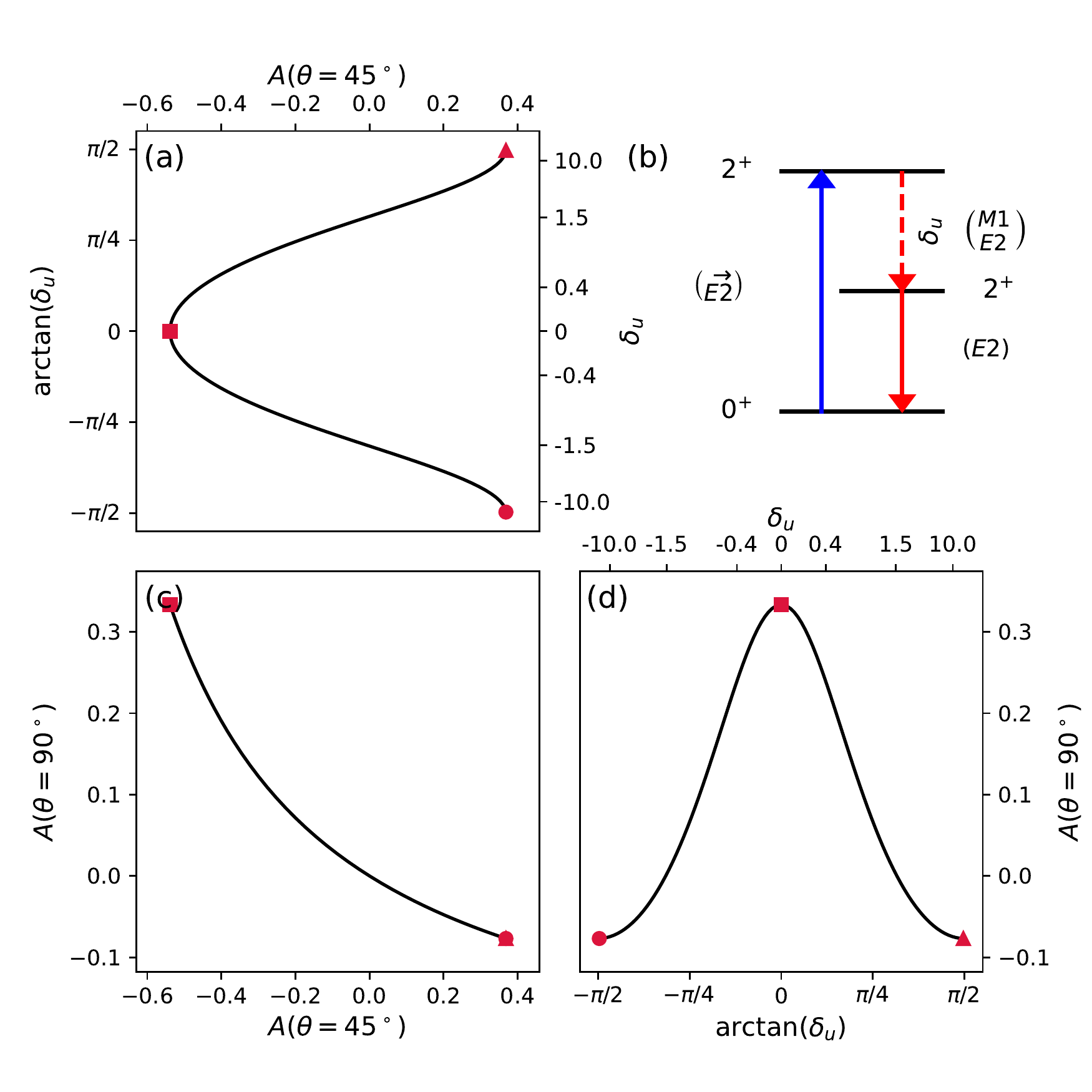} 
\caption{(Color online) Dependence of the analyzing powers, $A\left(\theta =  45^\circ \right)$ and $A \left(\theta = 90^\circ \right)$, on the multipolarity mixing ratio $\delta_u$ for the sequence $0^+$ $\rightarrow$ $2^+$ $\xrightarrow{\text{U}}$ $2$ $\rightarrow$ $0$. The symbol ``U'' indicates that the intermediate $\gamma$ ray in the sequence is unobserved.
}
\label{fig:ana_0220}
\end{figure*}

\begin{figure*}
\centering
\includegraphics[width=1.2\columnwidth]{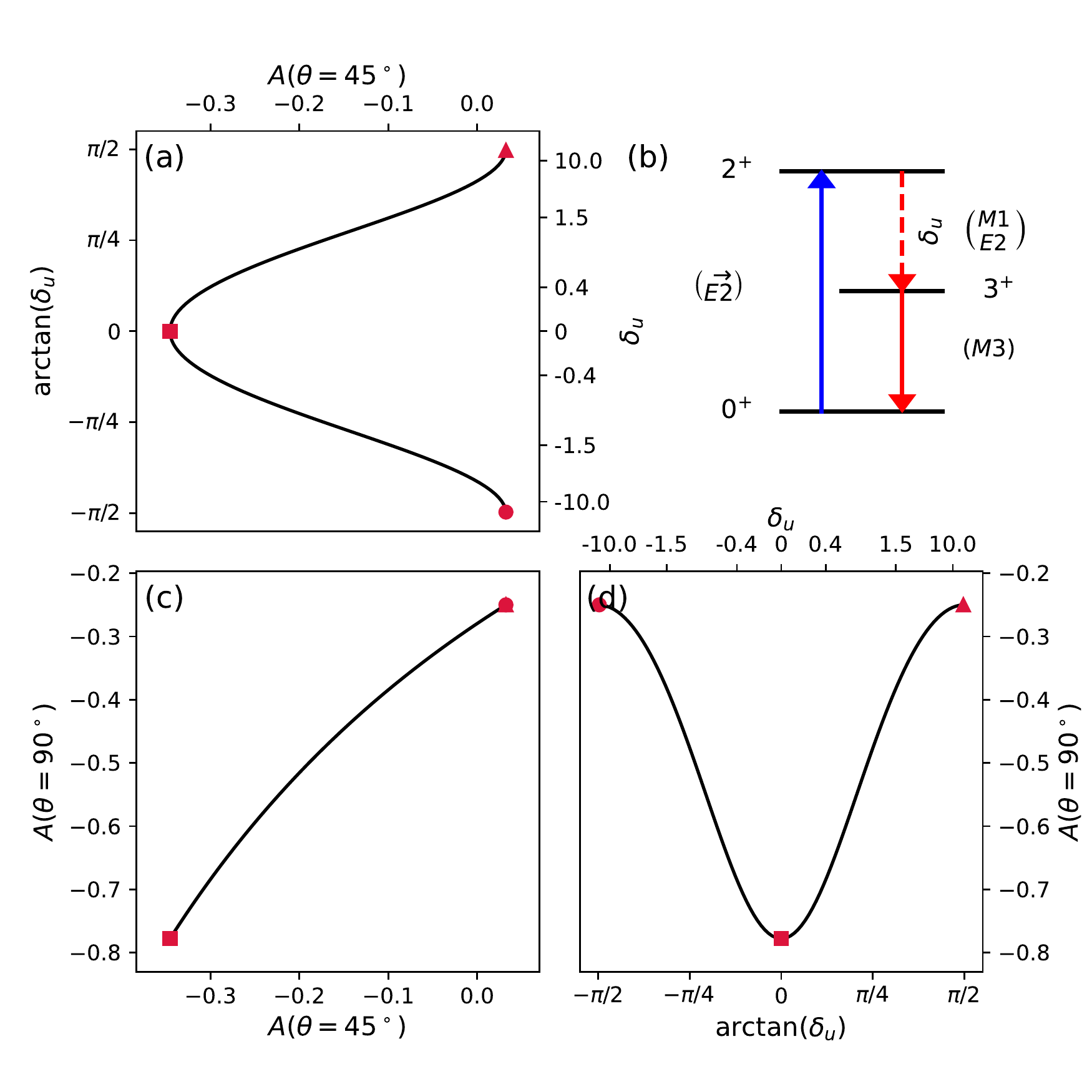} 
\caption{(Color online) Dependence of the analyzing powers, $A\left(\theta =  45^\circ \right)$ and $A \left(\theta = 90^\circ \right)$, on the multipolarity mixing ratio $\delta_u$ for the sequence $0^+$ $\rightarrow$ $2^+$ $\xrightarrow{\text{U}}$ $3$ $\rightarrow$ $0$. The symbol ``U'' indicates that the intermediate $\gamma$ ray in the sequence is unobserved.
}
\label{fig:ana_0230}
\end{figure*}

Figures~\ref{fig:ana_0110}--\ref{fig:ana_0230} show the analyzing powers for the case that the intermediate $\gamma$ ray in the sequence is unobserved. Note that the graphs in panels (c) depict the same values for a positive or negative sign of the mixing ratio, $\delta_u$, because the multipolarities of the unobserved intermediate radiation mix {\it incoherently} (see Eq.~(\ref{eq:wtotalu})). For the symmetry properties of the patterns, see Sect.~\ref{sec:unobs}.

%

\end{document}